\documentclass[12pt]{article}
\usepackage{a41}
\usepackage{eucal}
\usepackage{cite}
\usepackage{epsfig}
\usepackage{amsmath}
\usepackage{amssymb}
\usepackage{float}

\newcommand{\ep}{\varepsilon}

\newcommand{\Li}{{\rm Li}}
\newcommand{\DD}{{\mathcal D}}
\newcommand{\MOM}{{\sf MOM}}
\newcommand{\N}{{\nonumber}}
\newcommand{\Ahathat}{\hat{\hspace*{-0.3mm}\hat{A}}}
\newcommand{\BLCB}{\left\{ \phantom{\rule{0.1mm}{0.57cm}} \! \right.}
\newcommand{\BLB}{\left[ \phantom{\rule{0.1mm}{0.56cm}} \! \right.}
\newcommand{\BLP}{\left( \phantom{\rule{0.1mm}{0.56cm}} \! \right.}

\newcommand{\BRCB}{\left. \! \phantom{\rule{0.1mm}{0.56cm}} \right\}}
\newcommand{\BRB}{\left. \! \phantom{\rule{0.1mm}{0.56cm}} \right] }
\newcommand{\BRP}{\left. \! \phantom{\rule{0.1mm}{0.56cm}} \right) }

\begin{document}
\thispagestyle{empty}

\mbox{}
\begin{flushleft}
DESY  10--220 \hfill 
\\
DO-TH 11/13\\
SFB/CPP-11/42 \\
LPN 11/41\\
July 2011\\
\end{flushleft}
\vspace*{\fill}
\begin{center}
{\LARGE\bf Two-loop QED Operator Matrix Elements with} 

\vspace*{3mm}
{\LARGE\bf Massive External Fermion Lines} 

\vspace*{20mm}

\large

\vspace{7mm}\noindent
Johannes Bl\"umlein$^a$, Abilio De 
Freitas$^{a,b,}$\footnote{Alexander-von-Humboldt Fellow.}, and Wilhelmus 
van 
Neerven$^c,$~\footnote{deceased.}

\vspace{2em}

\normalsize
{\it $^a$DESY, Zeuthen, Platanenallee 6, D--15735 Zeuthen,  Germany}

\vspace{3mm}
{\it $^b$
Departamento de
F\'{i}sica, Universidad Sim\'{o}n Bol\'{i}var,}\\ {\it Apartado Postal 
89000,
Caracas 1080-A, Venezuela.}

\vspace{3mm}
{\it $^c$Instituut--Lorentz, Universiteit Leiden,}\\
{\it
P.O. Box 9506, 2300 HA 
Leiden, The Netherlands}\\
\today
\end{center}
\vspace*{\fill}
\begin{abstract}
\noindent
The two-loop massive operator matrix elements for the fermionic local twist--2
operators with external massive fermion lines in Quantum Electrodynamics (QED)
are calculated up to the constant terms in the dimensional parameter $\varepsilon
= D - 4$. We investigate the hypothesis of Ref.~\cite{BBN} that the 2--loop QED 
initial state corrections to $e^+e^-$ annihilation into a virtual neutral gauge
boson,  except power corrections of $O((m_f^2/s)^k),~~k \geq 1$, can be represented 
in terms of these matrix elements and the massless 2--loop Wilson coefficients 
of the Drell--Yan process.
\end{abstract}
\vspace*{\fill}
\newpage
\noindent

\section{Introduction}
\label{sec:1}

\vspace*{1mm}
\noindent
The QED corrections for differential distributions in $e^+e^-$ annihilation and 
other high energy reactions in which electrons or positrons participate, are 
particularly large due to the presence of physical logarithms $\ln(M^2/m_e^2)$, 
with $M$ a characteristic scale of the process and $m_e$ the electron mass.\footnote{
This also applies to the QED corrections in $eN$ scattering, cf. \cite{EP}.} 
Therefore, it is necessary to account for the QED initial state corrections up to 
$O(\alpha^2)$  for precision measurements in the various energy regimes in $e^+e^-$ annihilation
having been explored so far and those which are planned to be investigated in the future, 
cf.~\cite{ADONE,SPEAR,PETRA,LEP,SLC,ILC,CLIC,ILC:PHYS,DAFNE,ACC}. 
For the corrections 
to the inclusive Born cross section $\sigma(s)$, with $s$ the center of mass (cms) 
energy squared, power corrections $\propto (m_e^2/s)^k, k \geq 1$ can be safely 
disregarded. 
While the $O(\alpha)$ corrections are known for a large number of reactions, cf.
\cite{OLD}, the corrections beyond the universal contributions $O((\alpha 
L)^k), 1 \leq k \leq 5$ \cite{UNIV1,UNIV2}, to higher orders, were only 
calculated analytically once at 2--loop order in Ref.~\cite{BBN}.~\footnote{There 
are only a few other complete analytic $O(\alpha^2)$ two-loop calculations with 
massive particles available. Examples are the heavy quark fragmentation functions 
\cite{FRAG}, the $\mu$-lepton decay spectrum~\cite{MUDEC} to $O(\alpha^2 L)$, while 
the $O(\alpha^2)$ contribution was given numerically, and the heavy flavor Wilson 
coefficients for large virtualities \cite{HEAV1,BBK}.}
Besides the logarithmic 
orders $O(\alpha^2 L^2, \alpha^2 L)$ with $L = \ln(s/m_e^2)$, the constant 
terms $O(\alpha^2)$ are of interest. 

In Ref.~\cite{BBN} it has been proposed that the 2-loop corrections can be 
calculated using a factorization-representation of the scattering cross section 
in terms of Mellin convolutions of massive local operator matrix elements (OMEs) and the 
corresponding massless Wilson coefficients, which are those of the massless Drell--Yan 
process \cite{DY1,DY2}. The operator matrix elements are formed by the local twist--2
fermionic operators, being obtained in the light cone formalism \cite{LCE},  between massive 
on--shell electron states. These operator matrix elements bear all the mass dependence, 
$\mu^2/m_e^2$, and are universal quantities. The massless Wilson coefficients account 
for the process dependence and are functions of the ratio $\hat{s}/\mu^2$, with $\hat{s}$ 
the sub-system cms energy squared. Here, $\mu^2$ denotes a factorization scale.

In case of the heavy flavor corrections to deep-inelastic scattering the above method was 
used to calculate the massive Wilson coefficients in the region $Q^2 \gg m_Q^2$, with 
$Q^2$ the virtuality of the exchanged gauge boson and $m_Q$ the heavy quark mass. It has been 
shown that this description yields all but the power suppressed contributions 
in Refs.~\cite{HEAV1,BBK} comparing to the complete semi-analytic calculation \cite{Laenen}
at $O(\alpha_s^2)$~\footnote{For a fast and precise numerical implementations of these 
corrections in Mellin space cf. \cite{AB}.}. In this case the massive OMEs are formed 
between {\it massless} on--shell quark and gluon states. In Refs.~\cite{HQCD,BBK2,BBK3} 
2-- and 3--loop heavy flavor corrections for different unpolarized and polarized nucleon 
structure functions and transversity, respectively for their moments, have been calculated.  

In the present paper we compute the $O(\alpha^2)$ local OMEs with massive external 
fermions in QED up to the constant part. As a by-product we obtain the QED contributions 
to the 2--loop non--singlet and pure--singlet anomalous dimensions, within a massive 
calculation. We investigate, to which extent the decomposition, having been proposed in Ref.~\cite{BBN}, 
in terms of massive local OMEs and massless Wilson coefficients is possible. 

The paper is organized as follows. In Section~2  we summarize the decomposition 
of the initial state corrections to the inclusive $e^+e^-$ annihilation cross section into 
neutral vector bosons to $O(\alpha^2)$ using the renormalization group method, cf. \cite{BBN}.
The renormalization of the OMEs is described in Section~3. In Section~4 we compute the OMEs
to $O(\alpha)$. The details of the calculation of the $O(\alpha^2)$ corrections to the OMEs are
given in Section~5. We discuss the structure of the contributions to the 
differential scattering cross section $d \sigma_{e^+ e^-}/d \hat{s}$ and compare to the results
in the literature. Section~6 contains the conclusions. Technical aspects are summarized in the 
appendices.

\section{The Renormalization Group Method}

\vspace{1mm}
\noindent
The $O(\alpha^2)$ QED initial state corrections to the $e^+e^-$ annihilation cross section into
virtual neutral gauge bosons $(\gamma^*, Z^*$) in the limit $s \gg m_e^2$ can be expressed in the 
following form \cite{BBN}~:
\begin{equation}
\label{eq:XS}
\frac{d{\sigma}_{e^+e^-}}{ds'} = \frac{d{\sigma}_{e^+e^-}^{\rm I}}{ds'}
+ \frac{d{\sigma}_{e^+e^-}^{\rm II}}{ds'} + \frac{d{\sigma}_{e^+e^-}^{\rm III}}{ds'}~.
\end{equation}
Here, $s'$ denotes the invariant mass of the virtual vector boson 
and $s$ the cms energy of the process, 

\begin{eqnarray} 
\label{kin1}
s' =  x s,~~~x \in [0,1]~.
\end{eqnarray}
The term I refers to the photon radiation contributions, 
II to the flavor non--singlet contribution due to fermion pair production, and  III to the corresponding 
flavor pure--singlet contribution.~\footnote{In Ref.~\cite{BBN} four contributions were considered 
dividing those to process I into two pieces according to the genuine $2 \rightarrow 3$ particle 
scattering cross sections.} One may represent the scattering cross section in Mellin space by applying 
the integral transform
\begin{eqnarray} 
\label{mel1}
\widehat{f}(N) = \int_0^1~dx~x^{N-1}~f\left(x = \frac{s'}{s}\right)~,
\end{eqnarray}
with
\begin{eqnarray} 
\widehat{\frac{d\sigma}{ds'}}(N) &=& \int_0^1~dx~x^{N-1}~\frac{d\sigma}{ds'}(xs)~,\\ 
\widehat{\sigma_0}(N)            &=& \int_0^1~dx~x^{N-1}~\sigma_0(xs)~.
\end{eqnarray}

Since in the present calculation power corrections of $O(m_e^2/s)$ are disregarded, the following 
principle structure \cite{BBN} with respect to the scales $m_e^2, s$ and a factorization scale
$\mu^2$ is obtained, \cite{BBN}~:
\begin{alignat}{2} 
\widehat{\frac{d\sigma}{ds'}}(N) = \frac{1}{s} \widehat{\sigma_0}(N)
                                     \times \Biggl[ \hspace*{1mm}
                          & \Gamma_{e^+e^+}\left(N,\frac{\mu^2}{m_e^2}\right) 
                          & 
\tilde{\sigma}_{e^+e^-}\left(N,\frac{s'}{\mu^2}\right) 
                          & \Gamma_{e^-e^-}\left(N,\frac{\mu^2}{m_e^2}\right)
                         \nonumber\\  
                         + & \Gamma_{\gamma e^+}\left(N,\frac{\mu^2}{m_e^2}\right) 
                           & 
\tilde{\sigma}_{e^- \gamma }\left(N,\frac{s'}{\mu^2}\right) 
                           & \Gamma_{e^-e^-}\left(N,\frac{\mu^2}{m_e^2}\right) 
                         \nonumber\\   
                         + & \Gamma_{e^+e^+}\left(N,\frac{\mu^2}{m_e^2}\right) 
                           & 
\tilde{\sigma}_{e^+ \gamma}\left(N,\frac{s'}{\mu^2}\right) 
                           & \Gamma_{e^- \gamma}\left(N,\frac{\mu^2}{m_e^2}\right) 
                         \nonumber\\   
                         + & \Gamma_{e^+\gamma}\left(N,\frac{\mu^2}{m_e^2}\right) 
                           & 
\tilde{\sigma}_{\gamma \gamma}\left(N,\frac{s'}{\mu^2}\right) 
                           & \Gamma_{e^- \gamma}\left(N,\frac{\mu^2}{m_e^2}\right) 
\Biggr]~.
\label{decomp1}
\end{alignat}
To $O(\alpha^2)$ the last process in (\ref{decomp1}) does not contribute.
Here $\sigma_0(s)$ denotes the Born cross section for $e^+e^-$ annihilation into a virtual gauge 
boson $(\gamma, Z)$ which decays 
into a fermion pair $f \overline{f}$, see e.g. \cite{BDJ}, 
\begin{eqnarray} 
\frac{d\sigma^{(0)}(s)}{d \Omega} &=& \frac{\alpha^2}{4 s}
                                        N_{C,f} \sqrt{1 - \frac{4 m_f}{s}} 
\times
\nonumber
\end{eqnarray}
\begin{eqnarray}
& & \times
\left[\left(1+ \cos^2\theta + \frac{4 m_f^2}{s} \sin^2\theta \right) G_1(s)
- \frac{8 m_f^2}{s} G_2(s) 
+ 2 \sqrt{1-\frac{4m_f^2}{s}} \cos\theta G_3(s)\right]\\ 
\sigma^{(0)}(s) &=& \frac{4 \pi \alpha^2}{3 s} 
                                        N_{C,f} \sqrt{1 - \frac{4 m_f}{s}} 
\left[\left(1 + \frac{2 m_f^2}{s} \right) G_1(s)
- 6 \frac{m_f^2}{s} G_2(s)\right]~. 
\end{eqnarray}
Here $\alpha$ denotes the fine structure constant, with $\alpha = 4 \pi a$,
$N_{C,f}$ is the number of colors of the final state fermion, with $N_{C,f} = 1$ for colorless 
fermions, $s$ is the cms energy, $\Omega$ is the spherical angle, 
$\theta$ the cms scattering angle, and the effective couplings $G_i(s)|_{i=1...3}$ read
\begin{eqnarray}
G_1(s) &=& Q_e^2 Q_f^2 + 2 Q_e Q_f v_e v_f {\sf Re}[\chi_Z(s)]
          +(v_e^2+a_e^2)(v_f^2+a_f^2)|\chi_Z(s)|^2\\
G_2(s) &=& (v_e^2+a_e^2) a_f^2 |\chi_Z(s)|^2 \\
G_3(s) &=& 2 Q_e Q_f a_e a_f {\sf Re}[\chi_Z(s)] + 4 v_e v_f a_e a_f |\chi_Z(s)|^2.
\end{eqnarray}
The reduced $Z$--propagator is given by
\begin{eqnarray}
\chi_Z(s) = \frac{s}{s-M_Z^2 + i M_Z \Gamma_Z},
\end{eqnarray}
where $M_Z$ and  $\Gamma_Z$ are the mass and the with of the $Z$--boson and $m_f$ is the mass of the final 
state fermion. $Q_{e,f}$ are the electromagnetic charges of the electron $(Q_e = -1)$ 
and the final state fermion, resp., and
the electroweak couplings $v_i$ and $a_i$ read
\begin{eqnarray}
v_e &=& \frac{1}{\sin\theta_w \cos\theta_w}\left[I^3_{w,e} - 2 Q_e 
\sin^2\theta_w\right]\\ 
a_e &=& \frac{1}{\sin\theta_w  \cos\theta_w} I^3_{w,e} \\
v_f &=& \frac{1}{\sin\theta_w \cos\theta_w}\left[I^3_{w,f} - 2 Q_f 
\sin^2\theta_w\right]\\ 
a_f &=& \frac{1}{\sin\theta_w  \cos\theta_w} I^3_{w,f}~, 
\end{eqnarray}
where $\theta_w$ is the weak mixing angle, and $I^3_{w,i} = \pm 1/2$ the third component 
of the weak isospin for up and down particles, respectively. 

The factorization mass cancels in the physical cross section in each order of the
coupling constant. The initial
state fermion mass dependence is solely encoded in $\Gamma_{li}$. 
The operator matrix elements are given by
\begin{alignat}{5} 
\label{decomp4}
&\Gamma_{e^+ e^+}(N)     &~~=~~& \Gamma_{e^-e^-}(N)      &~~=~~& \langle e |O_F^{\sf NS,S}| e      
\rangle 
\\  
&\Gamma_{e^+ \gamma}(N)           &~~=~~& \Gamma_{e^-\gamma}(N)            &~~=~~& \langle \gamma|O_F^{\sf S}| \gamma 
\rangle \\  
&\Gamma_{\gamma e^+}(N)           &~~=~~& \Gamma_{\gamma e^-}(N)           &~~=~~& \langle e |O_V^{\sf S}|e \rangle~,   
\end{alignat}
where $O_F^{\sf NS,S}$ and $O_V^{\sf S}$ are the local twist--2 fermion and photon operators, 
\begin{eqnarray}
\label{COMP1}
O^{\sf NS,S}_{F;\mu_1, \ldots, \mu_N} &=& i^{N-1} {\bf S} [\overline{\psi}
\gamma_{\mu_1} D_{\mu_2} \ldots D_{\mu_N}
\psi] - {\rm trace~terms}~, \\
\label{COMP3}
O^{\sf S}_{V;\mu_1, \ldots, \mu_N} &=& 2 i^{N-2} {\bf S} [F_{\mu_1 \alpha}
D_{\mu_2} \ldots D_{\mu_{N-1}} F_{\mu_N}^{\alpha}] - {\rm trace~terms}~,
\end{eqnarray}
for the fermionic non--singlet {\sf (NS)}, singlet {\sf (S)}, and photonic case,~\cite{Geyer:1977gv}.
Here, {$\bf S$} is the symmetrization operator of the Lorentz indices
$\mu_1, \ldots, \mu_N$, 
$\psi$ denotes the electron field, $F_{\mu\nu}$ the photon
field--strength tensor, and $D_{\mu} = \partial_\mu - i e A_{\mu}$ the covariant derivative,
with $e = \sqrt{(4\pi)^2 a}$ the electric charge and $A_{\mu}$ the 4--potential. 
We consider only one fermion species. For
(\ref{decomp4}) both  the flavor non--singlet (NS) and pure singlet (PS) terms contribute.
It turns out that also in the present case the contributing functions both for the massive OMEs 
and the massless Wilson coefficients \cite{BRK} can be related to nested harmonic sums 
\cite{HSUM1,HSUM2} and multiple zeta values \cite{MZV}. These structures simplify further applying algebraic and structural 
relations \cite{REL}. One may perform the calculation in Mellin space completely and represent the result in 
$x$-space analytically resp. numerically performing the inverse Mellin transformation \cite{REL,ANCONT}.

The following representations apply 
\begin{eqnarray}
\Gamma_{li}\left(N,\frac{\mu^2}{m^2_e}\right) &=& \sum_{r=0}^\infty a^r(\mu^2) 
\sum_{n=0}^r a_{nr}(N) \ln^n\left(\frac{m^2_e}{\mu^2}\right)~, \\
\tilde{\sigma}_{lk}\left(N,\frac{s'}{\mu^2}\right) &=& \sum_{r=0}^\infty a^r(\mu^2) 
\sum_{n=0}^r b_{nr}(N) \ln^n\left(\frac{s'}{\mu^2}\right)~.
\end{eqnarray}
Here $i,j$  denote the external particles in the scattering process. The coefficients $a_{nr}$ 
and $b_{nr}$ of the above series adjust such, that the physical cross section is independent of $\mu$.

$\Gamma_{li}$, $\tilde{\sigma}_{lk}$ and the scattering cross sections $\sigma_{ij}$ obey the 
renormalization group equations \cite{RGE}~:
\begin{eqnarray}
\left[\left(\mu\frac{\partial}{\partial \mu} + \beta(g) \frac{\partial}{\partial g} \right) \delta_{al} + 
\gamma_{al}(N,g) \right] \Gamma_{li}\left(N,\frac{\mu^2}{m^2_e},g(\mu^2_e)\right) &=& 0 
\\
\left[\left( \mu\frac{\partial}{\partial \mu} + \beta(g) \frac{\partial}{\partial g} \right) \delta_{la}  
\delta_{kb} - \gamma_{la}(N,g) \delta_{kb} - \gamma_{kb}(N,g) \delta_{la}\right] 
\tilde{\sigma}_{lk}\left(\frac{s'}{m^2_e},g(\mu^2)\right) &=& 0 \\
\left[ \mu\frac{\partial}{\partial \mu} + \beta(g) \frac{\partial}{\partial g} \right] 
\sigma_{ij} \left(\frac{s'}{\mu^2}, g(\mu^2)\right) &=& 0~.
\end{eqnarray}
The $\beta$--function $\beta(g)$ is given by
\begin{eqnarray}
\label{eqbet}
\beta(g) = - \sum_{k=0}^\infty \beta_k \frac{g^{2k+3}}{(16 \pi^2)^{k+1}}~, 
\end{eqnarray}
with $g$ the electromagnetic coupling. In QED the first expansion coefficients are~\cite{BET}
\begin{eqnarray}
\beta_0 = - \frac{4}{3}, \hspace{2cm} 
\beta_1 = - 4~,
\end{eqnarray}
in the case of one light fermion. The running coupling
$a(\mu^2) = g^2(\mu^2)/(16 \pi^2)$ in the $\overline{\rm MS}$
scheme is obtained as the solution of
\begin{eqnarray}
\frac{d a(\mu^2)}{d \ln(\mu^2)} = - \sum_{k=0}^\infty \beta_k a^{k+2}(\mu^2)~,
\end{eqnarray}
which yields
\begin{eqnarray}
\label{ALPH}
a(\mu^2) &=& \frac{a_0}
{1 + a_0 \beta_0 \ln(\mu^2/m_e^2) }
= a_0 \left[ 1 + \frac{4}{3} a_0 
\ln\left(\frac{\mu^2}{m_e^2}\right)\right] + O(a_0^3)~,
\end{eqnarray}
with $a_0 = a(m_e^2)$.
We rewrite the renormalization group equations replacing
$\mu \partial/\partial \mu$ by $2 \partial/\partial {L}$, 
resp. $- 2 \partial/\partial {\lambda}$,  with 
\begin{eqnarray}
L = \ln\left(\frac{\mu^2}{m_e^2}\right),~~~~~~~~~~\lambda = \ln\left(\frac{s'}{\mu^2}\right)~,
\end{eqnarray}
one obtains
\begin{eqnarray}
\left[\frac{\partial}{\partial {L}} - \beta_0 a^2 \frac{\partial}{\partial a}
+ \frac{1}{2} \gamma_{ee}(N,a)\right] \Gamma_{ee}\left(N,a,\frac{\mu^2}{m^2_e}\right)
+ \frac{1}{2} \gamma_{e\gamma}(N,a) \Gamma_{\gamma e}\left(N,a,\frac{\mu^2}{m^2_e}\right) &=& 0\\
\left[\frac{\partial}{\partial {\lambda}} + \beta_0 a^2 \frac{\partial}{\partial a}
+ \gamma_{ee}(N,a)\right] \tilde{\sigma}_{ee}\left(N,a,\frac{s'}{\mu^2}\right)
+  \gamma_{\gamma e}(N,a) \tilde{\sigma}_{e \gamma}\left(N,a,\frac{s'}{\mu^2}\right) &=& 0~,
\end{eqnarray}
where $a = a(\mu^2)$. Here the anomalous dimensions $\gamma_{ij}(N,a)$ have the series expansion
\begin{eqnarray}
\gamma_{ij}(N,a) = \sum_{k=0}^\infty a^{k+1} \gamma_{ij}^{(k)}~.
\end{eqnarray}
For later convenience we also introduce the splitting functions in $N$--space,
\begin{eqnarray} 
P_{ij}^{(l)}(N) = \int_0^1 d z z^{N-1} P_{ij}^{(l)}(z) 
= - \gamma_{ij}^{(l)}(N)~.
\end{eqnarray} 
The higher expansion coefficients of the OMEs, $\Gamma_{ij}^{0}, \overline{\Gamma}_{ij}^{0}$ 
and $\Gamma_{ij}^{1}$, (\ref{Aij1}, \ref{eq:GA1}--\ref{eq:GA3}),
are defined such that they do not flip sign under the Mellin transform.

For the process under consideration one obtains to $O(a^2)$~:
\begin{eqnarray}
\Gamma_{ee}\left(N,a,\frac{\mu^2}{m^2_e}\right)
&=& 1 + a\left[-\frac{1}{2} \gamma_{ee}^{(0)}(N) L + \Gamma_{ee}^{(0)}(N)\right]\nonumber\\
& & + a^2 \Biggl[\left\{\frac{1}{8} \gamma_{ee}^{(0)}(N)\left(\gamma_{ee}^{(0)}(N) - 2 \beta_0\right)
+\frac{1}{8} \gamma_{e\gamma}^{(0)}(N) \gamma_{\gamma e}^{(0)}(N) \right\} L^2 \nonumber\\ & & 
+ \frac{1}{2} \left\{-\gamma_{ee}^{(1)}(N) + 2 \beta_0 \Gamma_{ee}^{(0)} - \gamma_{ee}^{(0)}(N)
\Gamma_{ee}^{(0)}(N) - \gamma_{e \gamma}^{(0)} \Gamma_{\gamma e}^{(0)}(N) \right\} L
\nonumber\\ & & + \Gamma_{ee}^{(1)} \Biggr] + O(a^3)~, \\  
\tilde{\sigma}_{ee}\left(N,a,\frac{s'}{\mu^2}\right)
&=& 1 + a\left[- \gamma_{ee}^{(0)}(N) \lambda + \tilde{\sigma}_{ee}^{(0)}(N)\right]\nonumber\\
& & + a^2 \Biggl[\left\{\frac{1}{2} \gamma_{ee}^{(0)}(N)\left(\gamma_{ee}^{(0)}(N) + \beta_0\right)
+\frac{1}{4} \gamma_{e\gamma}^{(0)}(N) \gamma_{\gamma e}^{(0)}(N) \right\} \lambda^2 \nonumber\\ & & 
+ \left\{-\gamma_{ee}^{(1)}(N) - \beta_0 \tilde{\sigma}_{ee}^{(0)} - \gamma_{ee}^{(0)}(N)
\tilde{\sigma}_{ee}^{(0)}(N) - \gamma_{\gamma e}^{(0)} \tilde{\sigma}_{e 
\gamma}^{(0)}(N) \right\} \lambda
\nonumber\\ & & + \tilde{\sigma}_{ee}^{(1)} \Biggr] + O(a^3)~, \\
\Gamma_{\gamma e}\left(N,a,\frac{\mu^2}{m^2_e}\right) &=& a \left[-\frac{1}{2} \gamma_{\gamma 
e}^{(0)}(N) L +
\Gamma_{\gamma e}^{(0)}\right] + O(a^2) \\ 
\tilde{\sigma}_{e \gamma}\left(N,a,\frac{s'}{\mu^2}\right) &=& a \left[-\frac{1}{2} \gamma_{e 
\gamma}^{(0)}(N) 
\lambda +
\tilde{\sigma}_{e \gamma}^{(0)}\right] + O(a^2)~, 
\end{eqnarray}
with the 1-- and 2--loop splitting functions given in Refs.~\cite{PijLO,PijNLO}.
We express the coupling constant $a = a(\mu^2)$ by (\ref{ALPH})
and assemble the differential scattering cross section (\ref{eq:XS})
in terms of the three contributions: the flavor non-singlet terms with a 
single fermion line (I), those with an additional closed fermion line (II), and the
pure-singlet terms (III). Here we corrected Eq.~(4.19, 4.22, 4.26) in 
Ref.~\cite{BBN}~\footnote{The contributions to process II have been also calculated in Ref.~\cite{KUHN}.}.
\begin{eqnarray} 
\label{eqMA1a}
\widehat{
\frac{d\sigma_{e^+e^-}^{\rm I}}{ds'}} &=&
\frac{1}{s} \widehat{\sigma^{(0)}} 
   \Biggl\{
   1 + a_0 \left[ P_{ee}^{(0)}  {\bf L}
   +\left(\tilde{\sigma}^{(0)}_{ee} + 2 \Gamma_{ee}^{(0)}\right)\right] 
   \nonumber\\
& & \hspace{21mm}
   + a_0^2\Biggl\{
   \frac{1}{2} {P_{ee}^{(0)}}^2 {\bf L}^2
   +\Biggl[P_{ee}^{(1),{\rm I}} 
   + P_{ee}^{(0)} \left( \tilde{\sigma}_{ee}^{(0)} + 
   2 \Gamma_{ee}^{(0)}\right) \Biggr]{\bf L}
   \nonumber\\
& & \hspace{21mm} 
   + \left(2 \Gamma_{ee}^{(1),{\rm I}} + 
\tilde{\sigma}_{ee}^{(1),{\rm I}}\right) + 2 
\Gamma_{ee}^{(0)} 
\tilde{\sigma}_{ee}^{(0)} 
+ {\Gamma_{ee}^{(0)}}^2 \Biggr\} \Biggr\}  
\\
\label{eqMA1b}
\widehat{\frac{d\sigma_{e^+e^-}^{\rm II}}{ds'}} &=&
\frac{1}{s} \widehat{\sigma^{(0)}} 
   a_0^2\Biggl\{- \frac{\beta_0}{2} P_{ee}^{(0)} {\bf L}^2
   +\Biggl[P_{ee}^{(1), {\rm II}} - \beta_0 \tilde{\sigma}_{ee}^{(0)} 
     \Biggr] {\bf L}
   + \left(2 \Gamma_{ee}^{(1),{\rm II}} + 
\tilde{\sigma}_{ee}^{(1),{\rm II}}\right) \Biggr\}   
\\
\label{eqMA1c}
\widehat{\frac{d\sigma_{e^+e^-}^{\rm III}}{ds'}} &=&
\frac{1}{s} \widehat{\sigma^{(0)}} 
   a_0^2\Biggl\{\frac{1}{4}
   P_{e \gamma}^{(0)}  P_{\gamma e}^{(0)} {\bf L}^2
   +\Biggl[P_{ee}^{(1),{\rm III}} + P_{\gamma e}^{(0)} 
\tilde{\sigma}_{e 
\gamma}^{(0)}
   + \Gamma_{\gamma e}^{(0)} P_{e \gamma}^{(0)} \Biggr] {\bf L}
   \nonumber\\
& & \hspace{21mm} 
   + \left( 2 \Gamma_{ee}^{(1),{\rm III}} + \tilde{\sigma}_{ee}^{(1),{\rm 
III}}\right)+ 2 \tilde{\sigma}_{e \gamma}^{(0)} 
   \Gamma_{\gamma e}^{(0)} 
\Biggr\}   
\end{eqnarray} 
with 
\begin{equation}
\label{eqL1}
{\bf L} =
{\bf \hat{L}} + \ln(z),~~~~ 
{\bf \hat{L}} 
= \ln\left(\frac{s}{m^2_e}\right)~.
\end{equation}
We will later rewrite Eqs.~(\ref{eqMA1a}--\ref{eqMA1c}) in terms of $\bf \hat{L}$. 
\section{The Renormalization of the Operator Matrix Elements}
\label{sec:3} 

\vspace*{1mm}
\noindent
Since the massless sub-system scattering cross sections $\tilde{\sigma}_{ab}(z,s/\mu^2)$ to $O(a_0^2)$
are known, the corresponding massive operator matrix elements for $e^{\pm} \rightarrow e^{\pm}, 
e^{\pm} \rightarrow \gamma$ and $\gamma \rightarrow e^{\pm}$ transitions have to be calculated. The 
latter two processes contribute only to first order of the coupling constant. Here, the 
external fermion is a massive particle, contrary to the cases studied in \cite{HQCD,BBK2,BBK3}.

The bare OMEs are given by
\begin{eqnarray}
\Ahathat _{ij}\left(\frac{{m}^2_e}{\mu^2},\ep,N\right) = \delta_{ij} + \sum_{k=1}^\infty
\hat{a}^k \Ahathat_{ij}^{(k)}\left(\frac{{m}^2_e}{\mu^2},\ep,N\right)~,
\end{eqnarray}
with $\hat{a}$ the unrenormalized coupling constant. The computation is performed 
in $D= 4 + \varepsilon$ dimensions. The electron mass is renormalized on-shell
\begin{eqnarray}
p^2 = m_e^2(\mu = m_e) 
\end{eqnarray}
with $p$ the momentum of the external fermion. Thus renormalization concerns the wave 
function, the charge renormalization, and the ultraviolet singularities 
of the local operators. For process I counter terms emerge at $O(\hat{a}^2)$.
Due to the finite fermion mass, no collinear singularities emerge.

\vspace*{2mm}\noindent
{\sf i) Wave function renormalization.}\\
The bare wave function $\psi_0$ is renormalized by 
\begin{eqnarray}
\psi_0 &=&  \sqrt{Z_2(\ep)}~\psi~.
\end{eqnarray}
The $Z$--factors are obtained from the fermion self--energy, see Figure~1, 
\begin{eqnarray}
\Sigma\!\!\!\!/~(p,m_e)  = m_e\Sigma_1(p^2,m_e)  + (p\!\!\!/ - m_e)  \Sigma_2(p^2,m_e)~.
\end{eqnarray}
Expanding $\Sigma\!\!\!\!/~(p,m_e)$ around $p^2 = m^2_e$ one obtains
\begin{eqnarray}
\frac{1}{Z_2} &=& 1 + \left. 
2 m^2_e \frac{\partial}{\partial{p^2}}\Sigma_1(p^2,m_e)\right|_{p^2=m^2_e} 
+\left. \Sigma_2(p^2,m_e)\right|_{p^2=m^2_e}~.
\end{eqnarray}
\restylefloat{figure}
\begin{figure}[H]
\begin{center}
\centerline{\epsfxsize 6 in \epsfbox{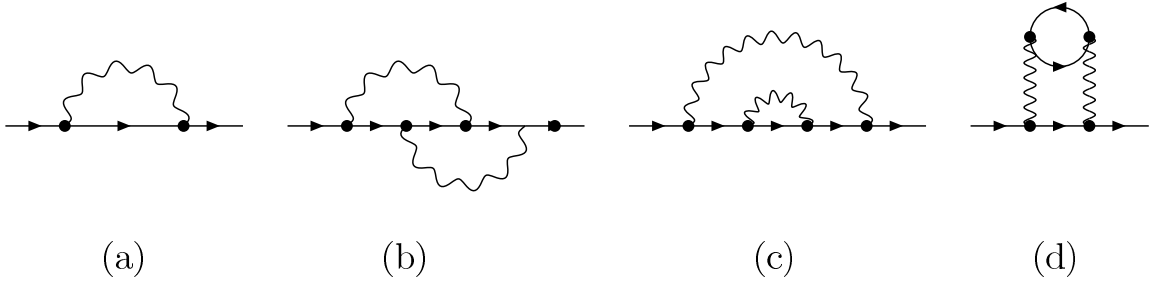}}
\end{center}
\vskip -.9 cm 
\caption{
\small The self-energy diagrams. \label{SelfEnergy}}
\end{figure}
\noindent
To $O(\hat{a}^2)$ $Z_2$  is given by \cite{MZ2}
\begin{eqnarray}
Z_2 &=& 1 + \sum_{k=1}^\infty \hat{a}^k Z_2^{(k)} = 1 + \hat{a} S_\varepsilon 
\left(\frac{m_e^2}{\mu^2}\right)^{\varepsilon/2}
        \left[\frac{6}{\varepsilon}
              -4 +
              \left(4 + \frac{3}{4} \zeta_2\right) \varepsilon 
\right]\nonumber\\ &&
+ \hat{a}^2 S_\varepsilon^2 \left(\frac{m_e^2}{\mu^2}\right)^{\varepsilon}
\Biggl\{
\left[18 \frac{1}{\varepsilon^2} - \frac{51}{2} \frac{1}{\varepsilon}
+ \left(\frac{433}{8} - \frac{147}{2} \zeta_2 + 96 \zeta_2 \ln(2) - 24 
\zeta_3\right) \right]_{\rm I}
\nonumber\\
& & \hspace{2.5cm} +
\left[16 \frac{1}{\varepsilon^2} - \frac{38}{3} \frac{1}{\varepsilon}
+ \left(\frac{1139}{18} - 28 \zeta_2 \right) \right]_{\rm II}\Biggr\}+ O(\hat{a}^3)~,
\label{Z2equation}
\end{eqnarray}
with the spherical factor
\begin{eqnarray}
S_\varepsilon = \exp\left[\frac{\varepsilon}{2}\left(\ln\left(4\pi\right)
- \gamma_E\right)\right]
\end{eqnarray} 
and $\zeta_k$ denotes Riemann's $\zeta$-function at integer values $k \geq 2$. In the $\overline{\rm MS}$ scheme the factors 
$S_\varepsilon$ are set to one
at the end of the calculation. In (\ref{Z2equation}) we separated the terms 
contributing to processes I and II. At 2-loop order also counter terms (CT) contribute with the 
$Z$-factor $Z_{\rm CT}$ to the OME
$A_{ee}^{\rm I}$, which will be calculated in Section~\ref{sec:5}.

After wave function renormalization and accounting for counter terms
the OME  is denoted by $\hat{A}_{ee}$. Up to  
$O(\hat{a}^2)$ it is given by~:
\begin{eqnarray}
\label{eq:hat1}
\hat{A}_{ee}^{\rm I}\left(\frac{m_e^2}{\mu^2},\ep,N\right) &=& 1 
+ \hat{a}~\left[\Ahathat_{ee}^{(1),\rm I}\left(\frac{m_e^2}{\mu^2},\ep,N\right) + Z_2^{(1)}\right]
\nonumber\\ &&
+ \hat{a}^2 \Biggl[\Ahathat_{ee}^{(2),\rm I}\left(\frac{m_e^2}{\mu^2},\ep,N\right)
+ Z_{2, \rm I}^{(2)} + Z_{\rm CT}^{(2)} +
Z_2^{(1)} \Ahathat_{ee}^{(1),\rm I }\left(\frac{m_e^2}{\mu^2},\ep,N\right) \Biggr]
+ O(\hat{a}^3)~.
\nonumber\\
\label{eq:hat2}
\hat{A}_{ee}^{\rm II}\left(\frac{m_e^2}{\mu^2},\ep,N\right) &=& 
\hat{a}^2 \Biggl[\Ahathat_{ee}^{(2),\rm II}\left(\frac{m_e^2}{\mu^2},\ep,N\right)
+ Z_{2, \rm II}^{(2)} 
\Biggr] + O(\hat{a}^3)~.
\nonumber\\
\label{eq:hat3}
\hat{A}_{ee}^{\rm III}\left(\frac{m_e^2}{\mu^2},\ep,N\right) &=& 
 \hat{a}^2 \Biggl[\Ahathat_{ee}^{(2),\rm III}\left(\frac{m_e^2}{\mu^2},\ep,N\right)
 \Biggr]
+ O(\hat{a}^3)~.
\end{eqnarray}

\vspace*{2mm}\noindent
{\sf ii) Charge renormalization.}\\
The bare coupling $\hat{a}$ and the renormalized coupling in the $\overline{\rm MS}$ scheme 
are related by
\begin{eqnarray}
\hat{a} &=& {Z_g^{\overline{\rm MS}}}^2(\ep,n_f)) a^{\overline{\rm MS}}(\mu^2)\nonumber\\
        &=& a^{\overline{\rm MS}}(\mu^2)\left[1 + \delta a_1^{\overline{\rm MS}}(\ep,n_f) a^{\overline{\rm MS}}(\mu^2)\right]
            +O({a^{\overline{\rm MS}}}^3)~,
\end{eqnarray}
with
\begin{eqnarray}
\delta a_1^{\overline{\rm MS}}(\ep,n_f) &=& \frac{2}{\ep} \beta_0(n_f) \\
\beta_0(n_f) &=& - \frac{4}{3} n_f~.
\end{eqnarray}
The above relations would apply to $n_f$ manifestly massless fermions. Since the fermion lines
are all massive the coupling constant is first being obtained in a {\sf MOM}-scheme,
which is defined by
\begin{eqnarray}
\hat{\Pi}_{H,\rm {\small BF}}(0, m_e^2) + Z_{A,H} &=& 0 \\
Z_g^2   &=& Z_{A,H}^{-1}~,
\end{eqnarray}
cf.~\cite{BBK2}. Here $\hat{\Pi}_{H,\rm {\small BF}}(0, m_e^2)$ denotes the on-shell
vacuum polarisation calculated using the background-field method \cite{BF},
\begin{eqnarray}
\hat{\Pi}_{H,\rm {\small BF}}^{\mu\nu}(p^2,m_e^2,\mu^2,\ep, \hat{a}) &=& i \left(-g^{\mu\nu} p^2 + p^\mu p^\nu \right)
\hat{\Pi}_{H,\rm {\small BF}}(p^2,m_e^2,\mu^2,\ep,\hat{a}) 
\\
\hat{\Pi}_{H,\rm {\small BF}}(p^2,m_e^2,\mu^2,\ep,\hat{a}) &=& \hat{a} \frac{2 \beta_{0,H}}{\ep} 
\left(\frac{m_e^2}{\mu^2}\right)^{\ep/2} \exp \left( \sum_{k=2}^\infty \frac{\zeta_k}{k} \left(\frac{\ep}{2}\right)^k\right)
+ O(\hat{a}^2)~,
\end{eqnarray}
with
\begin{eqnarray}
\beta_{0,H} &=& - \frac{4}{3}~.
\end{eqnarray}
One obtains
\begin{eqnarray}
{Z_g^{\rm MOM}}^2 =  1 + a^{\rm MOM}(\mu^2) \beta_{0,H} \left(\frac{m_e^2}{\mu^2}\right)^{\ep/2}
\exp \left(\sum_{k=2}^\infty \frac{\zeta_k}{k} \left(\frac{\ep}{2}\right)^k\right) 
+ O\left({a^{\rm MOM}(\mu^2)}^2\right)~.
\end{eqnarray}
Finally, we transform back to the $\overline{\rm MS}$ scheme using
\begin{eqnarray}
{Z_g^{\rm MOM}}^2 a^{\rm MOM}(\mu^2)
= {Z_g^{\overline{\rm MS}}}^2 a^{\overline{\rm MS}}(\mu^2),
\end{eqnarray}
which implies
\begin{eqnarray}
\label{eq:aTR}
a^{\rm MOM}
= a^{\overline{\rm MS}} - \beta_{0,H} \ln\left(\frac{m_e^2}{\mu^2}\right) {a^{\overline{\rm MS}}}^2 + O\left(
{a^{\overline{\rm MS}}}^3\right)~.
\end{eqnarray}

\vspace*{2mm}\noindent
{\sf iii) Renormalization of the composite operators.}\\

\noindent
We express the inverse $Z$-factors in the {\sf MOM}-scheme~: 
\begin{eqnarray}
   Z_{ij}^{-1}(a^{\MOM},n_f+1,\mu)&=&
      \delta_{ij}
     -a^{\MOM}\frac{\gamma_{ij}^{(0)}}{\ep}
     +{a^{\MOM}}^2\Biggl[
          \frac{1}{\ep}\Bigl(
                       -\frac{1}{2}\gamma_{ij}^{(1)}
                       -\delta a^{\MOM}_{1}\gamma_{ij}^{(0)}
                       \Bigr)
\N
\end{eqnarray}
\begin{eqnarray}
&&
         +\frac{1}{\ep^2}\Bigl(
                        \frac{1}{2}\gamma_{il}^{(0)}\gamma_{lj}^{(0)}
                       +\beta_{0}\gamma_{ij}^{(0)}
                        \Bigr)
         \Biggr] + O({a^{\sf MOM}}^3)~,
\end{eqnarray}
with
\begin{eqnarray}
\delta a_{1}^{\sf MOM} =  S_\ep \frac{2 \beta_{0,H}}{\ep} 
\left(\frac{m^2_e}{\mu^2}\right)^{\ep/2} \exp\left[ \sum_{i=2}^\infty \frac{\zeta_i}{i}
\left(\frac{\ep}{2}\right)^i\right]~.
\end{eqnarray}
The renormalized OMEs are given by
\begin{eqnarray}
A_{ij}^{\sf MOM} = \delta_{ij} 
                   + a^{\sf MOM} \left[\hat{A}_{ij}^{(1)} + Z_{i,j}^{-1,(1)} \right]
                   + {a^{\sf MOM}}^2 
                     \left[\hat{A}_{ij}^{(2)} + Z_{i,j}^{-1,(2)} 
+ Z_{i,j}^{-1,(1)} \hat{A}_{ij}^{(1)}\right] +O({a^{\sf MOM}}^3)~.
\end{eqnarray}
The transformation to the $\overline{\rm MS}$-scheme is obtained by (\ref{eq:aTR}).
We split the OME into the parts I--III. The corresponding $Z$-factors are given by 
\begin{eqnarray}
\left[Z_{ee}^{\rm I}(\ep, N)\right]^{-1} &=& 1 + a^{\sf MOM} S_\ep \frac{1}{\ep} 
P_{ee}^{(0)}(N)
+ {a^{{\sf MOM}}}^2 S_\ep^2 \Biggl\{\frac{1}{2 \ep^2} {P_{ee}^{(0)}}^2(N) 
+ \frac{1}{2\ep} P_{ee}^{(1),\rm NS}(N)                            
\Biggr\} 
\nonumber\\ &&
+ O({a^{{\sf MOM}}}^3) \\
\left[Z_{ee}^{\rm II}(\ep, N)\right]^{-1} &=& {a^{{\sf MOM}}}^2\Biggl\{
- \frac{1}{2 \ep^2} \beta_{0} P_{ee}^{(0)}
+ \frac{2}{\ep^2} \beta_{0,H} \left(\frac{m^2_e}{\mu_2}\right)^{\ep/2} \exp\left[ \sum_{i=2}^\infty \frac{\zeta_i}{i}
\left(\frac{\ep}{2}\right)^i\right]
+ \frac{1}{2\ep} P_{ee}^{(1),\rm II} \Biggr\}
\nonumber\\ &&
+ O({a^{{\sf MOM}}}^3) \\
\left[Z_{ee}^{\rm III}(\ep, N)\right]^{-1} &=& 
{a^{{\sf MOM}}}^2 S_\ep^2 \left\{
\frac{1}{2\ep^2} P_{e \gamma}^{(0)}(N) P_{\gamma e}^{(0)}(N) 
+ \frac{1}{2\ep} \left[P_{ee}^{(1),\rm III}(N) 
\right]                             
\right\}
+ O({a^{{\sf MOM}}}^3) \\
\left[Z_{e\gamma}(\ep, N)\right]^{-1} &=& a^{{\sf MOM}} S_\ep
\frac{1}{\ep}
P_{e\gamma}^{(0)}(N) + O({a^{{\sf MOM}}}^2)
\\
\left[Z_{\gamma e}^{\rm NS}(\ep, N)\right]^{-1} &=&  a^{{\sf MOM}} S_\ep
\frac{1}{\ep}
P_{\gamma e}^{(0)}(N) + O({a^{{\sf MOM}}}^2)~.
\end{eqnarray}
The OMEs to two--loop order after wave function and charge renormalization
are given by
\begin{eqnarray}
\label{unA2I}
\hat{A}_{ee}^{\rm I} &=& a^{\sf MOM}  S_\varepsilon \left(\frac{m^2_e}{\mu^2}\right)^{\ep/2}
\left[-\frac{1}{\ep} P_{ee}^{(0)} + \Gamma_{ee}^{(0)} + \ep \overline{\Gamma}^{(0)}_{ee}\right] 
\nonumber\\ &&
+ {a^{\sf MOM}}^2  S_\varepsilon^2 
                       \left(\frac{m^2_e}{\mu^2}\right)^{\varepsilon} 
\left\{\frac{1}{2 \varepsilon^2} P_{ee}^{(0)} \otimes P_{ee}^{(0)} 
- \frac{1}{2 \varepsilon} \left[P_{ee}^{(1), \rm I} + 2 \Gamma_{ee}^{(0)}
\otimes P_{ee}^{(0)}\right] + \hat{\Gamma}_{ee}^{(1), \rm I}\right\}
\\
\label{unA2II}
\hat{A}_{ee}^{\rm II} &=& {a^{\sf MOM}}^2  S_\varepsilon^2 
                       \left(\frac{m^2_e}{\mu^2}\right)^{\varepsilon} 
\left\{\frac{1}{ 2\varepsilon^2}  2\beta_0 P_{ee}^{(0)} 
- \frac{1}{2\varepsilon}  \left[ 
P_{ee}^{(1), \rm II} + 4 \beta_0 \Gamma_{ee}^{(0)} \right] + 
\hat{\Gamma}_{ee}^{(1), \rm II}\right\}
\\
\label{unA2III}
\hat{A}_{ee}^{\rm III} &=& {a^{\sf MOM}}^2  S_\varepsilon^2 
                       \left(\frac{m^2_e}{\mu^2}\right)^{\varepsilon} 
                       \left\{\frac{1}{2 \varepsilon^2} P_{e 
\gamma}^{(0)} \otimes P_{\gamma e}^{(0)} - \frac{1}{2 \varepsilon} 
\left[ P_{ee}^{(1), \rm III} + 2 \Gamma_{\gamma e}^{(0)} \otimes 
P_{e \gamma}^{(0)} \right]  
+ \hat{\Gamma}_{ee}^{(1), \rm III} \right\} \,\, .
\end{eqnarray}

\vspace{2mm}\noindent
The renormalized OMEs in the $\overline{\rm MS}$-scheme are finally given by
\begin{eqnarray}
\label{eq:R1}
A_{ee}^{\overline{\rm MS},\rm I}(N) &=& a^{\overline{\rm MS}}
\Biggl[-\frac{1}{2} P_{ee}^{(0)}(N) \ln\left(\frac{m^2_e}{\mu^2}\right) 
+ \Gamma_{ee}^{(0)}(N) \Biggr] \nonumber\\
&& + 
{a^{\overline{\rm MS}}}^2  \Biggl[\frac{1}{8} {P_{ee}^{(0)}}^2(N) \ln^2\left(\frac{m^2_e}{\mu^2}\right)
- \frac{1}{2}  \left[P_{ee}^{(1), \rm I}(N) + P_{ee}^{(0)}(N) \Gamma_{ee}^{(0)}(N) \right] 
\ln\left(\frac{m^2_e}{\mu^2}\right)  
\nonumber
\end{eqnarray}
\begin{eqnarray}
&& + \hat{\Gamma}_{ee}^{(1),\rm I}(N) + P_{ee}^{(0)}(N) \overline{\Gamma}_{e e}^{(0)}(N) \biggr] 
+O({a^{\overline{\rm MS}}}^3)
\\
\label{eq:R2}
A_{ee}^{\overline{\rm MS},\rm II}(N) &=& {a^{\overline{\rm MS}}}^2 
\Biggl[\frac{\beta_{0,H}}{4} P_{ee}^{(0)}(N) 
\ln^2\left(\frac{m^2_e}{\mu^2}\right)
- \left[ \frac{1}{2} P_{ee}^{(1), \rm II}(N) 
+ \beta_{0,H} \Gamma_{ee}^0 \right]
\ln\left(\frac{m^2_e}{\mu^2}\right) 
\nonumber\\ &&
+ \hat{\Gamma}_{ee}^{(1),\rm II}(N) 
+ 2 \beta_{0,H} \bar{\Gamma}_{ee}^{(0)}
\Biggr] +O({a^{\overline{\rm MS}}}^3)
\\
\label{eq:R3}
A_{ee}^{\overline{\rm MS},\rm III}(N) 
&=& {a^{\overline{\rm MS}}}^2 
\Biggl[
\frac{1}{8} P_{e \gamma}^{(0)}(N) P_{\gamma e}^{(0)}(N) \ln^2\left(\frac{m^2_e}{\mu^2}\right)
- \frac{1}{2} \Bigl[P_{ee}^{(1), \rm III}(N)  
\nonumber\\ &&
+ P_{e\gamma}^{(0)}(N) \Gamma_{\gamma e}^{(0)}(N)\Bigr]
\ln\left(\frac{m^2_e}{\mu^2}\right) 
+ \hat{\Gamma}_{ee}^{(1),\rm III}(N)
+ P_{e \gamma}^{(0)}(N) \overline{\Gamma}_{\gamma e}^{(0)}(N)
\Biggr]+O({a^{\overline{\rm MS}}}^3)~. 
\nonumber\\
\end{eqnarray}
Here $\Gamma_{ij}^{(0)}$ and $\overline{\Gamma}_{ij}^{(0)}$ (\ref{Gee0}--\ref{Gee1}) denote the constant and linear term in 
$\ep$
of the unrenormalized one-loop massive OMEs, and $\hat{\Gamma}_{ij}^{(1)}$ (\ref{eq:GA1}--\ref{eq:GA3}) the corresponding 
constant part of the two-loop OMEs. 

Also the OMEs $A_{e \gamma}(N)$ and $A_{\gamma e}(N)$  have to be calculated
to $O(a)$. Here only the operators have to be renormalized.
\begin{eqnarray}
\label{eq:R2a}
A_{e \gamma}^{\overline{\rm MS}}(N) &=& a^{\overline{\rm MS}} \left[\Ahathat_{e \gamma}(N,\ep) + Z_{e \gamma}^{-1} \right]
= a^{\overline{\rm MS}}
\Biggl[- \frac{1}{2} P_{e \gamma}^{(0)}(N) \ln\left(\frac{m^2_e}{\mu^2}\right) + \Gamma_{e \gamma}^{(0)}(N) \Biggr] 
\\
\label{eq:R3a}
A_{\gamma e}^{\overline{\rm MS}}(N) &=& a^{\overline{\rm MS}} \left[\Ahathat_{\gamma e}(N,\ep) + Z_{\gamma e}^{-1} \right]
= a^{\overline{\rm MS}}
\Biggl[- \frac{1}{2} P_{\gamma e}^{(0)}(N) \ln\left(\frac{m^2_e}{\mu^2}\right) + \Gamma_{\gamma e}^{(0}(N) \Biggr]~. 
\end{eqnarray}
\section{\boldmath The $O(a)$ Operator Matrix Elements}
\label{sec:4}

\begin{figure}[h]
\begin{center}
\epsfig{file=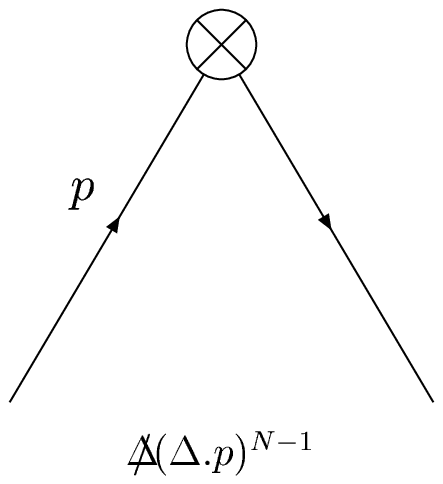,width=0.15\linewidth} \hspace*{15mm}
\epsfig{file=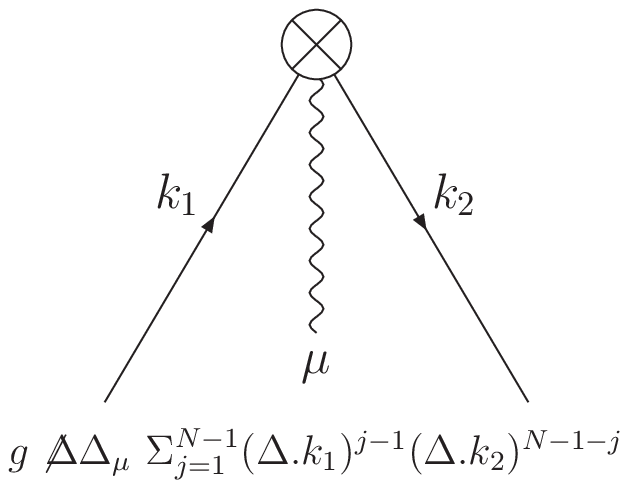,width=0.23\linewidth} \\
\vspace*{4mm}
\epsfig{file=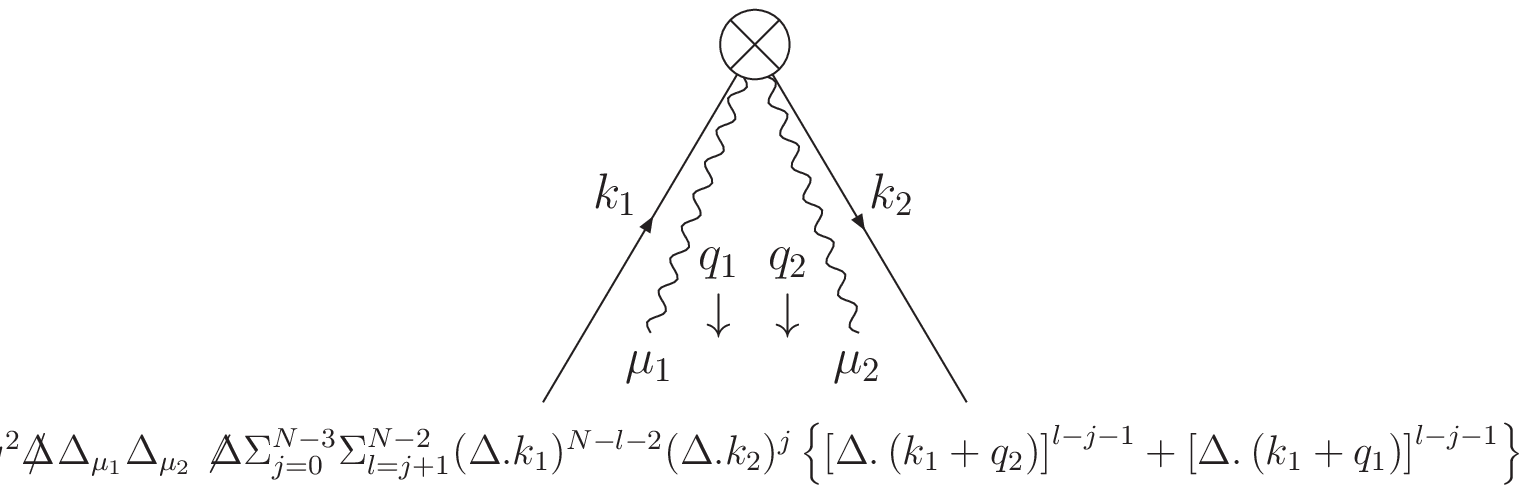,width=0.7\linewidth} 
\end{center}
\caption{\label{F:FR}
\small Feynman rules. \label{FeynmanRules}}
\end{figure}

\vspace*{1mm}
\noindent
The massive operator matrix element for process I, $A_{ee}^{(1)}$, emerges at $O(a)$ and is given by 
(\ref{eq:R1}). In the calculation of process III to $O(a^2)$ also the OMEs
$A_{e \gamma}^{(1)}$ and $A_{\gamma e}^{(1)}$, (\ref{eq:R2}, \ref{eq:R3}) contribute.
The Feynman rules for the operator insertions are given in Figure~\ref{F:FR}, cf.~\cite{BBK2}.
The external lines are taken on-shell, i.e. $p^2 = m^2_e$ for the fermion and $p^2 = 0$ 
for the photon lines, and the vector $\Delta$ is light-like with $\Delta \cdot \Delta = 0$.

\begin{figure}[h]
\begin{center}
\epsfig{file=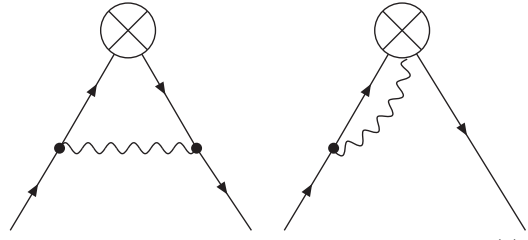,width=0.4\linewidth}
\end{center}
\vskip -.7 cm 
\caption{\label{F:pic1}
\small  Diagrams contributing to $A_{ee}^{(1)}$.}
\end{figure}

\begin{figure}[h]
\begin{center}
\epsfig{file=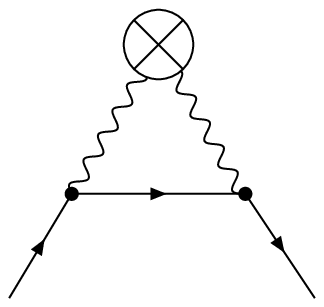,width=0.185\linewidth}~~ 
\epsfig{file=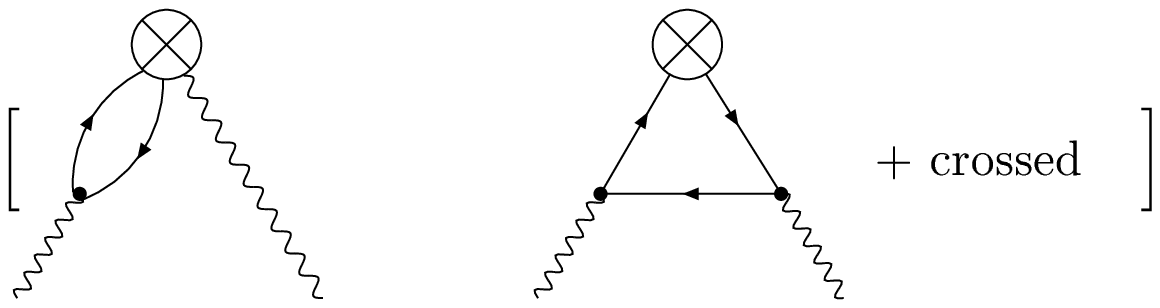,width=0.7\linewidth}
\end{center}
\vskip -.7 cm 
\caption{\label{F:pic23}
\small  Diagrams contributing to 
$A_{\gamma e}^{(1)}$ and $A_{e \gamma}^{(1)}$.}
\end{figure}

\restylefloat{figure}
%
%
In the following we present the OMEs in $x$--space. The unrenormalized 
OMEs are given by
\begin{eqnarray}
\hat{A}_{ij}^{(1)}(x,\ep) = \hat{a} \left(\frac{m^2_e}{\mu^2}\right)^{\varepsilon/2} 
S_\varepsilon\left[
- \frac{1}{\varepsilon} P^{(0)}_{ij}(x) + \Gamma_{ij}^{(0)}(x) + 
\varepsilon \overline{\Gamma}_{ij}^{(0)}(x)  + O(\varepsilon^2) \right]~,
\label{Aij1}
\end{eqnarray}
where for $\Ahathat_{ee}^{(1) \rm I}(x,\ep)$ the contributions up to $O(\ep)$ are needed.
In Figures~\ref{F:pic1}--\ref{F:pic23} the diagrams are shown, except self-energy diagrams, which 
contribute to the $O(a)$ OMEs, cf. Figure~1.
The expansion coefficients are the leading order splitting functions $P_{ij}^{(0)}$
\cite{PijLO}, $\Gamma_{ij}^{(0)}$, and $\overline{\Gamma}_{ij}^{(0)}$, 
respectively, with 
\begin{eqnarray}
\label{Pee0}
P^{(0)}_{ee}(x) &=& 
8 \DD_0(x) 
-4(1+x) + 6\delta(1-x) =
4 \left [\frac{1+x^2}{1-x} \right]_+~, \\
\label{Peg0}
P^{(0)}_{e \gamma}(x) &=& 4 \left [x^2 + (1-x)^2 \right]~,\\
\label{Pge0}
P^{(0)}_{\gamma e}(x) &=& 4 \left [\frac{1+(1-x)^2}{x}\right]~.
\end{eqnarray}
We define
\begin{eqnarray}
\DD_k(x) = \left(\frac{\ln^k(1-x)}{1-x}\right)_+~.
\end{eqnarray}
The $+$-prescription, used to regularize some of the terms, reads
\begin{eqnarray}
\int_0^1 dx \left[f(x)\right]_+ g(x) =
\int_0^1 dx f(x) \left [g(x) - g(1)
\right],
\label{plus}
\end{eqnarray}
and $g(x)~\epsilon~{\cal D}[0,1]$ denotes a test function, 
\cite{DIST}. 

The splitting functions obey the well--known relations
\begin{eqnarray}
P_{ee}^{(0)}(x)       &=& P_{\gamma e}^{(0)}(1-x),~~~x < 1\\
P_{e \gamma}^{(0)}(x) &=& P_{e \gamma}^{(0)}(1-x)
\end{eqnarray}
and
\begin{eqnarray}
\label{fnum}
\int_0^1 dx P_{ee}^{(l),\rm NS -}(x) &=& 0,~~~\forall l \in \mathbb{N} 
\\ \label{enmo}
\int_0^1 dx x\left[P_{ee}^{(0)}(x) + P_{\gamma e}^{(0)}(x)\right] &=& 0~. 
\end{eqnarray}
Eq.~(\ref{fnum}) derives from fermion number conservation with
\begin{eqnarray}
P_{ee}^{(0),\rm NS -}(x) 
&\equiv&
P_{ee}^{(0)}(x)~. 
\end{eqnarray}
Eq.~(\ref{enmo}) results from the conservation of 4--momentum. 

%
%
%
%
The $O(\varepsilon^0)$ terms are
\begin{eqnarray}
\label{Gee0}
\Gamma_{ee}^{(0)}(x) &=& 
- 8 \DD_1(x) - 4 \DD_0(x) + 4 \delta(1-x)
+ 2(1+x) \left[2 \ln(1-x) +1\right]
\nonumber\\
&=&
-  4 \left  [\frac{1 + x^2}{1 - x} \left \{
\ln(1 - x) + \frac{1}{2} \right \} \right]_+ \\
\label{Geg0}
\Gamma_{e\gamma}^{(0)}(x) &=& 0 \\
\label{Gge0}
\Gamma_{\gamma e}^{(0)}(x) &=& - 2 \frac{1+(1-x)^2}{x}[2 \ln(x) + 1]~, 
\end{eqnarray}
for $\Gamma_{e\gamma}^{(0)}(x)$ cf.~\cite{BBK,HEAV1}. 
Here we corrected Eq.~(4.28c) in \cite{BBN}.
The linear term in $\varepsilon$, $\overline{\Gamma}_{ee}^{(0)}(x)$, reads
\begin{eqnarray}
\label{Gee1}
\overline{\Gamma}_{ee}^{(0)}(x) &=& -\Biggl\{
4 \DD_2(x) + 4 \DD_1(x) + \zeta_2 \DD_0(x)  +\left(4+\frac{3}{4} 
\zeta_2\right) \delta(1-x) \nonumber\\ & &
- 2(1+x)\left[\ln^2(1-x) + \ln(1-x) + \frac{1}{4} \zeta_2\right] \Biggr\} \nonumber\\
&=&
 - 2 \left  [\frac{1 + x^2}{1 - x} \left \{
\ln^2(1 - x) + \ln(1-x) + \frac{1}{4} \zeta_2 \right \} \right]_+~. 
\end{eqnarray}

In the differential cross sections (\ref{eqMA1a}--\ref{eqMA1c}) different 
convolutions 
of the expansion coefficients of the leading order OMEs and the leading order Drell-Yan
scattering cross sections occur, see also Appendix~A.
Unlike the expansion coefficients of the massive OMEs, the coefficient functions
$\sigma_{ij}^{(0)}$ are process--dependent quantities. In case of the $e^+ 
e^-$ annihilation process the massless Wilson coefficients can be obtained
from those of the QCD Drell--Yan process given in \cite{DY1,HVN} to 
$O(a^2)$, adjusting the color factors. 
The $O(a)$ Wilson coefficients read~\footnote{Note a 
typographical error in Table~6, Ref.~\cite{HVN}.}:
\begin{eqnarray}
\label{see0}
\sigma_{ee}^{(0)}(x) &=& 
16 \DD_1(x) 
- 8 (2-\zeta_2) \delta(1-x)
- 8 \frac{\ln(x)}{1-x} -  4(1+x) \left[2\ln(1-x) - 
\ln(x)\right] \nonumber\\
\label{sge0}
\sigma_{\gamma e}^{(0)}(x) &=& 
\frac{1}{2} P_{e \gamma}^{(0)}(x) \left[2 \ln(1-x) - \ln(x) \right] + 1 + 6x - 
7 x^2~.
\end{eqnarray}
The combination
\begin{eqnarray}
\label{Gscomb}
2 \Gamma_{ee}^{(0)}(x) + \sigma_{ee}^{(0)}(x) &=& 
- 8 \DD_0(x) 
+ 8(\zeta_2-1) \delta(1-x)  
- 8 \frac{\ln(x)}{1-x} + 4(1+x)(\ln(x)+1)
\nonumber \\
&=&- 4 \frac{1+x^2}{1-x} \left[1 + 
\ln(x) \right] - 8(1- \zeta_2) \delta(1-x)
\end{eqnarray}
occurs in (\ref{eqMA1a}--\ref{eqMA1c}). Here the logarithm $\bf L$ still bears a $x$--dependence. 
Referring to $\bf \hat{L}$ instead, the $O(a_0)$ contribution to the annihilation cross
section (\ref{eqMA1a}) is given by
\begin{eqnarray}
\label{Res:a1}
T_{ee}^{1,\rm I} &=& T_{ee}^{1,1;\rm I} \hat{\bf L} + T_{ee}^{1,0;\rm I} 
\nonumber\\
&=& \left[P_{ee}^{(0)}(x) \hat{\bf L} - P_{ee}^{(0)}(x) + 2(4\zeta_2 - 1) \delta(1-z)\right]~,
\end{eqnarray}
which resembles the well-known behaviour of the splitting function $P_{ee}^{(0)}(x)$ contributing 
through $(\hat{\bf{L}} - 1)$. 

\section{\boldmath The $O(a^2)$ operator matrix elements}
\label{sec:5}

\vspace*{1mm}
\noindent
In the following we discuss the $O(a_0^2)$ contributions to operator matrix elements for the processes 
I--III individually. Further, we investigate their contribution to the differential scattering cross 
sections (\ref{eqMA1a}--\ref{eqMA1c}).
\subsection{\boldmath The OME $A_{ee}^{(2), \rm I}$}
\label{sec:5.1}

\begin{figure}[h]
\begin{center}
\epsfig{file=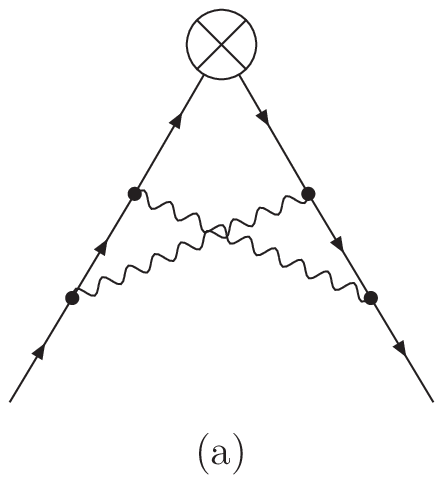,width=0.15\linewidth}~~~
\epsfig{file=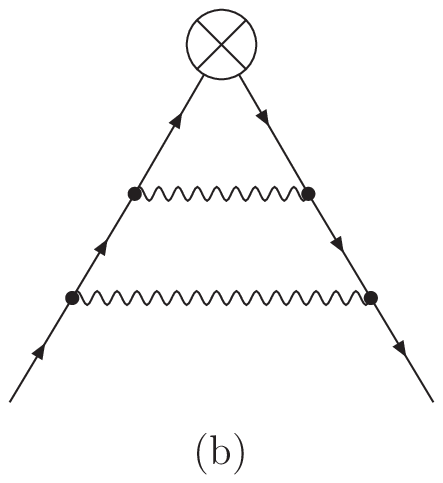,width=0.15\linewidth}~~~
\epsfig{file=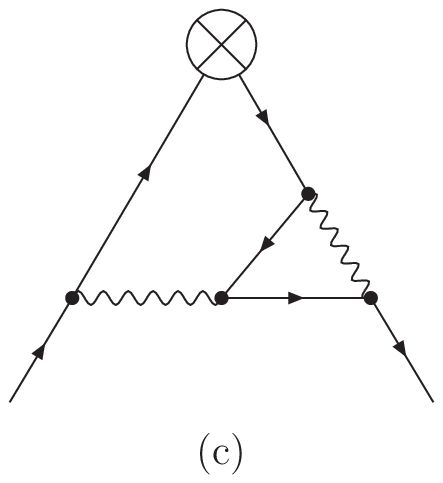,width=0.15\linewidth}~~~
\epsfig{file=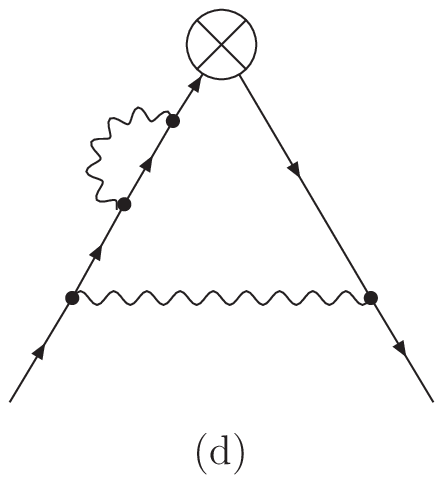,width=0.15\linewidth}~~~
\epsfig{file=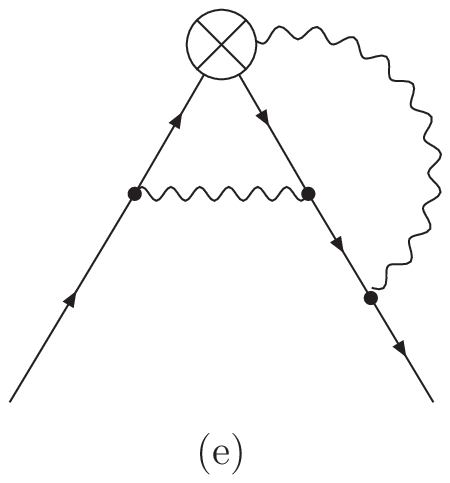,width=0.15\linewidth}
\\
\epsfig{file=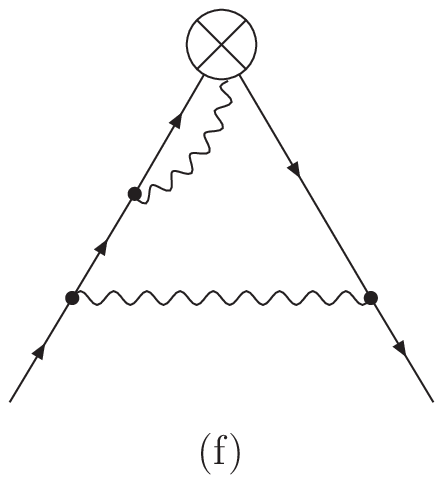,width=0.15\linewidth}~~~
\epsfig{file=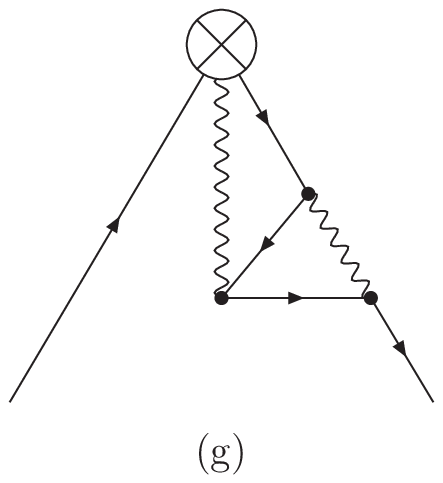,width=0.15\linewidth}~~~
\epsfig{file=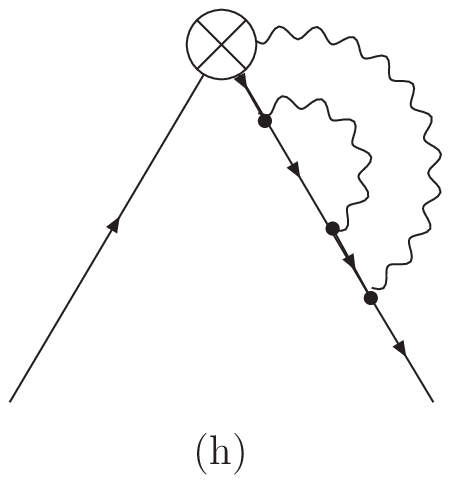,width=0.15\linewidth}~~~
\epsfig{file=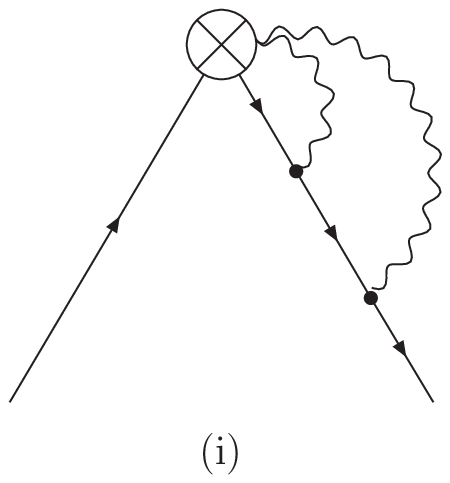,width=0.15\linewidth}~~~
\epsfig{file=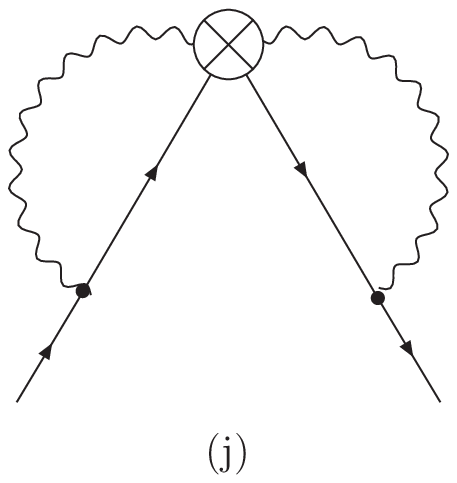,width=0.15\linewidth}
\end{center}
\vskip -.7 cm 
\caption{
\small Feynman diagrams for the calculation of the massive two-loop operator
matrix elements $A^{(2), {\rm I}}_{ee}$. \label{DiagramsI}}
\end{figure}

\begin{figure}[H]
\begin{center}
\epsfig{file=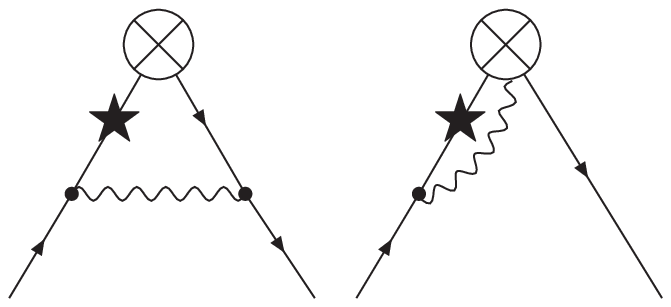,width=0.3\linewidth}~~~
\epsfig{file=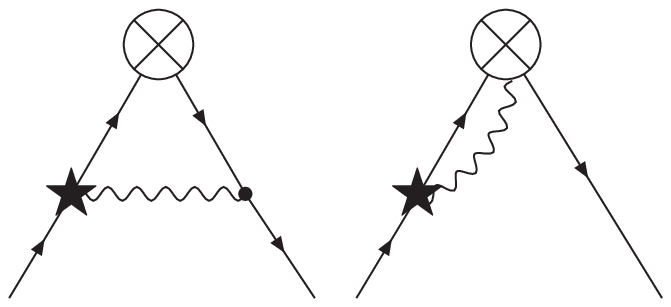,width=0.3\linewidth}
\end{center}
\vskip -.7 cm 
\caption{
\small Counterterm diagrams. The black squares represent the counterterm vertices, cf. \cite{PESC}.
\label{CounterTermDiags}}
\end{figure}

\vspace*{1mm}
\noindent
The Feynman diagrams contributing to $A_{ee}^{(2), \rm I}$ are shown in Figure~\ref{DiagramsI},  except those
contributing to the wave function renormalization, cf. Section~3. 
%
Furthermore, the counter terms shown  in Figure~\ref{CounterTermDiags}
contribute. The corresponding $Z$-factor is given by
\begin{eqnarray}
Z_{\rm CT}^{(2)}(x) &=& \left(\frac{m_e^2}{\mu^2}\right)^\ep \Biggl\{
-\frac{72}{\varepsilon^2} \delta(1-x)-\frac{24}{\varepsilon}\DD_0(x)
-24\DD_1(x)+16\DD_0(x)
\nonumber \\ 
& & 
-\left( 64+18\zeta_2 \right) \delta(1-x) 
-12 - N \left[ \left( - \frac{24}{\varepsilon} -8 \right) \delta(1-x) 
-24\DD_0(x) \right]\Biggr\}~.
\label{CTequation}
\end{eqnarray}
%
\restylefloat{figure}
%
%
It contains a term $\propto N$, which cancels a corresponding contribution in
$\Ahathat_{ee}^{(2),\rm I}$. It can be rewritten, cf.~\cite{DIST}, using
\begin{eqnarray}
\int_0^1 dx N~x^{N-1}  \delta(1-x) = 
\int_0^1 dx \left(\frac{d}{dx} x^{N}\right) \delta(1-x) = 
- \int_0^1 dx x^{N-1} \left[x \delta'(1-x)\right]
\label{eq:CT1}
\end{eqnarray}
and similar relations.

The two-loop diagrams shown in Figure \ref{DiagramsI} are calculated using the Feynman rules given in
Figure~\ref{FeynmanRules}. 
For example, the ladder diagram in Figure \ref{DiagramsI}b yields~:
\begin{eqnarray}
\int \frac{d^D k_1}{(2\pi)^D} \frac{d^D k_2}{(2\pi)^D} 
\frac{\gamma_{\mu} (\not k_1+m)
  \gamma_{\nu} (\not k_2+m) \not \!\! \Delta (\not k_2+m) \gamma^{\nu} (\not k_1
  +m) \gamma^{\mu}}{D_1^{\nu_1} D_2^{\nu_2} D_3^{\nu_3} D_4^{\nu_4} D_5^{\nu_5}}
(\Delta \cdot k_2)^n~,
\end{eqnarray}
where $n = N-1$. We introduce  the following short-hand notation for the denominators~:
\begin{eqnarray}
&& D_1 = k_1^2-m^2 \,\, , \quad D_2 = k_2^2-m^2 \,\, , 
\quad D_3 = (k_1-p)^2 \,\, , \nonumber \\
&& D_4 = (k_1-k_2)^2 \,\, , \quad D_5 = (k_2-k_1+p)^2-m^2 \,\, .
\label{props} 
\end{eqnarray}
The same propagator structure can be obtained for all diagrams from \ref{DiagramsI}a
to \ref{DiagramsI}h by choosing the flow of momenta appropriately, while
conveniently keeping all of the dot products coming from the
Feynman rules in Figure \ref{FeynmanRules} as simple as possible, i.e., as
powers of $\Delta \cdot k_1$, $\Delta \cdot k_2$ and $\Delta \cdot p$. The remaining diagrams
are calculated directly. The Dirac-structure of the numerators is projected multiplying by 
\begin{eqnarray}
\frac{1}{4} \left(\not \! p+m \right)
\end{eqnarray}
and taking the trace. We applied {\tt FORM} \cite{FORM} for this calculation. This produces a linear combination of products 
of 
all possible dot 
products of $\Delta$, $k_1$, $k_2$ and $p$. After canceling as much as possible these dot 
products against the propagators, and choosing $k_2 \cdot p$ as the only remaining 
irreducible numerator not involving $\Delta$, diagrams \ref{DiagramsI}a to \ref{DiagramsI}d 
can be expressed as linear combinations of the following type of integrals
\begin{eqnarray}
A^{a,b}_{\nu_1\nu_2\nu_3\nu_4\nu_5} &=&
\int \frac{d^D k_1}{(2\pi)^D} \frac{d^D k_2}{(2\pi)^D} 
\frac{ (\Delta \cdot k_1)^a (\Delta \cdot k_2)^b
}{D_1^{\nu_1}D_2^{\nu_2}D_3^{\nu_3}D_4^{\nu_4}D_5^{\nu_5}} \,\, ,
\label{A_type} \\
B^{a,b}_{\nu_1\nu_2\nu_3\nu_4\nu_5} &=&
\int \frac{d^D k_1}{(2\pi)^D} \frac{d^D k_2}{(2\pi)^D} 
\frac{ k_2 \cdot p (\Delta \cdot k_1)^a (\Delta \cdot k_2)^b
}{D_1^{\nu_1}D_2^{\nu_2}D_3^{\nu_3}D_4^{\nu_4}D_5^{\nu_5}} \,\,~.
\label{B_type} 
\end{eqnarray}
Diagrams \ref{DiagramsI}e and \ref{DiagramsI}f can be written as linear
combinations of integrals with the structure
\begin{equation}
E^{a,b}_{\nu_1\nu_2\nu_3\nu_4\nu_5} =
\int \frac{d^D k_1}{(2\pi)^D} \frac{d^D k_2}{(2\pi)^D} 
\frac{ (\Delta \cdot k_1)^a (\Delta \cdot k_2)^b 
}{D_1^{\nu_1}D_2^{\nu_2}D_3^{\nu_3}D_4^{\nu_4}D_5^{\nu_5}}
\sum_{j=0}^{n-1} (\Delta \cdot k_1)^j (\Delta \cdot k_2)^{n-1-j} \,\, ,
\label{E_type}
\end{equation}
and diagrams \ref{DiagramsI}g and \ref{DiagramsI}h are given by
linear combinations of integrals of the form
\begin{equation}
F^{a,b}_{\nu_1\nu_2\nu_3\nu_4\nu_5} =
\int \frac{d^D k_1}{(2\pi)^D} \frac{d^D k_2}{(2\pi)^D} 
\frac{ (\Delta \cdot k_1)^a (\Delta \cdot k_2)^b 
}{D_1^{\nu_1}D_2^{\nu_2}D_3^{\nu_3}D_4^{\nu_4}D_5^{\nu_5}}
\sum_{j=0}^{n-1} (\Delta \cdot p)^j (\Delta \cdot k_1)^{n-1-j} \,\, .
\label{F_type}
\end{equation}
We list now the results for diagrams \ref{DiagramsI}a to
\ref{DiagramsI}h in terms of the integrals above~:

\vspace*{2mm}
\noindent
Diagram \ref{DiagramsI}a:
\begin{eqnarray}
&&
{1 \over 2} (D-2) (D-4) 
\left[ A^{1, n}_{01111}
+A^{1, n}_{11101}-A^{1, n}_{11110}
-A^{1, n}_{11011}+2 A^{0, 1+n}_{02110} 
-2 A^{0, 1+n}_{12001} \right. \nonumber 
\end{eqnarray}
\begin{eqnarray}
&& \left.
-\left( \Delta \cdot p \right) A^{0, n}_{01111} \right] 
-2 (D-4) m^2_e \left[ \left( \Delta \cdot p \right) A^{0, n}_{11111}-A^{0, 1+n}_{11111} \right]
+8 m^2_e \left[ A^{0, 1+n}_{12011}+A^{0, 1+n}_{12101} \right]
\nonumber \\ &&
+{1 \over 2} (D-2) (D-8) 
\left[ A^{0, 1+n}_{11011}
-A^{0, 1+n}_{01111}
+\left( \Delta \cdot p \right) \left( A^{0, n}_{10111} 
-A^{0, n}_{11101} \right) \right]
\nonumber \\ && 
-2 (D-2) 
\left[ A^{0, 1+n}_{11101}
-\left( \Delta \cdot p \right) A^{0, n}_{11011}
-A^{0, 1+n}_{11110}
+2 B^{0, 1+n}_{12101}+2 B^{0, 1+n}_{12011} \right] 
\nonumber \\ &&
-4 m^2_e \left[ A^{0, 1+n}_{02111}+A^{0, 1+n}_{12110} \right]
-16 m^4_e A^{0, 1+n}_{12111}
\label{DiagramA}
\end{eqnarray}
Diagram \ref{DiagramsI}b:
\begin{eqnarray}
&&
(D-2)^2
\left[ A^{1, n}_{11110} 
-A^{0, 1+n}_{11110}
-\left( \Delta \cdot p \right) A^{0, n}_{11110}
-A^{0, 1+n}_{02110}
+2 B^{0,1+n}_{12110} \right] 
\nonumber \\ &&
+4 (D-2) m^2_e
\left[ A^{0, 1+n}_{21110}-A^{1, n}_{21110} \right] 
-2 (D^2-4 D+8) m^2_e A^{0, 1+n}_{12110}
-16 m^4_e A^{0, 1+n}_{22110}
\end{eqnarray}
Diagram \ref{DiagramsI}c:
\begin{eqnarray}
&&
(D-2) (D-8) 
\left[ \left( \Delta \cdot p \right) A^{n, 0}_{01111}
-\left( \Delta \cdot p \right) A^{n, 0}_{11101}
+A^{1+n, 0}_{11101}
+A^{1+n, 0}_{11011} 
-A^{1+n, 0}_{01111}
\right. \nonumber \\ && \left.
-A^{n, 1}_{11011}
+A^{n, 1}_{01111} \right] 
+D (D-2) A^{1+n, 0}_{11110}
+8 m^2_e \left[ 
2 A^{1+n, 0}_{21011}+2 A^{1+n, 0}_{21101}
-A^{1+n, 0}_{21110}
\right. \nonumber \\ && \left.
-A^{1+n, 0}_{20111} \right]
-(D-2) (D-4) 
\left[ 2 A^{1+n, 0}_{21001}
+\left( \Delta \cdot p \right) A^{n, 0}_{10111} \right] 
-32 m^4_e A^{1+n, 0}_{21111}
\nonumber \\ &&
+4 (D-2) 
\left[ A^{1+n, 0}_{10111}
+A^{n, 1}_{11101}
+\left( \Delta \cdot p \right) \left( A^{n, 0}_{11011}
-A^{n, 0}_{11110} \right)
-A^{1+n, 0}_{20101}
-A^{n, 1}_{11110} 
\right. \nonumber \\ && \left.
-2 B^{1+n, 0}_{21101}-2 B^{1+n, 0}_{21011} \right]
+4 (D-4) m^2_e
\left[ 2 A^{1+n, 0}_{11111}
-\left( \Delta \cdot p \right) A^{n, 0}_{11111}
-A^{n, 1}_{11111} \right] 
\end{eqnarray}
Diagram \ref{DiagramsI}d:
\begin{eqnarray}
&&
(D-2)^2 \left[
2 A^{1+n, 0}_{21100}
-2 A^{1+n, 0}_{11110}
+A^{n, 1}_{11110}
+\left( \Delta \cdot p \right) \left( A^{n, 0}_{11110}
-A^{n, 0}_{21100} \right)
+2 B^{1+n, 0}_{21110} \right]
\nonumber \\ &&
-2 (D-4)^2 m^2_e A^{1+n, 0}_{21110}
-4 (D-2) m^2_e \left[
\left( \Delta \cdot p \right) A^{n, 0}_{21110}
+2 A^{1+n, 0}_{31100}
+A^{n, 1}_{21110} \right] 
\nonumber \\ &&
-32 m^4_e A^{1+n, 0}_{31110}
\label{DiagramD}
\end{eqnarray}
Diagram \ref{DiagramsI}e:
\begin{eqnarray}
&&
2 (D-2) \left( \Delta \cdot p \right)^2 \left[
E^{0, 0}_{10111}
-E^{0, 0}_{11101}
+E^{0, 0}_{01111} \right]
-2 D \left( \Delta \cdot p \right) E^{1, 0}_{10111}
+4 E^{0, 2}_{11011}
\nonumber \\ &&
+2 (D-2) \left( \Delta \cdot p \right) \left[
E^{1, 0}_{11011}
+E^{0, 1}_{11011}
-E^{1, 0}_{11110} \right]
+2 (D-6) \left[ E^{1, 1}_{11011}
-E^{1, 1}_{01111} \right]
\nonumber \\ &&
+2 (D-4) \left[ E^{2, 0}_{01111}
-E^{2, 0}_{11011}+E^{2, 0}_{11110}-E^{2, 0}_{11101}
+\left( \Delta \cdot p \right) E^{0, 1}_{01111} \right]
-4 E^{0, 2}_{01111} 
\nonumber \\ &&
+4 (D-3) \left( \Delta \cdot p \right) \left[ E^{1, 0}_{11101}
-E^{1, 0}_{01111} \right]
+4 \left( \Delta \cdot p \right) \left[ E^{0, 1}_{10111}
-E^{0, 1}_{11101} \right]
+4 E^{1, 1}_{11110}
\nonumber \\ &&
-4 E^{1, 1}_{11101}
+8 m^2_e \left[ \left( \Delta \cdot p \right) 
\left( E^{0, 1}_{11111}
+E^{1, 0}_{11111} \right)
+E^{1, 1}_{11111}
-E^{2, 0}_{11111} \right]
\end{eqnarray}
Diagram \ref{DiagramsI}f:
\begin{eqnarray}
4 (D-2) \left[ \left( \Delta \cdot p \right) E^{0, 1}_{11110} 
-E^{1, 1}_{11110} \right]
+16 m^2_e E^{1, 1}_{21110}
\end{eqnarray}
Diagram \ref{DiagramsI}g:
\begin{eqnarray}
&&
2 (D-4) \left[ \left( \Delta \cdot p \right) F^{0, 1}_{01111}
-F^{2, 0}_{11011}
-F^{2, 0}_{11101}
+F^{2, 0}_{11110}
+F^{2, 0}_{01111} \right]
+2 D \left( \Delta \cdot p \right) F^{1, 0}_{10111}
\nonumber \\ &&
+2 (D-2) \left[ 
\left( \Delta \cdot p \right) \left( F^{0, 1}_{11101}
-F^{0, 1}_{10111} \right)
+F^{1, 1}_{11101}
-F^{1, 1}_{11110}
+F^{0, 2}_{01111}
-F^{0, 2}_{11011} \right]
\nonumber \\ &&
+4 (D-3)
\left[ F^{1, 1}_{11011}-F^{1, 1}_{01111} \right]
+2 (D-6) \left( \Delta \cdot p \right) 
\left[ F^{1, 0}_{11101}-F^{1, 0}_{01111} \right]
\nonumber \\ &&
+4 \left( \Delta \cdot p \right)^2 \left[ F^{0, 0}_{11101}
-F^{0, 0}_{10111}
-F^{0, 0}_{01111} \right]
+4 \left( \Delta \cdot p \right) \left[ F^{1, 0}_{11110}
-F^{1, 0}_{11011}
-F^{0, 1}_{11011} \right]
\nonumber \\ &&
+8 \left( \Delta \cdot p \right) m^2_e \left[ 
F^{1, 0}_{11111}
+F^{0, 1}_{11111} \right]
-8 m^2_e F^{2, 0}_{11111}
+8 m^2_e F^{1, 1}_{11111}
\end{eqnarray}
Diagram \ref{DiagramsI}h:
\begin{eqnarray}
4 (D-2) \left( \Delta \cdot p \right) \left[ 
F^{0, 1}_{11110} 
-F^{1, 0}_{11110}  
+F^{1, 0}_{21100} \right]
+16 \left( \Delta \cdot p \right) m^2_e F^{1, 0}_{21110}~.
\label{DiagramH}
\end{eqnarray}

Various of the integrals appearing in these expressions have only three or four 
propagators. The 4-propagator integrals can be represented in terms of up to three 
Feynman parameter integrals over the unit cube. In some cases, a direct calculation
will give integrals with the following structure
\begin{equation}
\label{FeynStruct1}
I(\ep,n) = \int_0^1 dx \int_0^1 dy \int_0^1 dz \,\, x^n f(x,y,z;\ep)~,
\end{equation}
while in other cases they will be of the form
\begin{equation}
\label{FeynStruct2}
I(\ep,n) = \int_0^1 dx \int_0^1 dy \int_0^1 dz \,\, x^n y^n f(x,y,z;\ep)~.
\end{equation}
In the first case, the integrals represent a Mellin transform, and only the $y$ and $z$ integrals
have to be performed. A mapping of Feynman parameters providing this case can be applied for 
(\ref{FeynStruct2}), cf.~\cite{HambergThesis}, by the following transformation
of the unit square into itself
\begin{equation}
x' = xy ,  \quad y' = \frac{x(1-y)}{1-xy}.
\label{map}
\end{equation}

For example, the integral
$A^{a,b}_{\nu_1\nu_2\nu_3\nu_4 0}$ can be calculated combining first
the propagators $D_2$ and $D_4$ by introducing a Feynman parameter, then performing
the integral in $k_2$. After this the result is combined with the remaining two
propagators and the $k_1$ integral is carried out. One obtains
\begin{eqnarray}
A^{a,b}_{\nu_1\nu_2\nu_3\nu_4 0} &=&
C \int^1_0 dx \int^1_0 dy \int^1_0 dz \,\, 
x^{b+\nu_{134}-3-{\ep \over 2}} (1-x)^{-\nu_4+1+{\ep \over 2}} 
y^{a+b+\nu_3-1} 
\nonumber \\ && \phantom{C \int^1_0 dx \int^1_0 dy \int^1_0 dz \,\,}
\times (1-y)^{-\nu_3+1+{\ep \over 2}}
z^{\nu_{24}-3-{\ep \over 2}} (1-z)^{\nu_1-1} 
\nonumber \\ && \phantom{C \int^1_0 dx \int^1_0 dy \int^1_0 dz \,\,}
\times \left[ z (1-x)+x (1-y) \right]^{4-\nu_{1234}+\ep} \,\, .
\label{FeynParExa1}
\end{eqnarray}
Here we use the notation
\begin{eqnarray}
\nu_{i,j \ldots k}=\sum_{l=\{i,j \ldots k\}} \nu_l
\end{eqnarray}
and  $C=(-1)^{\nu_{1234}}\Gamma(\nu_{1234}-4-\ep) (m^2_e)^{4-\nu_{1234}+\ep} (\Delta \cdot p)^{a+b}$. 
In the case $a = n, b = 0$, after interchanging $x$
and $y$, the integral (\ref{FeynParExa1}) is of the form given in 
(\ref{FeynStruct1}). On the other hand, if
$a = 0, b = n$, we obtain an expression of the form (\ref{FeynStruct2}). The change of
variables according to (\ref{map}) yields
\begin{eqnarray}
A^{0,n}_{\nu_1 \nu_2 \nu_3 \nu_4 0} &=&
C \int^1_0 dx' \int^1_0 dy' \int^1_0 dz \,\, x'^n
(x'+y'-x'y')^{\nu_{134}-4-\ep} x'^{\nu_3-1} y'^{-\nu_3+1+{\ep \over 2}}
\nonumber \\ && \phantom{C \int^1_0 dx \int^1_0 dy \int^1_0}
\times (1-x')^{-\nu_{123344}+7+2\ep} 
(1-y')^{-\nu_4+1+{\ep \over 2}} 
z^{\nu_{24}-3-{\ep \over 2}} 
\nonumber \\ && \phantom{C \int^1_0 dx \int^1_0 dy \int^1_0}
\times (1-z)^{\nu_1-1} 
\left[y'+z-y'z \right]^{4-\nu_{1234}+\ep} \,\, ,
\label{FeynParExa2}
\end{eqnarray}
which is of the form (\ref{FeynStruct1}) as desired.

The easiest way to calculate the integrals with five propagators is to write
them in terms of 4-propagator integrals, using integration by parts (IBP)
identities \cite{IBP}. For the 5-propagator $A$-type integrals which appear in expressions
(\ref{DiagramA}) to (\ref{DiagramD}), one obtains
\begin{eqnarray}
A^{n,0}_{11111} &=& \frac{1}{\ep} \left(
  A^{n,0}_{12101}-A^{n,0}_{02111}+A^{n,0}_{11102} \right)
\label{An011111_IBP}
\\
A^{0,n}_{11111} &=& \frac{1}{\ep} \left(
  A^{0,n}_{21011}+A^{0,n}_{11120}+A^{0,n}_{01121}-A^{0,n}_{10121}+A^{0,n}_{11012}-A^{0,n}_{10112} \right)
\label{A0n11111_IBP}
\end{eqnarray}
\begin{eqnarray}
\label{A21111_IBP}
A^{n,0}_{21111} &=& -\frac{1}{\ep} \left[ -A^{n,0}_{22101}-A^{n,0}_{21102}
                +\frac{1}{1-\ep} \left( -A^{n,0}_{12102}+2A^{n,0}_{03111}
                                        -2A^{n,0}_{13101} \right) \right]
\\
\label{A12111_IBP}
A^{0,n}_{12111} &=& -\frac{1}{\varepsilon} 
                     \left[ \phantom{\frac{|}{|}} 
                     -A^{0,n}_{12210}-A^{0,n}_{02211}-A^{0,n}_{12102}-A^{0,n}_{22101} 
                     \right. 
                     \nonumber \\ && 
                      \phantom{ \frac{1}{\varepsilon} \left[ \right. }
                     +\frac{1}{1-\varepsilon}
                     \left(
                     -2A^{0,n}_{31101} + 2A^{0,n}_{30111} - A^{0,n}_{21210} 
                     +2A^{0,n}_{20211} - A^{0,n}_{21102}
                     \right.
                     \nonumber \\  && 
                     \phantom{ \frac{1}{\varepsilon} \left[ \right. +\frac{1}{1+\varepsilon} \left(
                     \right. }
                     \left. \left.
                     - A^{0,n}_{21201} - 2A^{0,n}_{11310} + 2A^{0,n}_{10311} 
                     - 2A^{0,n}_{01311} -A^{0,n}_{11202} \right) \phantom{\frac{|}{|}} \right]~.
\end{eqnarray}
For the $E$-type integrals one finds
\begin{equation}
\label{E_IBP}
E^{a,b}_{11111} = \frac{1}{1-\ep} \left( -bA^{n+a,b-1}_{11111}+(n+b)A^{a,n+b-1}_{11111}
                -E^{a,b}_{12101}+E^{a,b}_{02111}-E^{a,b}_{11102} \right)~.
\end{equation}
In this way, the 5-propagator $E$-type integrals can be obtained from 4-propagator
integrals of the same type, together with the previously calculated $A^{a,b}_{11111}$.
In (\ref{E_IBP}) the factor of $n$ multiplying the integral $A^{a,n+b-1}_{11111}$
can be absorbed in the integrand using integration by parts 
\begin{eqnarray}
n \int^1_0 dx \,\, x^{n-1} f(x) = \int^1_0 dx \,\, x^n \left[ f(1)
  \delta(1-x)-f'(x) \right]  ,
\end{eqnarray}
where we have assumed that $x^n f(x) |_{x=0}=0$ and $f(x)$ is regular at $x=1$, which is
the case for the diagrams being considered.
Finally, we have
\begin{equation}
F^{a,b}_{11111}=\frac{1}{b+\ep} \left(
  bF^{a+1,b-1}_{11111}+F^{a,b}_{12101}-F^{a,b}_{02111}+F^{a,b}_{11102} \right) .
\label{F_IBP}
\end{equation}
This relation can be used recursively, that is, once $F^{a,0}_{11111}$ is
obtained from only 4-propagator integrals, we can use it to obtain
$F^{a-1,1}_{11111}$ and then $F^{0,2}_{11111}$. More details on the way to
obtain these equations can be found in Appendix~\ref{SectionIBP}.

The results for all of the required 4- and 5-propagator integrals, appearing
in expressions (\ref{DiagramA}--\ref{DiagramH}) can be found
in Appendix~\ref{Feyn_Int_Res_Section}. All of these integrals were checked
numerically for the first few moments using {\tt Tarcer}~\cite{TARCER}. It turns out that all 
of the integrals can be expressed in terms of Nielsen integrals
\cite{NIELS}, 
\begin{eqnarray}
S_{n,p}(x) &=& \frac{(-1)^{n+p-1}}{(n-1)! p!} \int_0^1 \frac{dz}{z} \ln^{n-1}(z) \ln^p(1-x z)
\\
\Li_n(x)    &=& S_{n-1,1}(x)~,
\end{eqnarray}
partly weighted by denominators $1/x$, $1/(1-x)^k$, $1/(1+x)^l$,
$k,l \leq 3$, as well as the distributions $\delta(1-x)$ and ${\cal D}_k(x)$.

The diagrams with two photons being attached to the operator vertex are more simple, 
since in this case only four propagators contribute. Furthermore,
three factors of $\not \!\!\! \Delta$ occur and only one term survives in the numerator after
taking the trace. For diagram \ref{DiagramsI}i we obtain
\begin{eqnarray}
I_{5,i} &=& S_\varepsilon^2 \int^1_0 dx \,\, x^n \BLCB
{32 \over \ep^2} \BLB -\ln(1-x) -{x \over 1-x} \ln(x) 
-1+(1-\zeta_2) \delta(1-x)+{\cal D}_0(x)
\nonumber\\ && \phantom{ \int^1_0 dx \,\, x^n \BLCB }
+{\cal D}_1(x)
\BRB
- {8 \over \ep} \BLB
  (4 {\cal D}_0(x)-4) \zeta_2-4-6 {\cal D}_2(x)-4 {\cal D}_1(x)
+4 {\cal D}_0(x) 
\nonumber\\ && \phantom{ \int^1_0 dx \,\, x^n \BLCB}
+{x \over 1-x} \left( 3 \ln^2(x)+2 \ln(x) \ln(1-x) 
-4 \ln(x)-2 {\rm Li}_2(1-x) \right)
\nonumber\\
&& \phantom{ \int^1_0 dx \,\, x^n \BLCB}
+2 \ln(1-x)+6 \ln^2(1-x)
+\left( 10-4 \zeta_2-6 \zeta_3 \right) \delta(1-x) \BRB
-32
\nonumber\\ && \phantom{ \int^1_0 dx \,\, x^n \BLCB}
+24 \zeta_2-64 \zeta_3+56 \zeta_2 \ln(1-x)
+48 \ln(1-x)-12 \ln^2(1-x)
\nonumber\\ && \phantom{ \int^1_0 dx \,\, x^n \BLCB}
-{112 \over 3} \ln^3(1-x)
+{8 x \over 1-x} \BLB \zeta_2 \ln(x)
 -8 {\rm S}_{1,2}(1-x)
+5 {\rm Li}_3(1-x)
\nonumber \\
&& \phantom{ \int^1_0 dx \,\, x^n \BLCB}
-3 \ln(1-x) {\rm Li}_2(1-x)
-2 \ln^2(x) \ln(1-x) 
-2 {\rm Li}_2(1-x)
\nonumber\\ && \phantom{ \int^1_0 dx \,\, x^n \BLCB}
-{1 \over 2} \ln^3(x)
+2 \ln(x) \ln(1-x)
+2 \ln^2(x)
-2 \ln(x) {\rm Li}_2(1-x)
\nonumber\\ && \phantom{ \int^1_0 dx \,\, x^n \BLCB}
-{5 \over 2} \ln(x) \ln^2(1-x)
\BRB
-32 {\rm Li}_2(1-x)
+(32+8 \zeta_2+64 \zeta_3) {\cal D}_0(x)
\nonumber \\
&& \phantom{ \int^1_0 dx \,\, x^n \BLCB}
-(56 \zeta_2+32) {\cal D}_1(x)
+16 {\cal D}_2(x)+{112 \over 3} {\cal D}_3(x)
\nonumber\\ && \phantom{ \int^1_0 dx \,\, x^n \BLCB}
+(152-120 \zeta_4-24 \zeta_2-32 \zeta_3) \delta(1-x) 
\BRCB m_e^{2\ep} (\Delta \cdot p)^n \,\, .
\end{eqnarray}
Diagram \ref{DiagramsI}j yields
\begin{eqnarray}
I_{5,j} 
&=& S_\varepsilon^2 \int^1_0 dx \,\, x^n \BLCB
{16 \over \ep^2} \BLB -1-2 \ln(1-x)-x+2 {\cal D}_0(x)+2 {\cal D}_1(x)
+\left( 1-\zeta_2 \right) \delta(1-x) \BRB
\nonumber\\ && \phantom{ \int^1_0 dx \,\, x^n \BLCB}
-{16 \over \ep} \BLB 2 x \ln(1-x)+3 \ln^2(1-x) 
-2 x-2 \zeta_2
+\left( 2+2 \zeta_2 \right) {\cal D}_0(x)
\nonumber\\ && \phantom{ \int^1_0 dx \,\, x^n \BLCB}
-2 {\cal D}_1(x) -3 {\cal D}_2(x)
+\left( 2-2 \zeta_3 \right) \delta(1-x) \BRB
+32-64 x-64 \zeta_3
\nonumber\\ && \phantom{ \int^1_0 dx \,\, x^n \BLCB}
-(20-12 x) \zeta_2
+16 (1-2 x) \left[ \ln^2(1-x)-2 \ln(1-x) \right]
+16 {\cal D}_2(x)
\nonumber\\ && \phantom{ \int^1_0 dx \,\, x^n \BLCB}
-{112 \over 3} \ln^3(1-x) 
+56 \zeta_2 \ln(1-x)
+\left( 32+8 \zeta_2+64 \zeta_3 \right) {\cal D}_0(x)
\nonumber\\ && \phantom{ \int^1_0 dx \,\, x^n \BLCB}
-\left( 32+56 \zeta_2 \right) {\cal D}_1(x)
+\left( 48+4 \zeta_2-{23 \over 45} \pi^4 \right) \delta(1-x)
+{112 \over 3} {\cal D}_3(x)
\nonumber\\ 
&& \phantom{ \int^1_0 dx \,\, x^n \BLCB}
+(-1)^n \BLB
{16 \over \ep^2} \BLP
{2 x \over 1+x} \ln(x)+1-x \BRP
-{16 \over \ep} \BLP
2 x \ln(x)+2-2 x
\nonumber \\ && \phantom{ \int^1_0 dx \,\, x^n \BLCB +(-1)^n \BLB}
-{x \over 1+x} \BLP
2 \zeta_2+\ln^2(x)+4 \ln(x) \ln(1+x)
+4 {\rm Li}_2(-x) \BRP \BRP
\nonumber 
\end{eqnarray}
\begin{eqnarray}
&& \phantom{ \int^1_0 dx \,\, x^n \BLCB +(-1)^n \BLB}
+64-64 x
-32 (1+x) \ln(x) \ln(1+x)
+64 x \ln(x)
\nonumber 
\\
&& \phantom{ \int^1_0 dx \,\, x^n \BLCB +(-1)^n \BLB}
+{8 x \over 1+x} \BLP
\zeta_2 \ln(x)
+8 \ln^2(1+x) \ln(x)
+8 \ln(x) {\rm Li}_2(-x)
\nonumber \\ && \phantom{ \int^1_0 dx \,\, x^n \BLCB +(-1)^n \BLB}
+4 \ln^2(x) \ln(1+x)
+8 \zeta_2 \ln(1+x)
+16 \ln(1+x) {\rm Li}_2(-x) 
\nonumber \\
&& \phantom{ \int^1_0 dx \,\, x^n \BLCB +(-1)^n \BLB}
+{2 \over 3} \ln^3(x)
+16 {\rm S}_{1,2}(-x)
-8 {\rm Li}_3(-x)
-8 \zeta_3 \BRP
-16 x \ln^2(x) 
\nonumber\\ 
&&
\phantom{ \int^1_0 dx \,\, x^n \BLCB +(-1)^n \BLB}
-(12+20 x) \zeta_2
-32 (1+x) {\rm Li}_2(-x)
\BRB
\BRCB m^{2\ep}_e (\Delta \cdot p)^n \,\, .
\end{eqnarray}

The result for the unrenormalized matrix element $\hat{\hat{A}}^{(2),{\rm I}}_{ee}$
is
\begin{eqnarray}
\hat{\hat{A}}^{(2),{\rm I}}_{ee}
&=& S_\ep^2 \int^1_0 dx \,\, x^n \BLCB
\frac{8}{\ep^2} \BLB
-\frac{1+3 x^2}{1-x} \ln(x)
-4 (1+x) \ln(1-x)
-8-4 x
+12 {\cal D}_0(x)
\nonumber \\ && \phantom{ \int^1_0 dx \,\, x^n \BLCB}
+8 {\cal D}_1(x)
+\left( 18-4 \zeta_2 \right) \delta(1-x) 
\BRB
-\frac{4}{\ep} \BLB
\frac{4 x^2+2 x-1}{1-x} \ln(x)
-7+3 x
\nonumber \\ && \phantom{ \int^1_0 dx \,\, x^n \BLCB}
+(1+x) \left( 12 \ln^2(1-x)-\frac{1}{2} \ln^2(x)
-4 {\rm Li}_2(1-x)-8 \zeta_2 \right)
-24 {\cal D}_2(x)
\nonumber \\ && \phantom{ \int^1_0 dx \,\, x^n \BLCB}
+(28+12 x) \ln(1-x)
+2 \frac{5 x^2+1}{1-x} \ln(x) \ln(1-x)
+(16 \zeta_2-2) {\cal D}_0(x)
\nonumber\\ 
&& \phantom{ \int^1_0 dx \,\, x^n \BLCB}
-40 {\cal D}_1(x)
-(10 \zeta_3-5 \zeta_2-24) \delta(1-x)
\BRB
%
+\left( {104 \over 3} x+16+{32 \over 1-x} \right) \zeta_2
\nonumber \\ && \phantom{ \int^1_0 dx \,\, x^n \BLCB}
+\frac{1+3 x^2}{1-x} \left[ 6 \zeta_2 \ln(x)
-8 \ln(x) {\rm Li}_2(1-x)-4 \ln^2(x) \ln(1-x) \right]
\nonumber \\
&& \phantom{ \int^1_0 dx \,\, x^n \BLCB}
-(24-18 x) \ln(1-x)
+16 \frac{1+x^2}{1-x} \left[ 2 {\rm Li}_3(-x)-\ln(x) {\rm Li}_2(-x) \right]
\nonumber \\ && \phantom{ \int^1_0 dx \,\, x^n \BLCB}
+\left( {22 \over 3} x+32+{64 \over 3 (1-x)^2}-{51 \over 1-x}
-{16 \over 3 (1-x)^3} \right) \ln^2(x)
+{80 \over 3 (1-x)}
\nonumber \\ && \phantom{ \int^1_0 dx \,\, x^n \BLCB}
+56 (1+x) \zeta_2 \ln(1-x)
-(104+32 x) \ln^2(1-x)
-\frac{1}{3} (1+x) \ln^3(x)
\phantom{ \int^1_0 dx \,\, x^n \BLCB}
\nonumber 
\\
&&
\phantom{ \int^1_0 dx \,\, x^n \BLCB}
+\left( {178 \over 3}-36 x+{64 \over 3 (1-x)^2}
-{140 \over 3 (1-x)}-{48 \over 1+x} \right) \ln(x) 
\nonumber \\ && \phantom{ \int^1_0 dx \,\, x^n \BLCB}
+4 \frac{x^2-8 x-6}{1-x} \ln(x) \ln(1-x) 
-2 \frac{1+17 x^2}{1-x} \ln(x) \ln^2(1-x)
\nonumber \\ && \phantom{ \int^1_0 dx \,\, x^n \BLCB}
 +32 \frac{1+x}{1-x} \left[ \ln(x) \ln(1+x)+{\rm Li}_2(-x) \right]
-\frac{112}{3} (1+x) \ln^3(1-x)
\nonumber \\ && \phantom{ \int^1_0 dx \,\, x^n \BLCB}
+4 \frac{5-11 x^2}{1-x} \left[ \ln(1-x) {\rm Li}_2(1-x)
-{\rm Li}_3(1-x)-2 \zeta_3 \right]
-30 x-{50 \over 3}
\nonumber \\
&& \phantom{ \int^1_0 dx \,\, x^n \BLCB}
-4 \frac{13 x^2+9}{1-x} {\rm S}_{1,2}(1-x)
+\frac{4 (16 x^2-10 x-27)}{3 (1-x)} {\rm Li}_2(1-x)
\nonumber \\ && \phantom{ \int^1_0 dx \,\, x^n \BLCB}
+\left( 16-40 \zeta_2+128 \zeta_3 \right) {\cal D}_0(x)
+\left( 24-112 \zeta_2 \right) {\cal D}_1(x)
+144 {\cal D}_2(x)
\nonumber 
\end{eqnarray}
\begin{eqnarray}
&& \phantom{ \int^1_0 dx \,\, x^n \BLCB}
+{224 \over 3} {\cal D}_3(x)
+\left[ 192-{67 \over 45} \pi^4
+\left( 128-144 \ln(2) \right) \zeta_2+82 \zeta_3 \right] \delta(1-x)
%
\nonumber\\ 
&& \phantom{ \int^1_0 dx \,\, x^n \BLCB}
+N \left[ \left( - {24 \over \ep}-8 \right) \delta(1-x)-24 {\cal D}_0(x) \right]
\nonumber \\ && \phantom{ \int^1_0 dx \,\, x^n \BLCB}
-(-1)^N \BLB
- \frac{4}{\ep} \BLP
\frac{1+x^2}{1+x} \left( 4 \ln(x) \ln(1+x)+2 \zeta_2
+4 {\rm Li}_2(-x)-\ln^2(x) \right)
\nonumber \\ && \phantom{ \int^1_0 dx \,\, x^n \BLCB +(-1)^n \BLB}
+4 x-4
-2 (1+x) \ln(x)
\BRP
+\frac{2 (1-x) (45 x^2+74 x+45)}{3 (1+x)^2}
\nonumber 
\end{eqnarray}
\begin{eqnarray}
&& \phantom{ \int^1_0 dx \,\, x^n \BLCB +(-1)^n \BLB}
+\frac{4 (x^2+10 x-3)}{3 (1+x)} \left( \zeta_2+2 {\rm Li}_2(-x)
+2 \ln(x) \ln(1+x) \right)
\nonumber \\ && \phantom{ \int^1_0 dx \,\, x^n \BLCB +(-1)^n \BLB}
+\frac{1+x^2}{1+x} \BLP
8 \zeta_2 \ln(x)-24 \zeta_2 \ln(1+x)+36 \zeta_3
-\frac{2}{3} \ln^3(x)
\nonumber\\ 
&& \phantom{ \int^1_0 dx \,\, x^n \BLCB +(-1)^n \BLB}
-4 \ln^2(x) \ln(1+x)-24 \ln(x) \ln^2(1+x)
-16 {\rm S}_{1,2}(1-x)
\nonumber \\ && \phantom{ \int^1_0 dx \,\, x^n \BLCB +(-1)^n \BLB}
-8 \ln(x) {\rm Li}_2(1-x)
-48 \ln(1+x) {\rm Li}_2(-x)
+40 {\rm Li}_3(-x)
\nonumber\\ 
&& \phantom{ \int^1_0 dx \,\, x^n \BLCB +(-1)^n \BLB}
-24 \ln(x) {\rm Li}_2(-x)
-48 {\rm S}_{1,2}(-x)
\BRP
+4 x \frac{1-x}{(1+x)^3} \ln^2(x)
\nonumber \\
&& \phantom{ \int^1_0 dx \,\, x^n \BLCB +(-1)^n \BLB}
+\frac{2 (9+12 x+30 x^2-20 x^3-15 x^4)}{3 (1+x)^3} \ln(x)
\nonumber\\  
&& \phantom{ \int^1_0 dx \,\, x^n \BLCB +(-1)^n \BLB}
-\frac{16 (x^4+12 x^3+12 x^2+8 x+3)}{3 (1+x)^3} {\rm Li}_2(1-x)
\nonumber \\ && \phantom{ \int^1_0 dx \,\, x^n \BLCB +(-1)^n \BLB}
-4 x^3 \frac{5-x}{(1+x)^3} \ln^2(x)
\BRB
\BRCB m^{2\ep}_e (\Delta \cdot p)^n
\label{BigA2Iunre}
\end{eqnarray}
Notice the presence of the term proportional to $N$, which is canceled by the counterterms. This term stems from diagrams
with one-loop insertions, such as  diagrams~\ref{DiagramsI}d and \ref{DiagramsI}h.
 
\subsection{\boldmath The OME $A_{ee}^{(2), \rm II}$}
\label{sec:5.2}

\begin{figure}[h]
\begin{center}
\epsfig{file=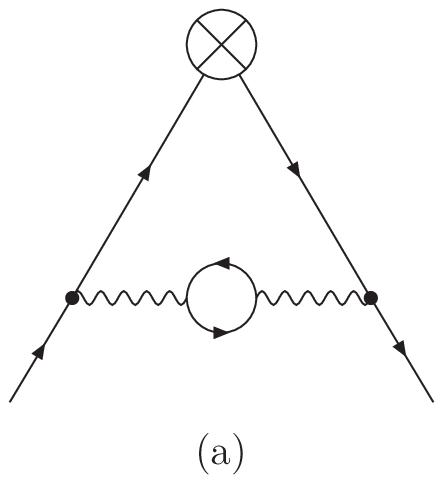,width=0.15\linewidth} \hspace*{15mm}
\epsfig{file=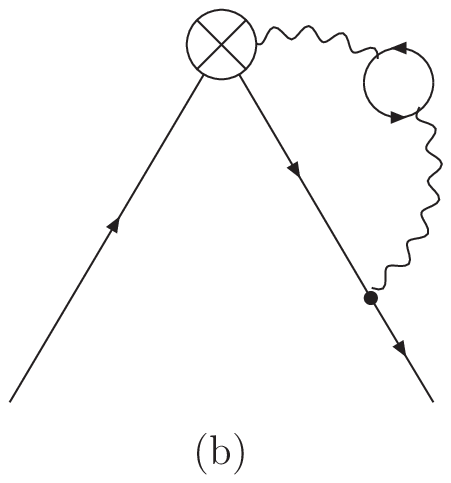,width=0.15\linewidth}
\end{center}
\vskip -.7 cm 
\caption{
\small Feynman diagrams for the calculation of the massive two-loop operator
matrix elements $A^{(2), {\rm II}}_{ee}$. \label{DiagramsII}}
\end{figure}

\vspace*{1mm}
\noindent
The matrix elements $\hat{\hat{A}}^{(2), {\rm II}}_{ee}$ are obtained from the fermionic one-loop
insertions shown in Figure~\ref{DiagramsII} supplemented by the corresponding self--energy
diagrams. 
%
%
%

\noindent
One may use the fermionic one-loop vacuum polarization 
\begin{eqnarray}
\Pi_2^{\mu \nu}(q) =-\frac{8 e^2}{(4\pi)^{D/2}}  
\left( q^2 g^{\mu \nu} - q^{\mu} q^{\nu} \right)
\int^1_0 dx  \frac{\Gamma\left( 2-D/2 \right) x(1-x)}{(m^2_e-x(1-x)q^2)^{2-D/2}} \,\, ,
\end{eqnarray}
in the corresponding one-loop diagrams, cf. 
Section~\ref{sec:4}. The result for $\hat{\hat{A}}^{(2), \rm II}_{ee}$ is
\begin{eqnarray}
\hat{\hat{A}}^{(2),{\rm II}}_{ee}
&=& S_\ep^2 \int^1_0 dx \,\, x^n \BLCB
{8 \over \ep^2} \left[ {2 \over 3} (1+x)
-{4 \over 3} {\cal D}_0(x)-3 \delta(1-x) \right]
-{8 \over \ep} \BLB {5 x-7 \over 9}
-{1+x^2 \over 3 (1-x)} \ln(x)
\nonumber \\ && 
\phantom{ \int^1_0 dx \,\, x^n \BLCB}
-{4 \over 3} (1+x) \ln(1-x)
+{8 \over 3} {\cal D}_1(x)
+{2 \over 9} {\cal D}_0(x)
-\left( {2 \over 3} \zeta_2+3 \right) \delta(1-x) \BRB
\nonumber\\ 
&& \phantom{ \int^1_0 dx \,\, x^n \BLCB}
+{76 \over 27} x-{572 \over 27}
-\left( 12 x+{4 \over 3}+{8 \over 1-x} \right) \ln(x)
+{128 \over 9 (1-x)^2}
+{80 \over 27 (1-x)}
\nonumber \\ && \phantom{ \int^1_0 dx \,\, x^n \BLCB}
-{64 \over 9 (1-x)^3}
-{32 \over 9} \left( {1 \over (1-x)^2}-{5 \over (1-x)^3}
+{2 \over (1-x)^4} \right) \ln(x)
\nonumber \\ && \phantom{ \int^1_0 dx \,\, x^n \BLCB}
+{16 \over 3} (1+x) \left( \ln(1-x)+\ln^2(1-x) \right)
-{2 (1+x^2) \over 3 (1-x)} \ln^2(x)
\nonumber \\ && \phantom{ \int^1_0 dx \,\, x^n \BLCB}
+\left( {8 \over 3} \zeta_3+18 \zeta_2
-{4420 \over 81} \right) \delta(1-x)
+\left( {224 \over 27}
-{8 \over 3} \zeta_2 \right) {\cal D}_0(x)
\nonumber \\ && \phantom{ \int^1_0 dx \,\, x^n \BLCB}
+{4 \over 3} (1+x) \zeta_2
-{32 \over 3} \left( {\cal D}_1(x)+{\cal D}_2(x) \right)
\BRCB m^{2 \ep}_e (\Delta \cdot p)^n  
\BRCB~.
\label{BigA2IIunre}
\end{eqnarray}

\subsection{\boldmath The OME $A_{ee}^{(2), \rm III}$}
\label{sec:5.3}

\begin{figure}[H]
\begin{center}
\epsfig{file=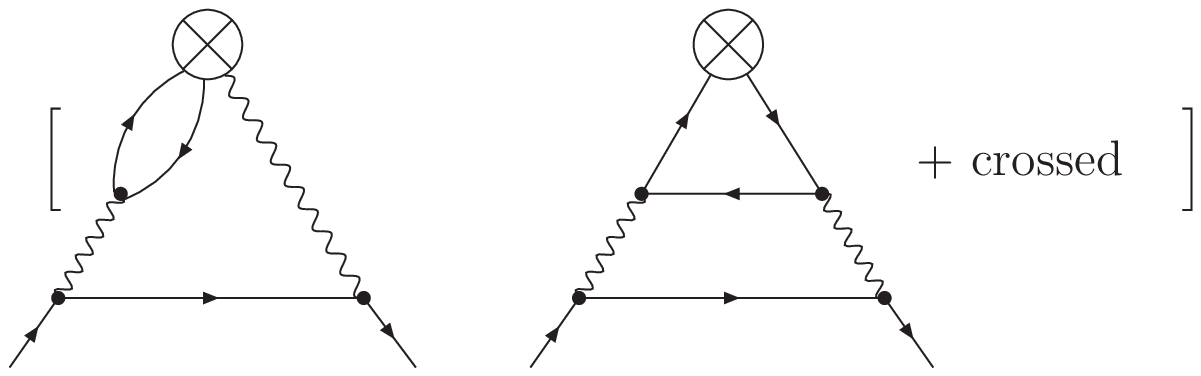,width=0.6\linewidth}
\end{center}
\vskip -.7 cm 
\caption{
\small Feynman diagrams for the calculation of the massive two-loop operator
matrix elements $A^{(2),{\rm III}}_{ee}$. \label{DiagramsIII}}
\end{figure}

\vspace*{1mm}
\noindent
The flavor pure--singlet diagrams of Figure~\ref{DiagramsIII} yield 
$\hat{\hat{A}}^{(2), {\rm III}}_{ee}$. They can be calculated using the
corresponding one-loop off-shell insertion of the diagrams in Figure~4, see \cite{BBK2}.
%
\restylefloat{figure}
%
%
\noindent
For the insertion in diagrams \ref{DiagramsIII}a,b one obtains
\begin{eqnarray}
I_{\mu \nu}^{(8,a)}(k) &=& 4 \frac{(\Delta \cdot k)^{n-1}}{(4\pi)^{D/2}} \Gamma(3-D/2) 
\int^1_0 dx \,\, x^{n+D/2-2} (1-x)^{-2+D/2} \BLB  \nonumber\\ &&
-2 x (1-x) \frac{g_{\mu \nu} k^2 -2 k_{\mu} k_{\nu}}{(-k^2+M^2)^{3-D/2}} 
(\Delta \cdot k)^2
-\frac{2 m^2_e g_{\mu \nu}}{(-k^2+M^2)^{3-D/2}} (\Delta \cdot k)^2 \nonumber
\\ 
&&
+(1-x) \frac{2(n+1)x-n}{2-D/2} 
\frac{k_{\mu} \Delta_{\nu}+k_{\nu} \Delta_{\mu}}{(-k^2+M^2)^{2-D/2}}(\Delta
\cdot k) \nonumber\\ &&
+(1-x) \frac{n(1-2x)-Dx}{2-D/2} \frac{g_{\mu \nu}}{(-k^2+M^2)^{2-D/2}} (\Delta
\cdot k)^2 \nonumber
\end{eqnarray}
\begin{eqnarray}
&&
-(1-x) n \frac{(n+1) (1-x)-1}{(1-D/2)(2-D/2)} \frac{\Delta_{\mu}
  \Delta_{\nu}}{(-k^2+M^2)^{1-D/2}} \BRB \,\, ,
\\
I_{\mu \nu}^{(8,b)}(k) &=& 8 \frac{(\Delta \cdot k)^{n-1}}{(4\pi)^{D/2}}
\Gamma(2-D/2) \int^1_0 dx \,\, (x-x^2)^{D/2-1} (x^n+(1-x)^n)  \nonumber\\ &&
\phantom{8 \frac{(\Delta \cdot k)^{n-1}}{(4\pi)^{D/2}}
\Gamma(2-D/2) \int^1_0 dx}
\times \frac{(\Delta \cdot k) \Delta_{\mu} k_{\nu} 
- k^2 \Delta_{\mu} \Delta_{\nu}}{(-k^2+M^2)^{2-D/2}}
\,\, ,
\end{eqnarray}
with $M^2 = m^2_e/[x(1-x)]$ and $k$ is the momentum of the photon line. Using these
results $\Ahathat^{(2), \rm III}_{ee}$  is given by
\begin{eqnarray}
\hat{\hat{A}}_{ee}^{(2), \rm III}
&=& S_\ep^2 \int^1_0 dx \,\, x^n {1+(-1)^N \over 2} \BLCB
{8 \over \ep^2} \left[ {1-x \over 3 x} (4 x^2+7 x+4)+2 (1+x) \ln(x) \right]
\nonumber \\ && \phantom{ \int^1_0 dx \,\, x^n \BLCB}
+{4 \over \ep} \BLB 5 (1+x) \ln^2(x) 
-{1+x \over 3 x} (8 x^2-17 x-16) \ln(x)
+{4 (1-x) \over 9 x} (5 x^2
\nonumber \\ && \phantom{ \int^1_0 dx \,\, x^n \BLCB}
+23 x+14) \BRB
+{2 \over x} (1-x) (4 x^2+13 x+4) \zeta_2
+{1 \over 3 x} (8 x^3+135 x^2+75 x
\nonumber \\ && \phantom{ \int^1_0 dx \,\, x^n \BLCB}
+32) \ln^2(x)
+\left[ {304 \over 9 x}-{80 \over 9} x^2-{32 \over 3} x+108
-{32 \over 1+x}-{64 (1+2 x) \over 3 (1+x)^3} \right] \ln(x)
\nonumber \\
&& \phantom{ \int^1_0 dx \,\, x^n \BLCB}
-{224 \over 27} x^2
+16 {1-x \over 3 x} (x^2+4 x+1) \left[ 2 \ln(x) \ln(1+x)
-{\rm Li}_2(1-x) \right.
\nonumber \\ && \phantom{ \int^1_0 dx \,\, x^n \BLCB}
\left.
+2 {\rm Li}_2(-x) \right]
+(1+x) \BLB
4 \zeta_2 \ln(x) 
+{14 \over 3} \ln^3(x)
-32 \ln(x) {\rm Li}_2(-x)
\nonumber \\ && \phantom{ \int^1_0 dx \,\, x^n \BLCB}
-16 \ln(x) {\rm Li}_2(x)
+64 {\rm Li}_3(-x)
+32 {\rm Li}_3(x)
+16 \zeta_3
\BRB 
-{182 \over 3} x+50
\nonumber \\ && \phantom{ \int^1_0 dx \,\, x^n \BLCB}
-{32 \over 1+x}
+{800 \over 27 x}
+{64 \over 3 (1+x)^2} 
\BRCB m^{2\ep} (\Delta \cdot p)^n~.
\label{BigA2IIIunre}
\end{eqnarray}
\noindent
The factor $[{1+(-1)^N}]/{2}$ occurs because of the different directions of the momentum flow in the 
external 
fermion lines in the diagrams of Figure~\ref{DiagramsIII}. For all operator matrix elements, which are 
not the same in the unpolarized and polarized case factors of this type appear \cite{CROSS}. In the 
present case $N$ is even.
\subsection{\boldmath $O(a^2)$ Results}
\label{sec:5.4}

\vspace{1mm}
\noindent
From the pole-terms of the unrenormalized OMEs (\ref{BigA2Iunre}, \ref{BigA2IIunre}, \ref{BigA2IIIunre}),
resp. the renormalized OMEs (\ref{eq:R1}--\ref{eq:R3}), one may determine the 2-loop splitting functions
$P_{ee}^{(1), \rm I}(x)$ to $P_{ee}^{(1), \rm III}(x)$ and the constant parts of the unrenormalized OMEs 
$\hat{\Gamma}_{ee}^{(1), \rm 
I-III}(x)$. Here the splitting functions result from a massive calculation. 
The operator matrix element (\ref{BigA2Iunre}) contains two branches. 
The contributions to the present process result from the first branch. The corresponding
contribution to the 2-loop splitting function is found to be
\begin{eqnarray}
P_{ee}^{(1), F}(x) &\equiv& \frac{1}{2} P_{ee}^{(1), \rm I}(x) \nonumber\\
&=&  - 8 \left [\frac{1+x^2}{1-x} \left\{\ln (x) 
\ln(1 - x) +
\frac{3}{4} \ln(x) \right \} + \frac{1}{4} (1+x) \ln^2(x)
+ \frac{1}{4} (3+7x) \ln (x) \right. \nonumber\\
&&\left.~~~+ \frac{5}{2} (1 - x) \right]
+\frac{3}{2} \left( 1-8\zeta_2+16\zeta_3 \right) \delta(1-x)~.
\end{eqnarray}
The pole part of the second branch, labeled by the factor $(-1)^N$,  is given by
\begin{eqnarray}
P_{ee}^{(1),A}(x) &\equiv& P_{e\overline{e}}^{(1)}(x) \equiv \frac{1}{2} \bar{P}_{ee}^{(1),\rm I}(x) 
= 8 \left[\frac{1+x^2}{1+x} S_2(x) +(1+x) \ln(x) +2(1-x)\right]~,
\end{eqnarray}
where
\begin{eqnarray}
S_2(x) = - 2 \Li_2(-x) - 2 \ln(x) \ln(1+x) + \frac{1}{2} \ln^2(x) - \zeta_2,
\end{eqnarray}
cf.~\cite{PijNLO}.

From the OMEs II and III the splitting functions
\begin{eqnarray}
P_{ee}^{(1), N_f}(x) &\equiv& \frac{1}{2} P_{ee}^{(1), {\rm II}}(x) 
\nonumber\\
&=& -\frac{80}{9} \DD_0(x) - \frac{2}{3}\left(1+8\zeta_2\right) 
\delta(1-x)
    -\frac{16}{3} \frac{\ln(x)}{1-x} + \frac{8}{3} (1+x) \ln(x) 
    - \frac{8}{9}(1-11 x) \nonumber\\
&=& - 8 \left [ \frac{1+x^2}{1-x} 
\left(
\frac{1}{3} \ln x + \frac{5}{9} \right) + \frac{2}{3}(1-x) \right]_+, \\
P_{ee}^{(1), PS}(x) &\equiv& \frac{1}{2} P_{ee}^{(1), {\rm III}}(x) 
\nonumber
\end{eqnarray}
\begin{eqnarray}
&=& 8  \left [\left( \frac{1}{2} + 
\frac{5}{2} x +
\frac{4}{3} x^2 \right) \ln x - \frac{1}{2} (1+x) \ln^2 x
+ \frac{1-x}{9 x} \left (10 + x + 28 x^2 \right)\right],
\end{eqnarray}
are obtained in accordance with known results from Quantum Chromodynalmics 
\cite{PijNLO} setting $C_A = 0, C_F = T_F = 1$.
In the present calculation we choose $N_f = 1$. 

The unrenormalized operator matrix elements at the level of ${\hat{A}}_{ee}^{(2),\rm I-III}$ 
(\ref{unA2I}--\ref{unA2III}) obey fermion number conservation, i.e. their first moment vanishes.
The constant contributions to the unrenormalized OMEs 
(\ref{unA2I}--\ref{unA2III}) are given by
\begin{eqnarray}
\label{eq:GA1}
\hat{\Gamma}^{(1),\rm I}_{ee} &=&  
\frac{1+3 x^2}{1-x} \left[ 6 \zeta_2 \ln(x) -8 \ln(x) {\rm Li}_2(1-x)-4 \ln^2(x) \ln(1-x) \right]
\nonumber \\ &&
+\left( {122 \over 3} x+22+{32 \over 1-x} \right) \zeta_2
+16 \frac{1+x^2}{1-x} \left[ 2 {\rm Li}_3(-x)-\ln(x) {\rm Li}_2(-x) \right]
+{80 \over 3 (1-x)}
\nonumber \\ &&
+56 (1+x) \zeta_2 \ln(1-x)
+\left( {22 \over 3} x+32+{64 \over 3 (1-x)^2}-{51 \over 1-x}
-{16 \over 3 (1-x)^3} \right) \ln^2(x)
\nonumber \\ &&
-(92+20 x) \ln^2(1-x)
+\left( {178 \over 3}-36 x+{64 \over 3 (1-x)^2} -{140 \over 3 (1-x)}-{48 \over 1+x} \right) \ln(x) 
\nonumber\\ 
&&
-\frac{1}{3} (1+x) \ln^3(x)
+4 \frac{x^2-8 x-6}{1-x} \ln(x) \ln(1-x) 
-2 \frac{1+17 x^2}{1-x} \ln(x) \ln^2(1-x)
\nonumber \\ &&
-\frac{112}{3} (1+x) \ln^3(1-x)
 +32 \frac{1+x}{1-x} \left[ \ln(x) \ln(1+x)+{\rm Li}_2(-x) \right]
-22 x-{62 \over 3}
\nonumber\\ 
&&
-4 \frac{13 x^2+9}{1-x} {\rm S}_{1,2}(1-x)
+4 \frac{5-11 x^2}{1-x} \left[ \ln(1-x) {\rm Li}_2(1-x)
-{\rm Li}_3(1-x)-2 \zeta_3 \right]
\nonumber \\ &&
+\frac{4 (16 x^2-10 x-27)}{3 (1-x)} {\rm Li}_2(1-x)
+14 (x-2) \ln(1-x)
\nonumber \\ &&
+\left( 16-52 \zeta_2+128 \zeta_3 \right) {\cal D}_0(x)
+\left( 8-112 \zeta_2 \right) {\cal D}_1(x)
+120 {\cal D}_2(x)
+{224 \over 3} {\cal D}_3(x)
\nonumber 
\end{eqnarray}
\begin{eqnarray}
&&
+\left[ {433 \over 8}-{67 \over 45} \pi^4
+\left( {37 \over 2}-48 \ln(2) \right) \zeta_2+58 \zeta_3 \right] \delta(1-x)
\nonumber \\ &&
+(-1)^n \left\{
\frac{2 (1-x) (45 x^2+74 x+45)}{3 (1+x)^2}
+\frac{2 (9+12 x+30 x^2-20 x^3-15 x^4)}{3 (1+x)^3} \ln(x) \right.
\nonumber \\ && \phantom{+(-1)^n \BLB}
+\frac{4 (x^2+10 x-3)}{3 (1+x)} \left( \zeta_2+2 {\rm Li}_2(-x)
+2 \ln(x) \ln(1+x) \right)
\nonumber\\  
&& \phantom{+(-1)^n \BLB}
+\frac{1+x^2}{1+x} \BLB
8 \zeta_2 \ln(x)-24 \zeta_2 \ln(1+x)+36 \zeta_3
-\frac{2}{3} \ln^3(x)
+40 {\rm Li}_3(-x)
\nonumber\\ 
&& \phantom{+(-1)^n \BLB}
-4 \ln^2(x) \ln(1+x)
-24 \ln(x) \ln^2(1+x)
-24 \ln(x) {\rm Li}_2(-x)
\nonumber \\ && \phantom{+(-1)^n \BLB}
-48 \ln(1+x) {\rm Li}_2(-x)
-8 \ln(x) {\rm Li}_2(1-x)
-16 {\rm S}_{1,2}(1-x)
\nonumber \\ && \phantom{+(-1)^n \BLB}
-48 {\rm S}_{1,2}(-x)
\BRB
-\frac{16 (x^4+12 x^3+12 x^2+8 x+3)}{3 (1+x)^3} {\rm Li}_2(1-x)
\nonumber \\ && \phantom{+(-1)^n \BLB}
\left. +4 x \frac{1-x-5 x^2+x^3}{(1+x)^3} \ln^2(x) 
\right\}
\end{eqnarray}
\begin{eqnarray}
\label{eq:GA2}
\hat{\Gamma}^{(1),\rm II}_{ee} &=& 
{76 \over 27} x-{572 \over 27}
-\left( 12 x+{4 \over 3}+{8 \over 1-x} \right) \ln(x)
+{128 \over 9 (1-x)^2}
+{80 \over 27 (1-x)}
-{64 \over 9 (1-x)^3}
\nonumber \\ &&
-{32 \over 9} \left( {1 \over (1-x)^2}-{5 \over (1-x)^3}
+{2 \over (1-x)^4} \right) \ln(x)
-{2 (1+x^2) \over 3 (1-x)} \ln^2(x)
+{4 \over 3} (1+x) \zeta_2
\nonumber \\ &&
+{16 \over 3} (1+x) \left( \ln(1-x)+\ln^2(1-x) \right)
+\left( {224 \over 27} -{8 \over 3} \zeta_2 \right) {\cal D}_0(x)
\nonumber \\ &&
-{32 \over 3} \left( {\cal D}_1(x)+{\cal D}_2(x) \right)
+\left( {8 \over 3} \zeta_3-10 \zeta_2
+{1411 \over 162} \right) \delta(1-x)
\end{eqnarray}
\begin{eqnarray}
\label{eq:GA3}
\hat{\Gamma}^{(1),\rm III}_{ee} &=&
{2 \over x} (1-x) (4 x^2+13 x+4) \zeta_2
+{1 \over 3 x} (8 x^3+135 x^2+75 x+32) \ln^2(x)
+50
\nonumber \\ &&
+\left[ {304 \over 9 x}-{80 \over 9} x^2-{32 \over 3} x+108
-{32 \over 1+x}-{64 (1+2 x) \over 3 (1+x)^3} \right] \ln(x)
-{224 \over 27} x^2
-{182 \over 3} x
\nonumber \\ &&
+16 {1-x \over 3 x} (x^2+4 x+1) \left[ 2 \ln(x) \ln(1+x)
-{\rm Li}_2(1-x)+2 {\rm Li}_2(-x) \right]
+{800 \over 27 x}
\nonumber \\ &&
+(1+x) \BLB
4 \zeta_2 \ln(x) 
+{14 \over 3} \ln^3(x)
-32 \ln(x) {\rm Li}_2(-x)
-16 \ln(x) {\rm Li}_2(x)
\nonumber \\ &&
+64 {\rm Li}_3(-x)
+32 {\rm Li}_3(x)
+16 \zeta_3
\BRB 
-{32 \over 1+x}
+{64 \over 3 (1+x)^2}~. 
\end{eqnarray}
The constant contributions (\ref{eq:GA1}--\ref{eq:GA3}) enter the single
differential cross section linearly and all contain terms of the kind
$\sim 1/(1-x)^3$ and ${\ln(x)}/{(1-x)^4}$ etc. The massless Wilson coefficients
of the Drell-Yan process to 2-loops are free of such terms, cf. \cite{DY1,DY2}, like all the 
other Mellin convolutions which contribute.
For the processes II and III non of these terms were found in \cite{BBN} resp. \cite {KUHN},
while terms like $\ln^2(x)/(1-x)^2$ are present in the cross section for process I. We therefore conclude
that the use of standard matrix elements for local operators alone, (\ref{decomp4}--\ref{COMP3}),
are not sufficient to reproduce the constant terms. This aspect requires further investigation.
It occurs for massive 
external fermion lines in contrast to the case of {\it massless} external parton lines. There the constant terms
are correctly reproduced in the limit $s \gg m^2$.

Finally we present the $O(a_0^2)$ contributions up to the logarithmic term $\hat{\bf L}$ 
in (\ref{eqMA1a}--\ref{eqMA1c}), which have been computed based on the present 2-loop
calculation of the massive operator matrix elements with an external massive fermion line. 
They are given by
\begin{eqnarray}
\label{fres:1}
T_{ee}^{2,\rm Ia} &=& 16 a_0^2 \Biggl\{\Biggl[
4 \DD_1(x) 
+ 3 \DD_0(x) 
+ \left(\frac{9}{8} -2 \zeta_2\right) \delta(1-x)
- 2 \frac{\ln(x)}{1-x}
\nonumber\\ &&
\phantom{6 a_0^2} - (1+x) \left(2 \ln(1-x)
- \frac{3}{2} (\ln(x)  - 1)\right) - 1 + x \Biggr] \hat{\bf L}^2
\nonumber\\
 & & 
\phantom{6 a_0^2} 
+ \Biggl[ - 8 \DD_1(x) + \left(-7 + 4 \zeta_2 \right) \DD_0(x)
 -\frac{1}{4}(11 + 7x) \ln(x)
\nonumber\\ && \phantom{6 a_0^2}
+ \frac{1+x^2}{1-x} \left( - \ln^2(x) + \frac{11}{4} \ln(x) + \ln(1-x) \ln(x) \right)
\nonumber\\ && \phantom{6 a_0^2}
+ (1+x) \left( - 2 \zeta_2 + 4 \ln(1-x)+ \frac{1}{4} \ln^2(x)\right)  
+ 3 + 4x  \nonumber\\ && \phantom{6 a_0^2}
+ \left(-\frac{45}{16}+ \frac{11}{2} \zeta_2 + 3 \zeta_3  \right) \delta(1-x)
\Biggr] \hat{\bf L}\Biggr\}
\\
\label{fres:2}
T_{ee}^{2,\rm IIa} &=& 16 a_0^2 \Biggl\{
\Biggl[\frac{1}{3} \DD_0(x) - \frac{1}{6} (1+x) + \frac{1}{4} \delta(1-x) \Biggr] \hat{\bf L}^2
\nonumber\\ && \phantom{6 a_0^2}
+ \Biggl[\frac{1+x^2}{1-x} \left[\frac{2}{3} \ln(1-x) - \frac{1}{3} \ln(x) - \frac{5}{9}
\right] - \frac{2}{3} (1-x) - \frac{17}{12} \delta(1-x) \Biggr] \hat{\bf L} \Biggr\}
\\ 
\label{fres:3}
T_{ee}^{2,\rm IIIa} &=& 16 a_0^2 \Biggl\{\left[\frac{1}{2}\left(1+x\right) \ln(x) + \frac{1}{3x} 
+ \frac{1}{4} 
- \frac{1}{4} x
- \frac{1}{3} x^2 \right] \hat{\bf L}^2
\nonumber\\ &&
\phantom{16 a_0^2}
+ \Biggl[(1+x)\left(2\ln(x)\ln(1-x)-\ln^2(x) + \Li_2(1-x)\right)
\nonumber\\ && 
\phantom{16 a_0^2}
+ \left(\frac{4}{3x} + 1 - x - \frac{4}{3} x^2\right) \ln(1-x) 
-\left(\frac{2}{3x} +1 - \frac{1}{2}x - \frac{4}{3} x^2\right) \ln(x) 
\nonumber\\ &&
\phantom{16 a_0^2}
- \frac{8}{9x} - \frac{8}{3} + \frac{8}{3} x 
+ \frac{8}{9} x^2\Biggr] \hat{\bf L} \Biggr\}~. 
\end{eqnarray}
These terms agree with the corresponding terms in Ref.~\cite{BBN}, Eqs.~(2.30, 2.40, 2.42, 2.43) and
for (\ref{fres:2}) with Ref.~\cite{KUHN}.

\section{Conclusions}
\label{sec:7}

\vspace{1mm}\noindent
We have calculated the QED matrix elements with an external massive fermion for
the local operators of the light-cone expansion to $O(a^2)$ up to the constant term in
the dimensional parameter $\ep$ in the $\overline{\rm MS}$ scheme. The OMEs can be renormalized 
applying the technique having been developed recently in \cite{BBK2}. The renormalized OMEs
obey fermion number conservation. Various technical details of the calculation and intermediate
results are provided. We investigated the factorization of the $O(\alpha^2)$ QED 
initial state corrections to $e^+e^-$ annihilation into a virtual boson for large cms energies 
$s \gg m_e^2$ into massive OMEs and the massless Wilson coefficients of the Drell-Yan process 
adapting the color coefficients to the case of QED, as being proposed in Ref.~\cite{BBN}. 
We have shown by an explicit calculation, that the representation works at one-loop order 
and at two-loop order including all terms up to the linear order in $\ln(s/m_e^2)$. In 
the constant term in $O(a^2)$ the OMEs bear a few structural terms, which have not been obtained 
in previous direct calculations \cite{BBN,KUHN}, despite a large part of other terms appear 
as being expected by the factorization. Further studies are needed to reveal the reason for this.
This finding appears in contrast to the case of massless external fermion and boson lines,
where the corresponding cross sections have been shown to factorize including the constant terms
in \cite{Laenen, HEAV1, BBK, HQCD}.

\vspace*{5mm}\noindent
{\bf Acknowledgment.}\\ 
We would like to thank S. Klein, J.H. K\"uhn,  and R. Mertig for discussions and N. Hatcher for a {\tt C++} code 
for {\tt Latex} checks. This paper has been supported in part by the Alexander-von-Humboldt 
Foundation, by SFB-TR/9, and the EU TMR network LHCPhenoNet, PITN-GA-2010-264564.

\newpage
\appendix

\section{Convolutions of one--loop quantities}
\label{ConvolutionsAppendix}

\vspace*{1mm}\noindent
In Eqs.~(\ref{eqMA1a}--\ref{eqMA1c}) a series of convolutions of 
one-loop functions occurs. Some of them where given in the Tables 
in \cite{UNIV2,BRN}.~\footnote{The distribution-valued contributions 
to (\ref{eqD1D1}) were also given in \cite{NV}.} A few more convolutions are needed
which are given by~:
\begin{eqnarray}
\label{eqD1D1}
\DD_1(x) \otimes \DD_1(x) &=& \DD_3(x) - 2\zeta_2 \DD_1(x) + 2 \zeta_3 
\DD_0(x) - \frac{\zeta_4}{4} \delta(1-x) \nonumber\\
& & + \frac{2\left[\Li_3(x) - \zeta_3 \right]- \ln(x) \Li_2(x) 
- \ln(x) \ln^2(1-x)}{1-x}
\\
\DD_1(x) \otimes \ln(1-x) &=& 
\Li_3(1-x) + 2 \Li_3(x) + \left[\ln(1-x) - \ln(x)\right] \Li_2( x ) 
\nonumber\\ & &
+\frac{1}{2} \ln^3(1-x) 
- 2 \zeta_2 \ln( 1-x ) - \zeta_3 
\\
\DD_1(x) \otimes x \ln(1-x) &=& 
x \Li_3(1-x) + 2 x \Li_3(x) 
- \left[1 + x \left[\ln( x )  - \ln ( 1-x )\right]  \right] \Li_2(x) 
\nonumber\\ &&
+ \frac{x}{2}  \ln^3( 1-x )  
+ (1- x) \ln^2( 1-x ) 
- (1-x) \ln ( 1-x ) 
\nonumber\\ &&
- x \ln( x ) 
-(1-x)  \ln ( x ) \ln ( 1-x ) 
-2 x \zeta_2 \ln ( 1-x ) 
+ x (\zeta_2 -\zeta_3)
\nonumber\\
\\
\frac{\ln(x)}{x} \otimes x^2 &=& \frac{1}{9x} \left[3 \ln(x) + 1 - 
x^3\right]
\\
\ln(x) x \otimes x^2 &=& x \left[ \ln(x) + 1 -x \right]
\\
\ln(x) \otimes x^2   &=& \frac{1}{4} \left[ 2 \ln(x) + 1 -x^2 \right]
\\
\frac{\ln(x)}{x} \otimes x &=& \frac{1}{4x} \left[2 \ln(x) + 1 - 
x^2\right]
\\
\frac{\ln(x)}{x} \otimes 1 &=& \frac{1}{x} \left[ \ln(x) + 1 - 
x\right] \\
\ln(x) \otimes \ln(1-x) &=& \Li_3(x) - \zeta_2 \ln(x) - \zeta_3 \\
\ln(x) \otimes x \ln(1-x) &=& - \left[\Li_2(x) - \zeta_2 \right] + (1-x) 
\left[\ln(1-x) - 2\right] - \ln(x) 
\\
x \ln(x) \otimes \ln(1-x) &=& x \left[\Li_2(x) - \zeta_2\right] - (1-x) 
\ln(1-x) + x \ln(x) \left[\frac{1}{2} \ln(x) - 1 \right]\\
x \ln(x) \otimes x \ln(1-x) &=& x \left[ \Li_3(x) - \zeta_3 - \zeta_2 
\ln(x) \right]\\
\ln(1-x) \otimes \ln(1-x) &=& - \left\{2 S_{12}(x) + 2 S_{12}(1-x)
+ \ln(x) \left[\Li_2(1-x) - \zeta_2\right] - 2 \zeta_3 \right\} \\
\ln(1-x) \otimes x \ln(1-x) &=& (1-x) \left[\Li_2(1-x) + \ln^2(1-x)
- \ln(1-x)\right]  
\nonumber\\ & &
+ \zeta_2 (2x-1) - x \left[\Li_2(x) + \ln(x)\right]\\ 
x\ln(1-x) \otimes x\ln(1-x) &=& -x \left\{2 S_{12}(x) + 2 S_{12}(1-x)
+ \ln(x) \left[\Li_2(1-x) - \zeta_2\right] - 2 \zeta_3 \right\}
\nonumber\\
\end{eqnarray}
We finally list some Mellin convolutions of splitting- and related functions~:
\begin{eqnarray}
\label{eqPee02}
\frac{1}{2}
\left( P^{(0)}_{ee} \otimes P^{(0)}_{ee} \right )(x)  
&=&  
64 \DD_1(x) + 48 \DD_0(x) +\left(  18  - 32 \zeta_2 \right) \delta(1-x) 
\nonumber\\
& & - 32 \frac{\ln(x)}{1-x} - 16(1+x) \left[2 \ln(1-x) -\frac{3}{2} \ln(x) 
+ \frac{3}{2}\right] - 16 (1-x)
\nonumber\\
&=& 16
\left [
\frac{1+x^2}{1-x} \left\{2 \ln(1-x) - \ln x + \frac{3}{2} \right \}
+ \frac{1}{2} (1+x) \ln x - (1 -x) \right ]_+
\nonumber\\
\end{eqnarray}
\begin{eqnarray}
\label{eqPeg0Pge0}
\frac{1}{2}
\left( P^{(0)}_{e\gamma} \otimes P^{(0)}_{\gamma e} \right )(x)  &=&  16
\left  [(1+x) \ln x + \frac{1}{2} (1 - x) + \frac{2}{3} \frac{1}{x}
(1-x^3) \right]
\\
\left[P_{ee}^{(0)} \otimes \Gamma_{ee}^{(0)}\right](x)       &=& 
- 96 \DD_2(x) - 112 \DD_1(x) + 8 (1+ 8\zeta_2) \DD_0(x) 
\nonumber\\
& & + 8(-8 \zeta_3 + 4 \zeta_2 + 3) \delta(1-x)
+ 16 (1+x) \Li_2(x) 
\nonumber\\
& &
+ 48 (1+x) \left(\ln^2(1-x) - \zeta_2 \right) + 8(11+3x) \ln(1-x) 
\nonumber\\
&& - 4(1+x)
+ 32 \Bigl\{\frac{1+x^2}{1-x} \left[\ln(1-x) +\frac{1}{2}\right] 
- \frac{1}{4}(1-x) \Bigr\} \ln(x)
\nonumber\\
\\
\left[\Gamma_{ee}^{(0)} \otimes \Gamma_{ee}^{(0)}\right](x)       &=& 
64 \DD_3(x) + 96 \DD_2(x) - 32 \left(1 + 4 \zeta_2\right) \DD_1(x) 
\nonumber\\
& &
-32 \left(1 +2 \zeta_2 - 4 \zeta_3 \right) \DD_0(x)
\nonumber\\
& &
+ 16 \left(1 - \zeta_2 + 4 \zeta_3 - \zeta_4\right) \delta(1-x)
+ 8(3+x) 
\nonumber\\
& &
+ 64 \Biggl[ 1 + {\frac {\ln(x)}{1-x}} +  (1+x) \ln(1-x) 
- \frac{3}{4} (1+x) \ln(x)\Biggr] \zeta_2
\nonumber\\
& & 
- 64 \left(1+x\right) \zeta_3
+ 16 (1+x) \ln(1-x) 
- 32 (1+x) \ln^3(1-x)  
\nonumber\\ 
& &
- 16 (5+x) \ln^2(1-x)
+48 (1+x) \ln(x) \ln^2(1-x) 
\nonumber\\ 
& &
+ 16 (3+x) \ln(x) \ln(1-x)
+ 4(3+7 x) \ln(x) 
\nonumber\\ 
& &
- 16 {\frac {\ln  \left( x \right) }{1-x}} \left[4 \ln^2(1-x)
+ 4 \ln(1-x) +1
\right]
\nonumber\\ 
& &
+ 16 \Biggl[2 (1+x) \ln(1-x)  + 3 (1+x) \ln(x)   
- (1-x)
\nonumber\\ 
& &
- 4\,{\frac {\ln  \left( x \right) }{1-x}}\Biggr] \Li_2(1-x) 
- 32 \left(1+x\right) \Li_3(1-x)
\nonumber\\
& &
+ 32 \left[ 3(1+x) - \frac{4}{1-x} \right] S_{1,2}(1-x)
\\
\left[P_{e \gamma}^{0)} \otimes \Gamma_{\gamma e}^{(0)}\right](x) &=& 
-16 (1+x) \ln^2(x) - \left(\frac{64}{3x} + 48 + 32 x\right) \ln(x) 
-
\frac{304}{9x}\left(1-x^3\right)
\nonumber\\
&& -8(1-x)~. 
\\
\left[P_{ee}^{(0)} \otimes \left(2 \Gamma_{ee}^{(0)} + 
\tilde{\sigma}_{ee}^{(0)}\right)\right](x) &=& 
-128 \DD_1(x) - 16(7-4\zeta_2) \DD_0(x) - 16(3-7\zeta_2) \delta(1-x)
\nonumber\\
& &  - 8 \left(\frac{1+x^2}{1-x} \right) \ln(x) \left[ 4 \ln(1-x) - 2 \ln(x) -1 \right]
\nonumber
\\
& &  - 8 (1+x) \left[ \ln^2(x) - 8 \ln(1-x) + 4 \zeta_2 \right] 
\nonumber\\
& &
- 32 x \ln(x) + 88 + 24 x 
\\
\left[\Gamma_{ee}^{(0)} \otimes \tilde{\sigma}_{ee}^{(0)}\right](x)
&=& -128 \DD_3(x) -96 \DD_2(x) + 192 \left( 1 + \zeta_2 \right) \DD_1(x)
\nonumber\\
& & + 32 \left( 2 + \zeta_2 - 8 \zeta_3\right) \DD_0(x)
    + 32 \left(- 2 +\zeta_2 - 2 \zeta_3 + \zeta_4 \right) \delta(1-x)
\nonumber\\
& &
- 16 (1 + 3x) + 8(3-x) \ln(x) 
- 64(2+x) \ln(1-x)
\nonumber\\
& &
- 112 (1+x) \ln(x) \ln^2(1-x) 
+ 64  (1+x) \ln^3(1-x) 
\nonumber
\\
& &
+ 24  (1+x) \ln^2(x) \ln(1-x)
+  4(3 +  x) \ln^2(x) 
\nonumber\\
& &
- 16(5- x)  \ln(x) \ln(1-x)
+ 4 (28 - 4  x)  \ln^2(1-x)
\nonumber\\
& & 
+ 16 \frac{\ln(x)}{1-x} \Biggl[10 \ln^2(1-x)
 -2 \ln(x) \ln(1-x)                            
\nonumber
\end{eqnarray}
\begin{eqnarray}
&&
\hspace{2cm}
 + 6 \ln(1-x)
                             - \ln(x) - 2\Biggr] 
+ 128 (1+x) \zeta_3
\nonumber\\ 
& &
+ 16 \Biggl[2 + 4 {\frac{\ln(x)}{1-x}}
         -  (1+x) \left(3 \ln(x) +4 \ln(1-x)\right) \Biggr] 
  \Li_2(1-x)
\nonumber\\ 
& &
+ 64 \left( 1 + x \right) \Li_3(1-x) 
- 16 \left[7(1+x) -\frac{8}{1-x}\right] S_{1,2}(1-x) 
\nonumber\\
& &
+ \Biggl[96 (1+x)\left(\ln(x) - \ln(1-x)\right) + 48 x - 128 {\frac {\ln 
(x)}{1-x}} 
- 80 \Biggr] \zeta_2
\\
\left[\overline{\Gamma}_{\gamma e}^{(0)} \otimes P_{e \gamma}^{(0)}\right](x)
&=& 8 \Biggl\{\frac{2}{3} (1+x) \ln^3(x) + (3+2x) \ln^2(x) +
\left[(7x+8)+\frac{1}{2}(1+x) \zeta_2\right] \ln(x) 
\nonumber\\ &&
- (4 x^2 + 7x -11) + \frac{3}{4} (1-x) \zeta_2\Biggr\}
\\
\left[P_{\gamma e}^{(0)} \otimes \tilde{\sigma}_{e \gamma}^{(0)}\right](x)
&=& 
-{\frac {40}{3}}\,x+{\frac {272}{9}}\,{x}^{2}+{\frac {16}{9}}\, 
\frac{1}{x}
-{\frac {56}{3}} 
\nonumber\\ & &
+ \left( 16+{\frac {64}{3}}\,\frac{1}{x}-{\frac 
{64}{3}}\,{x}^{2}-16\,x 
\right) \ln
 \left( 1-x \right) 
\nonumber\\ & &
+ \left( 32+32\,x \right) \left[\Li_2\left(1-x 
\right) + \ln(x) \ln(1-x) \right] 
\nonumber\\ & &
+ \left( 16+{\frac {32}{3}}\,{x}^{2}+8\,x \right) \ln  \left( x \right)  
- 8\left(1+x\right) \ln^2  \left( x \right)
\\
\left[\Gamma_{\gamma e}^{(0)} \otimes \tilde{\sigma}_{e \gamma}^{(0)}\right](x)
&=&
\frac{8}{3}\left(1+x\right) \ln^3 \left( x \right) 
+8\, x \ln^2  \left( x \right)  \nonumber\\ &&
+\Biggl[ 
-{\frac {16}{9}}\,\frac{1}{x} 
-20 
+28 x
-{\frac {152}{9}}\,{x^2} 
- 32 \left(1+x\right) \zeta_2
\Biggr] \ln  \left( x \right) 
\nonumber\\ &&
+ \Biggl[ 
-{\frac {304}{9}}\,\frac{1}{x} 
-8
+8\,x
{\frac {304}{9}}\,{x}^{2}
\Biggr] \ln  \left( 1-x \right) 
\nonumber\\ &&
+ \left( 32
\,x+48+{\frac {64}{3}}\, \frac{1}{x} \right) \left[\Li_2\left(x \right)
- \zeta_2\right]
+ \left( 32+32\,x \right) \left[ \Li_3\left(x \right) - \zeta_3\right] 
\nonumber\\ && 
+{\frac {28}{27}}\,\frac{1}{x}
+{\frac {314}{9}}
+{\frac {232}{9}}\,x
-{\frac {1666}{27}}\,{x}^{2}
\\
\bar{\Gamma}^{(0)}_{ee}(x) \otimes P^{(0)}_{ee}(x) &=&
\frac{128}{3} {\cal D}_3 (x)
+72 {\cal D}_2(x)
+(24-48 \zeta_2) {\cal D}_1(x)
+\left( 64 \zeta_3-20\zeta_2+32 \right) {\cal D}_0(x)
\nonumber \\ &&
+4 (1+x) \left( 4 {\rm S}_{1,2}(x) +6 \zeta_2 \ln(1-x) -12 \zeta_3
-\frac{16}{3} \ln^3(1-x) -3 \ln(1-x)  \right)
\nonumber \\ &&
-2 \frac{1+3 x^2}{1-x} \zeta_2 \ln(x)
-16 \frac{1+x^2}{1-x} \ln(x) \left[ \ln(1-x)+\ln^2(1-x) \right]
\nonumber \\ &&
-8 (1-x) {\rm Li}_2(x)
-(52+20 x) \ln^2(1-x)
+(14+6 x) \zeta_2
-8 x \ln(x)
\nonumber \\ &&
-24-8 x
+\left( \frac{9}{2} \zeta_2+32 \zeta_3 - 84 \zeta_4 +24 \right) 
\delta(1-x)
\end{eqnarray}

\newpage
\section{Integrals}
\label{Integrals_Section}

\vspace{1mm}\noindent
In the following we present some decompositions of five--propagator integrals in terms of 
four-propagator integrals using integration by parts~\cite{IBP}. In some cases Mellin--Barnes 
representations are applied for checks. In a sample 
calculation we illustrate main steps of the computation of the massive OMEs. Furthermore, we list a 
series of complicated integrals and finally summarize the structure of the individual diagrams in terms
of Nielsen integrals.
\subsection{Integration by parts}
\label{SectionIBP}

\vspace{1mm}\noindent
Eqs. (\ref{An011111_IBP}--\ref{A12111_IBP}) are obtained from the
integration by parts relation
\begin{equation}
{\cal E}^{a,b}_{\nu_1 \nu_2 \nu_3 \nu_4 \nu_5}(q,l)=
\int \frac{d^D k_1}{(2\pi)^D} \frac{d^D k_2}{(2\pi)^D}
\frac{\partial}{\partial q^{\mu}} 
\left( l^{\mu} \frac{ (\Delta \cdot k_1)^a (\Delta \cdot k_2)^b}{
D_1^{\nu_1} D_2^{\nu_2} D_3^{\nu_3} D_4^{\nu_4} D_5^{\nu_5}} \right) = 0 \,\, ,
\label{IBP1}
\end{equation}
where $q^{\mu}=k_1^{\mu}, \, k_2^{\mu}$ and  $l^{\mu}=p^{\mu}, \, k_1^{\mu}, \, k_2^{\mu}$.
For example, (\ref{An011111_IBP}) results  subtracting 
${\cal E}^{n,0}_{11111}(k_2,k_1)$ from ${\cal E}^{n,0}_{11111}(k_2,k_2)$. On
the other hand, the most complicated equation of this family, namely, 
(\ref{A12111_IBP}) can be obtained from
\begin{eqnarray}
{\cal E}^{0,n}_{21111}(k_1,k_2) - {\cal E}^{0,n}_{21111}(k_1,k_1) +
{\cal E}^{0,n}_{11211}(k_1,k_2) - {\cal E}^{0,n}_{11211}(k_1,k_1) &&
\nonumber \\ 
 - (D-5) \left[ 
{\cal E}^{0,n}_{12111}(k_1,k_2) - {\cal E}^{0,n}_{12111}(k_1,k_1) \right] &=& 0
\,\, .
\end{eqnarray}
In the same way, it is possible to obtain relations for $A^{n,1}_{11111}$ and
$A^{1,n}_{11111}$, which are needed to obtain the $E$-type integrals from
(\ref{E_IBP}).

Sometimes it is even possible to use (\ref{IBP1}) to
reduce 4-propagator $A$-type integrals to a linear combination of integrals with just
three propagators. For example, it can be shown that
\begin{equation}
A^{n,0}_{31110}=\frac{1}{1+\ep} \left\{ 
A^{n,0}_{32100}-\frac{2}{\ep} \left[ 
A^{n,0}_{23100}+\frac{3}{1-\ep} \left(
A^{n,0}_{14100}-A^{n,0}_{04110} \right) \right] \right\} \,\, .
\end{equation}
It turns out to be much easier to calculate $A^{n,0}_{31110}$ from this
equation than by introducing Feynman parameters directly.

Equations for the $B$-type integrals can also be obtained by setting one of
the powers of the propagators in (\ref{IBP1}) to zero. For example, from
\begin{equation}
{\cal E}^{0,n}_{11110}(k_1,p) + {\cal E}^{0,n}_{11110}(k_2,p) = 0 \,\, ,
\end{equation}
one may obtain
\begin{equation}
B^{0,n}_{12110} = \frac{n}{2} (\Delta \cdot p) A^{0,n-1}_{11110}
-m^2 A^{0,n}_{21110} -\frac{1}{2} A^{0,n}_{01210} 
+\frac{1}{2} A^{0,n}_{21010} \,\, ,
\end{equation}
and from ${\cal E}^{n,0}_{11101}(k_1,k_2)=0$, one may get
\begin{eqnarray}
B^{n,0}_{21101} &=& \frac{1}{2} \left[ 
n A^{n-1,1}_{11101} -A^{n,0}_{21001} +A^{n,0}_{21100} -A^{n,0}_{20101}
+A^{n,0}_{11200} -A^{n,0}_{10201} \right.
\nonumber \\ && \phantom{\frac{1}{2} [} \left.
-A^{n,0}_{11002} +A^{n,0}_{10102}
+2 m^2 A^{n,0}_{11102} \right] \,\, .
\end{eqnarray}
These kind of equations can be used to check the results obtained directly 
using Feynman parameters for this type of integrals.

Let us consider now the $E$-type integrals. Using
\begin{equation}
\sum^{n-1}_{j=0} (\Delta \cdot k_2)^j (\Delta \cdot k_1)^{n-1-j} =
\frac{(\Delta \cdot k_2)^n - (\Delta \cdot k_1)^n}{\Delta \cdot k_2 - \Delta
  \cdot k_1} \,\, ,
\end{equation}
we can rewrite them as
\begin{equation}
E^{a,b}_{\nu_1 \nu_2 \nu_3 \nu_4 \nu_5} = 
J^{a,n+b}_{\nu_1 \nu_2 \nu_3 \nu_4 \nu_5} 
- J^{n+a,b}_{\nu_1 \nu_2 \nu_3 \nu_4 \nu_5} \,\, ,
\end{equation}
where
\begin{equation}
J^{a,b}_{\nu_1 \nu_2 \nu_3 \nu_4 \nu_5} =
\int \frac{d^D k_1}{(2\pi)^D} \frac{d^D k_2}{(2\pi)^D} 
\frac{ (\Delta \cdot k_1)^a (\Delta \cdot k_2)^b
}{D_{\Delta} D_1^{\nu_1}D_2^{\nu_2}D_3^{\nu_3}D_4^{\nu_4}D_5^{\nu_5}} \,\, ,
\end{equation}
with $D_{\Delta}=\Delta \cdot k_2 - \Delta \cdot k_1$. We define also
\begin{equation}
K^{a,b}_{\nu_1 \nu_2 \nu_3 \nu_4 \nu_5} =
\int \frac{d^D k_1}{(2\pi)^D} \frac{d^D k_2}{(2\pi)^D} 
\frac{ (\Delta \cdot k_1)^a (\Delta \cdot k_2)^b
}{D_{\Delta}^2 D_1^{\nu_1}D_2^{\nu_2}D_3^{\nu_3}D_4^{\nu_4}D_5^{\nu_5}} \,\, .
\end{equation}
Then, from 
\begin{equation}
\int \frac{d^D k_1}{(2\pi)^D} \frac{d^D k_2}{(2\pi)^D}
\frac{\partial}{\partial k_2^{\mu}} 
\left( (k_2-k_1)^{\mu} \frac{ (\Delta \cdot k_1)^a (\Delta \cdot k_2)^b}{
D_{\Delta} D_1 D_2 D_3 D_4 D_5} \right) = 0 \,\, ,
\label{k2k1IBP}
\end{equation}
and using
\begin{equation}
K^{a+1,b}_{\nu_1 \nu_2 \nu_3 \nu_4 \nu_5} 
- K^{a,b+1}_{\nu_1 \nu_2 \nu_3 \nu_4 \nu_5}
= J^{a,b}_{\nu_1 \nu_2 \nu_3 \nu_4 \nu_5} \,\, , 
\end{equation}
and
\begin{equation}
J^{a+1,b}_{\nu_1 \nu_2 \nu_3 \nu_4 \nu_5} 
- J^{a,b+1}_{\nu_1 \nu_2 \nu_3 \nu_4 \nu_5}
= A^{a,b}_{\nu_1 \nu_2 \nu_3 \nu_4 \nu_5} \,\, ,
\end{equation}
Eq.~(\ref{E_IBP}) is obtained. Similar manipulations lead to  (\ref{F_IBP}).

\subsection{Mellin-Barnes representation}
\label{MBSection}

\vspace{1mm}
\noindent
The 5-propagator integrals can be also calculated and/or checked using a
Mellin-Barnes representation \cite{Smirnov}. To do this, we follow the ideas developed in
\cite{BIER_WEIN}, where a Mellin-Barnes representation was used to study the
massless two-loop two-point function with arbitrary powers of the propagators.
In fact, these ideas have been already applied to the calculation of massive
operator matrix elements with external gluon lines \cite{BIER_BLU_KLEIN}. 

The presence of the mass and the Mellin variable $n$ make our integrals more
complicated compared with the massless two-point function, so in order to
obtain the Mellin-Barnes representation of lowest dimensionality, it is
necessary to choose the right momenta flow in the diagrams. 
Let us consider the integrals $A^{a,b}_{\nu_1 \nu_2 \nu_3 \nu_4 \nu_5}$.
It turns out that changing variables
\begin{eqnarray}
k_1 \rightarrow k_2+p \,\, , \quad k_2 \rightarrow k_1 \,\, ,
\end{eqnarray}
so that now the propagators in (\ref{props}) become
\begin{eqnarray*}
&& D_1 = (k_2+p)^2-m^2 \,\, , \quad D_2 = k_1^2-m^2 \,\, ,
\quad D_3 = k_2^2 \,\, , \\
&& D_4 = (k_2-k_1+p)^2 \,\, , \quad D_5 = (k_1-k_2)^2-m^2 \,\, ,
\end{eqnarray*}
one can obtain a two-dimensional Mellin-Barnes representation. This is
achieved combining the propagators $D_2$ and $D_5$ with a Feynman parameter,
and then combining the result with $D_4$ introducing another Feynman parameter. 
After completing squares and performing the $k_1$ integral, one obtains
\begin{eqnarray}
A^{a,b}_{\nu_1,\nu_2,\nu_3,\nu_4,\nu_5} &=&
\int^1_0 dx \int^1_0 dy \int \frac{d^D k_2}{(2\pi)^D} \,\,
i \frac{\Gamma(\nu_{245}-2-\ep/2)}{\Gamma(\nu_2)\Gamma(\nu_4)\Gamma(\nu_5)}
\frac{(-1)^{\nu_{245}}}{(4\pi)^{D/2}} 
\nonumber 
\end{eqnarray}
\begin{eqnarray}
&& 
\times x^{\nu_2-1} (1-x)^{\nu_5-1} y^{-\nu_4+1-\frac{\ep}{2}} (1-y)^{\nu_4-1} 
\nonumber \\
&& \times
\frac{ \left( \Delta \cdot k_2 + \Delta \cdot p \right)^a
\left[ (1-xy) \Delta \cdot k_2 + (1-y) \Delta \cdot p \right]^b}
{D_1^{\nu_1} D_3^{\nu_3} \left[ y m^2_e-x(1-y)D_1-xy(1-x)D_3
  \right]^{\nu_{245}-D/2}} \,\, .
\end{eqnarray} 
Now, the denominator can be split twice using
\begin{equation}
\frac{1}{(X+Y)^{\nu}} = \frac{Y^{-\nu}}{2\pi i\Gamma(\nu)} 
\int^{+i \infty}_{-i \infty} d\sigma \,\, X^{\sigma} Y^{-\sigma} 
\Gamma(-\sigma) \Gamma(\sigma+\nu) \,\, ,
\label{MBbasis}
\end{equation}
and we are left with only two propagators, which can be calculated
directly. One obtains
\begin{eqnarray}
\!\!\!\!\!\! A^{a,b}_{\nu_1,\nu_2,\nu_3,\nu_4,\nu_5} &\propto& 
\frac{(-1)^{\nu_{1 \ldots 5}+1}}{\Gamma(\nu_2)\Gamma(\nu_4)\Gamma(\nu_5)} \sum_{k=0}^b 
(-1)^k \frac{b!}{(b-k)!k!} 
\frac{1}{(2\pi i)^2}
\nonumber \\ &&
\times
\int_{-i\infty}^{+i\infty} d\sigma \int_{-i\infty}^{+i\infty} d\tau \,\,
\Gamma(-\sigma) \Gamma(-\tau)
\frac{\Gamma(\sigma+k+\nu_5)\Gamma(\sigma+\tau+\nu_2)}{\Gamma(2\sigma+\tau+k+\nu_{25})} 
\nonumber \\ &&
\phantom{ \int_{-i\infty}^{+i\infty} d\sigma \int_{-i\infty}^{+i\infty}} 
\!\!\!\!\!\! \times
\Gamma(\sigma+\tau+\nu_{245}-2-\ep/2)
\frac{\Gamma(-\sigma-\tau+\nu_{13}-2-\ep/2)}{\Gamma(\nu_3-\sigma)\Gamma(\nu_1-\tau)}
\nonumber \\ &&
\phantom{ \int_{-i\infty}^{+i\infty} d\sigma \int_{-i\infty}^{+i\infty}}
\!\!\!\!\!\! \times
\frac{\Gamma(2\sigma+\tau+k-\nu_1-2\nu_3+4+\ep)\Gamma(-\sigma+a+b-k+\nu_3)}{\Gamma(\sigma+\tau+a+b-\nu_{13}+4+\ep)}
\nonumber \\ &&
\phantom{ \int_{-i\infty}^{+i\infty} d\sigma \int_{-i\infty}^{+i\infty}}
\!\!\!\!\!\! \times
\frac{\Gamma(\tau+b-k+\nu_4)\Gamma(-\tau-\nu_{25}-2\nu_4+k+4+\ep)}{\Gamma(b-\nu_{245}+4+\ep)}
\,\, .
\label{MBint}
\end{eqnarray}
For this expression, we can use the package {\tt MB} \cite{MBCzakon} to check the
integrals numerically, which can be done for up to relatively large values of
$n$. This is the main advantage of this method over {\tt Tarcer} \cite{TARCER}, which allows
to check only the first three or four moments\footnote{{\tt Tarcer} can be
modified to handle larger values of $n$ on the expense of very long computational times.}.
One may try also to obtain analytic results starting
from equation (\ref{MBint}), but unfortunately, unlike the case where we have
an external massless particle \cite{BIER_BLU_KLEIN}, this turns out to be rather
complicated, although being possible in some cases.

This method can also be applied to check the 4-propagator integrals. For example, the
denominators in  (\ref{FeynParExa1}) and (\ref{FeynParExa2}) can be
split using equation (\ref{MBbasis}) just once, leading to a simple
one-dimensional Mellin-Barnes representation. In spite of this simplicity, on
occasions the {\tt MB} package cannot find a proper contour. This usually
happens for integrals with high powers of the propagators, like the
4-propagator integrals appearing in equations (\ref{A21111_IBP}) and
(\ref{A12111_IBP}), which have a highly singular structure.
As it is well--known, this problem can be cured introducing an additional
regularization parameter.

\subsection{Sample calculations}

\vspace{1mm}
\noindent
As it was mentioned before, the method of integration by parts has the
advantage that it allows to write all 5-propagator integrals in terms of
4-propagator ones, which are simpler to calculate because they can be parameterized using
only three Feynman parameters. In the case of Eq.~(\ref{A12111_IBP}), this
simplicity is somewhat  spoiled by the high powers of the propagators in
the 4-propagator integrals. Let us consider one of the integrals appearing in
that equation, namely,
\begin{eqnarray}
A^{0,n}_{11310} & \propto & 
\int^1_0 dx \int^1_0 dy \int^1_0 dz \,\, x^{n+2} (1-x)^{-3+2\ep} (x+y-xy)^{1-\ep}
\nonumber 
\end{eqnarray}
\begin{eqnarray}
&& \phantom{ \int^1_0 dx \int^1_0 dy \int^1_0 dz }
\times y^{-2+\ep/2} (1-y)^{\ep/2} z^{-1-\ep/2} (z+y-zy)^{-2+\ep} \,\, .
\end{eqnarray}
It is not difficult to see that the singularity structure in $\ep$ is
shared by the three Feynman parameters in a way that is not easy to
disentangle. If one tries to expand in $\ep$ directly, the integrals will be
ill-defined, and it is difficult to find suitable subtraction integrals 
to cure this. To solve the problem, we decomposed the integral as
\begin{equation}
A^{0,n}_{11310} \propto I_1 + I_2 - I_3 \,\, ,
\end{equation}
where
\begin{eqnarray}
I_1 &=& 
- \frac{2}{\ep} \int^1_0 dx \int^1_0 dy \,\, x^{n+2} (1-x)^{-3+2\ep}
(x+y-xy)^{1-\ep} \nonumber \\ && \phantom{\frac{2}{\ep} \int^1_0 dx \int^1_0 dy}
\times y^{-4+\ep/2} (1-y)^{\ep/2} \,\, , \\
I_2 &=& 
\int^1_0 dx \int^1_0 dy \int^1_0 dz \,\, x^{n+2} (1-x)^{-3+2\ep} (x+y-xy)^{1-\ep}
\nonumber \\ && \phantom{ \int^1_0 dx \int^1_0 dy \int^1_0 dz }
\times y^{-4+\ep/2} (1-y)^{1+\ep/2} z^{-\ep/2} (z+y-zy)^{-1+\ep} \,\, , \\ 
I_3 &=& 
\int^1_0 dx \int^1_0 dy \int^1_0 dz \,\, x^{n+2} (1-x)^{-3+2\ep} (x+y-xy)^{1-\ep}
\nonumber \\ && \phantom{ \int^1_0 dx \int^1_0 dy \int^1_0 dz }
\times y^{-3+\ep/2} (1-y)^{1+\ep/2} z^{-\ep/2} (z+y-zy)^{-2+\ep} \,\, .
\end{eqnarray}
Here, integral $I_1$ was obtained performing integration by parts in $z$. 
It turns out that the change of variables (\ref{map}), which was used before
to write the integrals as Mellin transforms, can be used now to obtain
expressions that can be safely expanded in $\ep$. 
For example, for the integrals $I_2$ and $I_3$, we change variables by
\begin{eqnarray}
y=y'z' \,\, , \quad z=\frac{y'(1-z')}{1-z'y'} \,\, ,
\end{eqnarray}
which leads to
\begin{eqnarray}
I_2 &=&
\int^1_0 dx \int^1_0 dy' \int^1_0 dz' \,\, x^{n+3-\ep} (1-x)^{-3+2\ep}
\left( 1+\frac{1-x}{x}y'z' \right)^{1-\ep}
\nonumber \\ && \phantom{\int^1_0 dx \int^1_0 dy \int^1_0 dz} 
\times y'^{-4+\ep} z'^{-4+\ep/2} (1-z')^{-\ep/2} (1-y'z')^{\ep} \,\, , \\
I_3 &=&
\int^1_0 dx \int^1_0 dy' \int^1_0 dz' \,\, x^{n+3-\ep} (1-x)^{-3+2\ep}
\left( 1+\frac{1-x}{x}y'z' \right)^{1-\ep} 
\nonumber \\ && \phantom{\int^1_0 dx \int^1_0 dy \int^1_0 dz}
\times y'^{-4+\ep} z'^{-3+\ep/2} (1-z')^{-\ep/2} (1-y'z')^{\ep} \,\, .
\end{eqnarray}
Now, the expansion in $\ep$ (not including the term $(1-x)^{-3+2\ep}$) will produce only logarithms that 
are regular at $x = 1$, and can be calculated using
\begin{eqnarray}
\int^1_0 dy \,\, y^{-a+b\ep} (1-y)^{c\ep} \ln^k(1+\chi y) &=&
\frac{1}{1-a+b\ep} \left[ -k \chi 
\int^1_0 dy \frac{y^{-a+1+b\ep}(1-y)^{c\ep}}{1+\chi y} \ln^{k-1}(1+\chi y) \right.
\nonumber \\ && \left. 
+c\ep \int^1_0 dy \,\, y^{-a+1+b\ep} (1-y)^{-1+c\ep} \ln^k(1+\chi y) \right]
\,\, ,
\end{eqnarray}
recursively to analytically continue the expressions by reducing 
the highly singular powers in the integration variables. A few other 
integrals which appear in (\ref{A12111_IBP}) can be performed in the same way.

The $E$- and $F$-type integrals can be written in terms of Feynman
parameters from the corresponding expressions for the $A$-type integrals
using
\begin{eqnarray}
E^{a,b}_{\nu_1 \nu_2 \nu_3 \nu_4 \nu_5} 
&=& \sum^{n-1}_{j=0} A^{a+j,b+n-1-j}_{\nu_1 \nu_2 \nu_3 \nu_4 \nu_5} \,\, , \\
F^{a,b}_{\nu_1 \nu_2 \nu_3 \nu_4 \nu_5} 
&=& \sum^{n-1}_{j=0} (\Delta \cdot p)^{n-1-j} 
A^{a+j,b}_{\nu_1 \nu_2 \nu_3 \nu_4 \nu_5} \,\, .
\end{eqnarray}
For example,
\begin{eqnarray}
E^{1,0}_{11101} & \propto &
\int^1_0 dx \int^1_0 dy \int^1_0 dz \sum^{n-1}_{j=0} x^{n-1-j-\ep/2}
(1-x)^{-\ep/2} y^{j+1-\ep/2} (1-y)^{n-1-j} 
\nonumber \\ && \phantom{aaaaaaaaaaaaa}
\times z^{-1-\ep/2} [yz+x(1-x)(1-y)^2]^{\ep} \nonumber \\
&=& 
\int^1_0 dx \int^1_0 dy \int^1_0 dz \BLB x^{n+1-\ep/2}
\frac{y^{-\ep/2}(1-y)^{-\ep/2}z^{-1-\ep/2}}{(x+y-xy)[xz+y(1-y)(1-x)^2]^{-\ep}}
\nonumber \\ && \phantom{\int^1_0 dx \int^1_0 dy \int^1_0 dz }
- x^n \frac{(x+y-xy)^{-1-\ep}(1-x)^2 y^{1-\ep/2}(1-y)^{-\ep/2}z^{-1-\ep/2}}
{(x^2+y-x^2y)[yz+x^2(1-y)]^{-\ep}} \BRB \,\,  , \nonumber\\
\end{eqnarray}
and
\begin{eqnarray}
E^{1,1}_{21110} & \propto &
\int^1_0 dx \int^1_0 dy \int^1_0 dz \sum^{n-1}_{j=0} x^{n+1-j-\ep/2}
(1-x)^{\ep/2} y^{\ep/2} (1-y)^{n+1} 
\nonumber \\ && \phantom{aaaaaaaaaaaaa}
\times z^{-1-\ep/2} (1-z) [z(1-x)+xy]^{-1+\ep} \nonumber \\
&=& I_a - I_b \,\, ,
\end{eqnarray}
where
\begin{equation}
I_a = \int^1_0 dx \int^1_0 dy \int^1_0 dz \,\, x^{n+1}
\frac{y^{-1+\ep/2} (1-y)^{2-\ep/2} z^{-1-\ep/2} (1-z)}
{(1-x)^{-\ep/2} [zy+(1-y)(1-x)]^{1-\ep}} \,\, ,
\end{equation}
and
\begin{equation}
I_b = \int^1_0 dx \int^1_0 dy \int^1_0 dz \,\, x^{n+1}
\frac{y^{\ep/2} (1-y)^{-1+\ep/2} z^{-1-\ep/2} (1-z)}
{(1-x)^{1-2\ep} (x+y-xy)^{\ep} (y+z-yz)^{1-\ep}} \,\, .
\end{equation}
Integral $I_a$ is particularly difficult. One way to perform it is to use 
the integral representation of the hypergeometric function \cite{HYPER}
\begin{equation}
_2F_1(\alpha,\beta;\gamma;z) =
\frac{1}{B(\beta,\gamma-\beta)} \int^1_0 dt \,
t^{\beta-1} (1-t)^{\gamma-\beta-1} (1-tz)^{-\alpha} \,\, ,
\end{equation}
for $\mbox{Re }\gamma>\mbox{Re }\beta>0$
and then use the following analytic continuation \cite{Gradshteyn}
\begin{equation}
_2F_1(\alpha,\beta;\gamma;z) =
(1-z)^{-\beta} {_2F_1} \left( \beta,\gamma-\alpha;\gamma;\frac{z}{1-z} \right) \,\, ,
\end{equation}
to write
\begin{equation}
I_a = I_{a1} + I_{a2} - I_{a3} \,\, ,
\end{equation}
with
\begin{eqnarray}
I_{a1} &=& K \int^1_0 dx \int^1_0 dy \int^x_0 dz \,\, x^{n+1}
\frac{(1-x)^{-1+\ep} y^{-1+\ep/2} z^{\ep/2}
  (1-z)^{-\ep}}{[1-x+(x-z)y]^{-\ep/2}} \,\, , \\
I_{a2} &=& K \int^1_0 dx \int^1_0 dy \int^1_x dz \,\, x^{n+1}
\frac{(1-x)^{-1+\ep} y^{-1+\ep/2} z^{\ep/2}
  (1-z)^{-\ep}}{[1-x+(x-z)y]^{-\ep/2}} \,\, , \\
I_{a3} &=& K \int^1_0 dx \int^1_0 dy \int^1_0 dz \,\, x^{n+1}
\frac{(1-x)^{-1+\ep} y^{\ep/2} z^{\ep/2}
  (1-z)^{-\ep}}{[1-x(1-y)-yz]^{-\ep/2}} \,\, , 
\end{eqnarray}
where $K=-\frac{2}{\ep}+\ep \zeta_2 +\frac{\ep^2}{2} \zeta_3+O(\ep^3)$. The integral
$I_{a3}$ can be performed straightforwardly. For integral $I_{a1}$ we change
variables $z=xz'$, while for $I_{a2}$ one substitutes $z=(1-x)z'+x$, which leads to
\begin{eqnarray}
I_{a1} &=& K \int^1_0 dx \int^1_0 dy \int^1_0 dz' \,\, x^{n+2+\ep/2}
\frac{(1-x)^{-1+\ep} y^{-1+\ep/2} z'^{\ep/2}
  (1-xz')^{-\ep}}{[1-x+(1-z')xy]^{-\ep/2}} \,\, , \\
I_{a2} &=& K \int^1_0 dx \int^1_0 dy \int^1_0 dz' \,\, x^{n+1}
\frac{(1-x)^{\ep/2} y^{-1+\ep/2}
  (1-z')^{-\ep}}{[(1-x)z'+x]^{-\ep/2} [1-z'y]^{-\ep/2}} \,\, .
\end{eqnarray}
Now integral $I_{a2}$ is easy to obtain, and integral
$I_{a1}$ can be done using another analytic continuation of the hypergeometric
function, namely
\begin{eqnarray}
_2F_1(\alpha,\beta;\gamma;z) &=&
\frac{\Gamma(\gamma)\Gamma(\beta-\alpha)}{\Gamma(\beta)\Gamma(\gamma-\alpha)}
\left(-\frac{1}{z}\right)^{\alpha} {_2F_1}\left(\alpha,\alpha+1-\gamma;\alpha+1-\beta;\frac{1}{z} \right)
\nonumber \\ && 
+\frac{\Gamma(\gamma)\Gamma(\alpha-\beta)}{\Gamma(\alpha)\Gamma(\gamma-\beta)}
\left(-\frac{1}{z}\right)^{\beta} {_2F_1}\left(\beta,\beta+1-\gamma;\beta+1-\alpha;\frac{1}{z} \right)
\end{eqnarray}
which yields
\begin{eqnarray}
I_{a1} &=& K \int^1_0 dx \int^1_0 dy \int^1_0 dz \BLB - x^{n+2+\ep}
(1-x)^{-1+\ep} y^{-1-\ep} z^{\ep/2} (1-z)^{\ep/2} \nonumber \\
&& \phantom{ K \int^1_0 dx \int^1_0 dy \int^1_0 dz}
\times (1-xz)^{-\ep} \left( 1 +\frac{1-x}{(1-z)x} y \right)^{\ep/2} 
\nonumber \\ && \phantom{ K \int^1_0 dx \int^1_0 dy \int^1_0 dz}
+\left( \frac{1}{\ep}+\frac{\ep}{2}\zeta_2+\frac{\ep^2}{4}\zeta_3 \right)
x^{n+2} (1-x)^{-1+2\ep} \nonumber \\
&& \phantom{ K \int^1_0 dx \int^1_0 dy \int^1_0 dz}
\times z^{\ep/2} (1-z)^{-\ep/2} (1-xz)^{-\ep} \BRB \,\, .
\end{eqnarray}
This integral can now be computed using known integrals and specific integrals given in the following 
Section.

\subsection{Polylogarithmic integrals and analytic continuations}

\vspace*{1mm}
\noindent
In the following we list a series of integrals over polylogarithms derived in
the present calculation beyond those which were given in \cite{DUDE}. They may be
of use in other higher order calculations. 
\begin{eqnarray}
\int^1_0 dy \,\,\, \frac{1-x}{1-(1-x)y} {\rm Li}_2 \left( -\frac{1-y}{x^2y} \right)  
&=& 
2 \zeta_2 \ln(x)+{7 \over 6} \ln^3(x)+4 \ln(x) {\rm Li}_2(-x)
\nonumber \\ 
& & 
+3 \ln(x) {\rm Li}_2(x)+2 {\rm Li}_3(x)-2 {\rm Li}_3(x^2) \,\, , 
\end{eqnarray}
\begin{eqnarray}
\int^1_0 dy \,\,\, \frac{x}{(1-(1-x)y)^2} {\rm Li}_2 \left( -\frac{1-y}{x^2y} \right) 
&=& -{\rm Li}_2(1-x)-{1 \over 2} \ln^2(x)-\zeta_2 \,\, , 
\\
\int^1_0 dy \,\,\, \frac{x}{1-xy} {\rm Li}_2 \left( -\frac{1-x}{x^2y(1-y)} \right) 
&=& 
2 {\rm S}_{1,2}(x)+4 {\rm S}_{1,2}(1-x)-4 \zeta_3 
\nonumber \\ 
&&
-\ln(x) \ln^2(1-x) \,\, , 
\\
\int^1_0 dy \,\,\, {\rm Li}_2 \left( \frac{x}{(1-y+xy)(x+y-xy)} \right) 
&=& 
{1+x \over 1-x} \left[ {\rm Li}_2(x^2)+2 \ln(x) \ln(1+x) \right] 
\nonumber \\ 
&& 
-4 \ln(1-x) -{2 x \over 1-x} \left[ \zeta_2+2 \ln(x) \right] \,\, , 
\\
\int^1_0 dy \,\,\, \frac{x}{1-xy} \ln(1-y) \ln(1-xy) 
&=& 
{\rm Li}_3(1-x)+\frac{1}{2} \ln(x) \ln^2(1-x) 
\nonumber \\ 
&&
-\frac{1}{3} \ln^3(1-x)-\zeta_2 \ln(1-x)-\zeta_3 \,\, , 
\\
\int^1_0 dy \,\,\, \frac{x(1-x)}{(1-xy)^2} \ln(1-y) \ln(1-xy) 
&=& 
{\rm Li}_2(x)+\ln^2(1-x)+\ln(1-x) \,\, , 
\\
\int^1_0 dy \,\,\, \frac{x}{1-xy} \ln(y) \ln(1-xy) 
&=& 
{\rm S}_{1,2}(x) \,\, , 
\\
\int^1_0 dy \,\,\, \frac{x}{(1-xy)^2} \ln(y) \ln(1-xy) 
&=& 
{\rm Li}_2(x)+\frac{1}{2} \ln^2(1-x)+\ln(1-x) \,\, .
\end{eqnarray}
In all of the results given above, it is assumed that $0 \leq x \leq 1$.

Some double integrals that were also used are shown below, where $u=(1-x)/x$,
\begin{eqnarray}
\int_0^1 dy \int_0^1 dz   \,\,\,
{\frac {z \ln  \left( 1-zy \right) \ln  \left( 1+uyz \right) }{1-zy}} 
&=& 
2 + \frac{1+u}{u}\left[\frac{1}{2} \ln^2(1+u) 
- \ln(1+u)
+ \Li_2(-u)\right]
\nonumber \\ 
&=&
2 + \frac{\ln(x) - \Li_2(1-x)}{1-x} 
\\
%
\int_0^1 dy \int_0^1 dz   \,\,\,
{\frac {z \ln  \left( 1-z \right) \ln  \left( 1+uyz \right) }{1-zy}}
&=&
3 + \frac{1+u}{u} \Biggl[\frac{1}{2} \ln^2(1+u) 
 - 2 \ln(1+u) + \Li_2(-u)\Biggr] 
\nonumber\\ 
&=&
3 + \frac{2 \ln(x) - \Li_2(1-x)}{1-x}
\\
%
\int_0^1 dy \int_0^1 dz   \,\,\,
{\frac {y{z}^{2} \ln^2  \left( 1+uyz \right)}{1-zy}}
&=&
\frac{1}{4} - \frac{3}{2u} +\frac{1}{2}\left(1- \frac{1}{u^2}\right) \ln^2(1+u)
\nonumber\\ 
&& 
-\left( \frac{1}{2}-\frac{1}{u}-\frac{3}{2u^2}\right)\ln(1+u)
\nonumber\\
&=&
\frac{1-2x}{2 (1-x)^2} \ln^2(x)
+\frac{1-4x}{2 (1-x)^2} \ln(x)
-\frac{7x-1}{4 (1-x)}
\\
%
\int_0^1 dy \int_0^1 dz   \,\,\,
{\frac {y {z}^{2} \ln  \left( z \right) \ln  \left( 1+uyz \right) }{1-zy}}
&=&
- \frac{5}{4u}+\frac{5}{2}- 2 S_{1,2}(-u) + \Li_3(-u)
\nonumber\\ 
&&
+ \left[-\frac{1}{2 u^2} +\frac{1}{u} - \ln(1+u)\right] \Li_2(-u) 
\nonumber\\ 
&&
+ \left[\zeta_2 -\frac{7}{4}+\frac{3}{4 u^2} -\frac{1}{u} \right] \ln(1+u) 
\nonumber\\
&=& - \frac{5x}{4(1-x)} + \frac{5}{2} - \zeta_3 - \ln(x) \Li_2(x)
\nonumber
\end{eqnarray}
\begin{eqnarray}
&&
- \frac{1}{2} \ln^2(x)\ln(1-x) + \Li_3(x) -\Li_3(1-x)
\nonumber\\ 
&&
+\left(\frac{x}{1-x}- \frac{3x^2}{4(1-x)^2} + \frac{7}{4} - \zeta_2 \right) \ln(x)
\nonumber\\ 
&&
+ \frac{3x^2-2x}{2(1-x)^2} \Li_2(1-x) 
+ \frac{3x^2-2x}{4(1-x)^2} \ln^2(x)
\\
%
\int_0^1 dy \int_0^1 dz   \,\,\, \frac {z \ln (z) \ln(1+uyz)}{1-zy}
&=&
3 - \ln(1+u) \Li_2(-u) + \ln(1+u) \zeta_2 + \frac{1}{u} \Li_2(-u) 
\nonumber\\ 
& &
-2 \left(1+\frac{1}{u}\right) \ln(1+u) + \Li_3(-u) -2 S_{1,2}(-u)   
\nonumber\\ 
&=& 
3 + \left(\frac{2}{1-x} - \zeta_2\right) \ln(x) 
- \frac{1}{2} \frac{x}{1-x}   \ln^2(x) 
\nonumber\\ 
& &
- \frac{x}{1-x} \Li_2(1-x) -\Li_3(1-x) - S_{1,2}(1-x)
\\
%
\int_0^1 dy \int_0^1 dz \,\,\, \frac {z \ln (y) \ln(1+uyz)}{1-zy}
&=&
-1 +\left(1+\frac{1}{u} - \zeta_2\right) \ln(1+u) + \ln(1+u) \Li_2(-u)
\nonumber\\ 
& &
- \Li_3(-u) + 2 S_{1,2}(-u) 
\nonumber\\ 
&=&
-1+\left(\zeta_2 - \frac{1}{1-x}\right) \ln(x) 
\nonumber\\  
& &
+ \Li_3(1-x) + S_{1,2}(1-x)
\\
%
\int_0^1 dy \int_0^1 dz   \,\,\, \frac {z \ln^2(1+uyz)}{1-zy}
&=&
2 + \left(1 + \frac{1}{u} \right) \left[\ln^2(1+u) - 2 \ln(1+u)\right]
\nonumber\\ 
&=&
2 + \frac{\ln^2(x)}{1-x} + 2 \frac{\ln(x)}{1-x}
\\
%
\int_0^1 dy \int_0^1 dz  \,\,\,
{\frac{y {z}^{3}\ln  \left( z \right) \ln  \left( 1- x y \right) }{1-zy}} 
&=&
- \frac{1}{9} + \frac{5}{4} x 
-\frac{1}{2} \left(x -\frac{1}{2} x^2\right)\zeta_2
+\frac{1}{2} x^2 \zeta_3
\nonumber\\ 
&&
+\left(\frac{3}{4}x - \frac{23}{36} -\frac{1}{9x}\right) \ln(1-x)
-\frac{1}{4}\left(1-x^2\right) \Li_2(x) 
\nonumber\\ 
&&
-\frac{1}{2} x^2\left[\Li_3(x) +S_{1,2}(x)\right] 
- \frac{1}{8}\left(1-x^2\right)\ln^2(1-x)
\nonumber\\
&& 
+ \frac{1}{2} \left(1- x^2\right) \ln(1-x) \zeta_2
\\
%
\int_0^1 dy \int_0^1 dz   \,\,\,
{\frac {{z}^{3}\ln  \left( 1- x y \right) \ln  \left( y 
\right) }{1-zy}} 
&=&
-\frac{5}{24}x -\frac{4}{9} x^2 +\frac{1}{3}x^3 \zeta_3
+\left(\frac{1}{3} x^2 + \frac{1}{6} x -\frac{x^3}{9}\right) \zeta_2
\nonumber\\ 
&&
+\frac{1}{3}\left(1-x^3\right) \ln(1-x) \left[\Li_2(x) -\zeta_2\right]  
\nonumber\\ 
&&
+\left(-\frac{x^2}{9} -\frac{x}{18} +\frac{1}{6}\right) \ln(1-x)
+\frac{1}{3}\left(2-x^3\right) S_{1,2}(x) 
\nonumber\\ 
&&
-\frac{1}{3}\left(1-2 x^3\right) \Li_3(x)
+ \frac{1}{18}\left(1-x^3\right) \ln^2(1-x)
\nonumber\\ 
&&
-\frac{1}{3}\left(x^2+\frac{x}{2}+\frac{x^3}{3}\right) \Li_2(x)
\\
%
\int_0^1 dy \int_0^1 dz   \,\,\,
{\frac {{z}^{3}\ln  \left( 1-y \right) \ln  \left( 1- x y \right) }{1-zy}}
&=&
\frac{x}{3} + \frac{x}{2}(1+x) \zeta_2 + \frac{2}{3} x^3 \zeta_3
-\frac{1}{6}(1-x) \ln(1-x)
\nonumber\\ 
&&
- \frac{1}{3}\left(1-2 x^3\right) \Li_3(x) + \frac{2}{3} \left(1-x^3\right) \ln(1-x) \Li_2(x)
\nonumber
\end{eqnarray}
\begin{eqnarray}
&&
+ \left(1- \frac{2}{3} x^3\right) S_{1,2}(x) + \frac{x}{6}(1-x) \Li_2(x)
\nonumber\\ 
&&
+ \frac{x}{12}(1-x) \ln^2(1-x) + \frac{1}{9} \left(1-x^3\right) \ln^3(1-x)
\\ 
%
\int_0^1 dy \int_0^1 dz   \,\,\,
{\frac {y {z}^{3}\ln  \left( 1- x y \right) \ln  \left( y \right) }{1-zy}}
&=&
-\frac{2}{3} - \frac{3}{4} x 
+ \frac{1}{2} x^2 \zeta_3 
-\frac{1}{2}\left(1-x^2\right) \ln(1-x) \zeta_2 
\nonumber\\ 
& &
+ \left(\frac{7}{12}
- \frac{x}{4} -\frac{1}{3x} \right) \ln(1-x)
+\left(\frac{x}{2}-\frac{x^2}{4}\right) \zeta_2
\nonumber\\ 
& &
-\left(\frac{x}{2}-\frac{1}{3x}+\frac{x^2}{4}\right) \Li_2(x)
+ \frac{1}{8} \left(1-x^2\right) \ln^2(1-x)
\nonumber\\ 
& &
+\left(x^2 - \frac{1}{2} \right) \Li_3(x) +\left(1 -\frac{x^2}{2}\right) S_{1,2}(x)
\nonumber\\ 
& &
+\frac{1}{2}\left(1-x^2\right)\ln(1-x)\Li_2(x)
\\ 
%
\int_0^1 dy \int_0^1 dz   \,\,\,
{\frac {y {z}^{3} \ln^2  \left( 1- x y \right) }{1-zy}}
&=&
-\frac{2}{3} +x^2 \zeta_2+ \left(1-x^2\right) \Li_2(x)\ln(1-x) 
\nonumber\\ 
& &
+ \frac{2}{3}\left(1-\frac{1}{x}\right) \ln(1-x) + S_{1,2}(x) 
+ x^2 \Li_3(x) 
\nonumber\\ 
& &
+ x(x-1) \Li_2(x)
+\left(\frac{1}{6} -x + \frac{1}{3 x} + \frac{x^2}{2} \right) \ln^2(1-x)
\nonumber\\ 
& &
+\frac{1}{3}\left(1-x^2\right) \ln^3(1-x)
\\
%
\int_0^1 dy \int_0^1 dz   \,\,\,
{\frac {{z}^{3}\ln  \left( z \right) \ln  \left( 1- x y \right) }{1-zy}} 
&=&
\frac{5x}{12} + \frac{7x^2}{9} 
+\left(\frac{x^3}{9}-\frac{x^2}{3}-\frac{x}{6}\right) \zeta_2
+\frac{x^3}{3} \zeta_3
\nonumber\\ 
& &
+\frac{1}{3}\left(1-x^3\right)\ln(1-x) \zeta_2 - \frac{x^3}{3}\left[S_{1,2}(x) + \Li_3(x)\right]
\nonumber
\\ 
& &
-\frac{1}{9}\left(1-x^3\right) \Li_2(x) - \frac{1}{18}\left(1-x^3\right) \ln^2(1-x)
\nonumber\\ 
& &
+\left(-\frac{7}{12}+\frac{5}{36}x+\frac{4}{9} x^2 \right) \ln(1-x)
\\
%
\int_0^1 dy \int_0^1 dz   \,\,\,
{\frac {{z}^{3} \ln^2  \left( 1- x y \right)}{1-zy}}
&=& 
\frac{x^2}{3} + x^3 \zeta_2 + \frac{2}{3}\left(1-x^3\right) 
\ln(1-x)\Li_2(x)
\nonumber\\ 
& &
-\frac{x}{3}(1-x) \ln(1-x) 
+ \frac{2}{9} \left(1-x^3\right) \ln^3(1-x)
\nonumber\\ 
& &
+ \left(-\frac{x}{3} -\frac{2}{3} x^2 +x^3 
\right) \Li_2(x) + \frac{2}{3} x^3 \Li_3(x) +\frac{2}{3} S_{1,2}(x)
\nonumber\\ 
& &
+ \left(\frac{1}{2} - \frac{x}{3} - \frac{2}{3} x^2 +\frac{1}{2} x^3\right) \ln^2(1-x)
\\
%
\int_0^1 dy \int_0^1 dz   \,\,\,
{\frac {{z}^{3} \ln  \left( 1-zy \right) \ln  \left( 1- x y \right) }{1-zy}}
&=&
\frac{3x}{8} + \frac{11x^2}{18}  
+\left(-\frac{3}{4} + \frac{5x}{36} +\frac{11x^2}{18}\right) \ln(1-x)
\nonumber\\ 
&&
+\left(-\frac{1}{9}-\frac{x}{6} -\frac{x^2}{3} +\frac{11x^3}{18}\right) \Li_2(x) + \frac{11}{18} x^3 \zeta_2
\nonumber\\ 
&&
-\frac{1}{3}\left(1-x^3\right) \left[\Li_3(x) - S_{1,2}(x)\right] + \frac{x^3}{3} \zeta_3
\nonumber\\ 
&&
+\left(1-x^3\right) \left[\frac{1}{18} \ln^3(1-x) + \frac{1}{3} \ln(1-x) \Li_2(x)\right] 
\nonumber
\end{eqnarray}
\begin{eqnarray}
&&
+\left(-\frac{1}{18}-\frac{x}{12} -\frac{x^2}{6} +\frac{11 x^3}{36}\right) \ln^2(1-x)
\\
%
\int_0^1 dy \int_0^1 dz   \,\,\,
{\frac {y {z}^{3} \ln  \left( 1-zy \right) \ln  \left( 1- x y \right) }{1-zy}}
&=& 
-\frac{7}{9} +\frac{11x}{12} 
+ \left(\frac{11}{12}x - \frac{17}{36}  - \frac{4}{9 x}\right) \ln(1-x)
\nonumber\\ 
& &
+ \left(\frac{11}{24}x^2 - \frac{3}{8}  -\frac{x}{4} + \frac{1}{6 
x}\right) \ln^2(1-x)
\nonumber\\ 
& &
+ \frac{11}{12} x^2 \zeta_2 +\frac{1}{12} \left(1-x^2\right) \ln^3(1-x)
\nonumber\\ 
& &
+\left(\frac{1}{3x} -\frac{3}{4} - \frac{x}{2} + \frac{11}{12} x^2 \right) \Li_2(x) 
+ \frac{1}{2} x^2 \zeta_3
\nonumber\\ 
& &
+\frac{1}{2}\left(1- x^2\right) \left[S_{1,2}(x)  - \Li_3(x) \right.
\nonumber\\
& &
+\frac{1}{2}\left(1- x^2\right) \left[ \right. 
\left. + \ln(1-x) \Li_2(x) \right]
\\
%
\int_0^1 dy \int_0^1 dz   \,\,\,
{\frac {{z}^{2}\ln  \left( z \right) \ln  \left( 1- x y \right) }{1-zy}}
&=&
\frac{5}{4} x - \frac{x}{2}\left(1-\frac{x}{2}\right) \zeta_2 + 
\frac{x^2}{2} \zeta_3 
\nonumber\\ 
& &
- \frac{3}{4} \left(1-x\right)\ln(1-x) - \frac{1}{4} \left(1 - x^2 \right) 
\Li_2(x)
\nonumber\\ 
& & 
+\frac{1}{2}\left(1-x^2\right) \ln(1-x) \zeta_2- \frac{1}{8} 
\left(1-x^2\right) 
\ln^2(1-x)
\nonumber\\ 
& & 
- \frac{1}{2} x^2 \left[\Li_3(x) +S_{1,2}(x) \right]
\\
%
\int_0^1 dy \int_0^1 dz   \,\,\,
{\frac {{z}^{2}\ln  \left( 1- x y \right) \ln  \left( y \right) }{1-zy}}
&=&
-\frac{3}{4} x 
- \frac{1}{2}\left(1-x^2\right) \ln(1-x) \zeta_2
\nonumber\\ 
& &
+ \frac{1}{8} \left(1-x^2\right) \ln^2(1-x)
+ \frac{x}{2}\left(1-\frac{x}{2} \right) \zeta_2
\nonumber\\ 
& &
+ \frac{1}{4} \left(1-x\right) \ln(1-x)
- \frac{x}{2} \left(1+\frac{x}{2}\right) \Li_2(x) 
\nonumber\\ 
& &
+ \left(1-\frac{x^2}{2}\right) S_{1,2}(x) 
+\left(x^2 - \frac{1}{2}\right) \Li_3(x)
+ \frac{1}{2} x^2 \zeta_3
\nonumber\\ 
& &
+ \frac{1}{2}\left(1-x^2\right) \ln(1-x) \Li_2(x)
\\
%
\int_0^1 dy \int_0^1 dz   \,\,\,
{\frac {{z}^{2}\ln  \left( 1-y \right) \ln  \left( 1- x y \right) }{1-zy}}
&=&
x \zeta_2 + x^2 \zeta_3 + \left(\frac{3}{2} - x^2\right) S_{1,2}(x)
\nonumber\\ 
& &
+\left(1-x^2\right) \ln(1-x) \Li_2(x) - \left(\frac{1}{2} - x^2\right) \Li_3(x)
\nonumber\\ 
& & 
+ \frac{1}{6}\left(1 - x^2\right) \ln^3(1-x)
\\
%
\int_0^1 dy \int_0^1 dz   \,\,\, {\frac {{z}^{2} \ln^2  \left( 1- x y \right) }{1-zy}}
&=&
x^2 \zeta_2 + \left(1-x^2\right)\ln(1-x) \Li_2(x) + S_{1,2}(x) + x^2 \Li_3(x)
\nonumber\\ 
& &
-x(1-x) \Li_2(x) +\left(\frac{1}{2} x^2 +\frac{1}{2} -x \right) \ln^2(1-x) 
\nonumber\\ 
& &
+ \frac{1}{3}\left(1-x^2\right) \ln^3(1-x)
\\
%
\int_0^1 dy \int_0^1 dz   \,\,\,
{\frac {{z}^{2} \ln  \left( 1-zy \right) \ln  \left( 1- x y \right) }{1-zy}} 
&=&
\frac{3}{4} x
+\left(\frac{3}{8} x^2 - \frac{1}{4}x -\frac{1}{8} \right) \ln^2(1-x) 
\nonumber
\end{eqnarray}
\begin{eqnarray}
& &
- \frac{3}{4}(1-x)\ln(1-x) 
+ \frac{1}{12}\left(1-x^2\right) \ln^3(1-x) 
\nonumber\\ 
& &
-\left( \frac{1}{4} +\frac{1}{2}x -\frac{3}{4} x^2 \right) \Li_2(x)
+\frac{3}{4} x^2 \zeta_2
+ \frac{1}{2} x^2 \zeta_3
\nonumber\\ 
& &
+\frac{1}{2} \left(1 - x^2 \right) \left[ S_{1,2}(x) -  \Li_3(x)  \right.
\nonumber\\ 
& &
+\frac{1}{2} \left(1 - x^2 \right) \left[ \right. 
\left. + \ln(1-x) \Li_2(x)\right]
\\
%
\int_0^1 dy \int_0^1 dz   \,\,\, \frac {z  \ln^2(1-xy)}{1-zy}
&=&
2 S_{1,2}(x) +2 x \Li_3(x) + 2(1-x) \ln(1-x) \Li_2(x) 
\nonumber\\ 
& &
+ \frac{2}{3} \left(1-x\right) \ln^3(1-x) 
\\
%
\int_0^1 dy \int_0^1 dz   \,\,\, \frac {y z^2  \ln^2(1-xy)}{1-zy}
&=&  
-1 +2 (1-x) \ln(1-x) \Li_2(x) 
+ 2 S_{1,2}(x) + 2 x \Li_3(x) 
\nonumber\\ 
& & 
+ \left(1-\frac{1}{x} \right)\ln(1-x) - \frac{1}{2}\left(1-\frac{1}{x}\right) \ln^2(1-x)
\nonumber\\ 
& & 
+\frac{2}{3} (1-x) \ln^3(1-x)
\\
%
\int_0^1 dy \int_0^1 dz   \,\,\, \frac {z  \ln(y) \ln(1-xy)}{1-zy}
&=&  
x \left( \zeta_3-\zeta_2 \right) -(1-2 x) \Li_3(x) 
- \zeta_2 (1-x) \ln(1-x) 
\nonumber\\ 
& &
+ (2-x) S_{1,2}(x) + (1-x) \ln(1-x) \Li_2(x) 
\nonumber\\ 
& &
+ \frac{1}{2} (1-x) \ln^2(1-x)
- x \Li_2(x) 
\\
%
\int_0^1 dy \int_0^1 dz \,\,\, \frac {z  \ln(1-y) \ln(1-xy)}{1-zy}
&=&  
2 x \zeta_3+2 (1-x) \ln(1-x) \Li_2(x) + (2 x-1) \Li_3(x) 
\nonumber\\ 
& &
+(3-2 x) S_{1,2}(x) + \frac{1}{3}(1-x) \ln^3(1-x)
\\
%
\int_0^1 dy \int_0^1 dz \,\,\, \frac {y z^3  \ln(1-y) \ln(1-xy)}{1-zy}
&=&  
x \zeta_2 - \frac{2}{3} + x^2 \zeta_3 +\left(1-x^2\right) \ln(1-x) \Li_2(x)
\nonumber\\ 
& &
+ \left(x^2-\frac{1}{2} \right) \Li_3(x) + \left(\frac{3}{2} -x^2\right) S_{1,2}(x)
\nonumber\\ 
& &
+ \frac{x-1}{3x} \left[\ln(1-x)-\Li_2(x) - \frac{1}{2} \ln^2(1-x) \right] 
\nonumber\\ 
& &
+ \frac{1}{6}\left(1-x^2\right) \ln^3(1-x) 
\\
%
\int_0^1 dy \int_0^1 dz \,\,\, \frac {z  \ln(1-z) \ln(1-xy)}{1-zy}
&=&  
 (1-x) \left[ 2 \zeta_2 \ln(1-x) -\frac{1}{2} \ln^2(1-x) - \Li_2(x)  \right.
\nonumber\\ 
& &
-2 \Li_3(1-x) +\ln(1-x) \Li_2(1-x)  
\nonumber\\ 
& &
\left. +\frac{1}{6} \ln^3(1-x) \right] + x \zeta_2 + 2 \zeta_3
\\
%
\int_0^1 dy \int_0^1 dz \,\,\, \frac {z^2  \ln(1-z) \ln(1-xy)}{1-zy}
&=&  
\frac{5x}{4} + x^2 \zeta_3 +\frac{1}{2}\left(1-x^2\right) \ln(1-x)\left[\zeta_2 +\Li_2(x)\right]
\nonumber\\ 
& &
-\frac{3}{4} (1-x) \ln(1-x) 
- \frac{1}{4} \left(3-2 x-x^2 \right) \Li_2(x) 
\nonumber\\ 
& &
+\left(1-x^2\right)
S_{1,2}(x) 
+ \frac{1}{2}\left(x +\frac{x^2}{2} \right) \zeta_2
\nonumber
\end{eqnarray}
\begin{eqnarray}
& &
+ \frac{1}{8} \left(2 x-3+x^2\right) \ln^2(1-x) 
\nonumber\\ 
& &
+ \frac{1}{12} \left(1-x^2\right) \ln^3(1-x) 
\\
%
\int_0^1 dy \int_0^1 dz \,\,\, \frac {z^3  \ln(1-z) \ln(1-xy)}{1-zy}
&=&  
\frac{5x}{6} + \frac{7x^2}{9} 
+ \left(-\frac{11}{12} + \frac{17}{36} x + \frac{4}{9} x^2\right) \ln(1-x)
\nonumber\\ 
& &
+\left(-\frac{11}{18} + \frac{x}{3} +\frac{x^2}{6} + \frac{x^3}{9} \right) \Li_2(x)
+ \frac{2}{3} x^3 \zeta_3
\nonumber\\ 
& &
+\frac{2}{3}\left(1-x^3\right) S_{1,2}(x) + \left(\frac{x}{3} + \frac{x^2}{6} + \frac{x^3}{9}\right) \zeta_2
\nonumber\\
& &
+\left(-\frac{11}{36} + \frac{x}{6} + \frac{x^2}{12} + \frac{x^3}{18} \right) \ln^2(1-x)
\nonumber\\ 
& &
+\frac{1}{3}\left(1-x^3\right) \left\{ \ln(1-x) \left[\Li_2(x) + \zeta_2\right] \phantom{\frac{1}{6}} \right.
\nonumber\\ 
& &
+\frac{1}{3}\left(1-x^3\right) \left\{ \right. 
\left. + \frac{1}{6} \ln^3(1-x) \right\}
\\
%
\int_0^1 dy \int_0^1 dz \,\,\, \frac {y z^2  \ln(1-z) \ln(1-xy)}{1-zy}
&=&  
2 x \zeta_3 - \frac{3}{4} + x \zeta_2 +\frac{3}{4} \left(1-\frac{1}{x} \right) \ln(1-x) 
\nonumber\\ 
& &
-\frac{1}{2} (1-x) \ln^2(1-x) + \frac{1}{6} (1-x) \ln^3(1-x)
\nonumber\\ 
& &
-(1-x) \Li_2(x) +2 (1-x) S_{1,2}(x) 
\nonumber\\ 
& &
+ (1-x) \ln(1-x) \left[\Li_2(x)+\zeta_2 \right]
\\
%
\int_0^1 dy \int_0^1 dz \,\,\, \frac {y z^3  \ln(1-z) \ln(1-xy)}{1-zy}
&=&  
\frac{5x}{4} - \frac{11}{18}+ x^2 \zeta_3  
+ \frac{1}{8} \left(2x-3+x^2\right) \ln^2(1-x) 
\nonumber\\ 
& &
+ \frac{1}{12} \left(1-x^2\right) \ln^3(1-x) 
+\left(1-x^2\right) S_{1,2}(x)
\nonumber\\ 
& &
+\frac{1}{2}\left(1-x^2\right) \ln(1-x)\left[\zeta_2 +\Li_2(x)\right]
\nonumber\\ 
& &
+\left(\frac{3}{4}x - \frac{5}{36} - \frac{11}{18x}\right) \ln(1-x) 
+ \frac{2 x + x^2}{4} \zeta_2 
\nonumber\\ 
& &
+ \frac{1}{4} \left(-3+2x+x^2\right) \Li_2(x) 
\\
%
\int_0^1 dy \int_0^1 dz \,\,\, \frac {z  \ln(1-yz) \ln(1-xy)}{1-zy}
&=&  
x \left(\zeta_3 + \zeta_2\right) -(1-x) \left[\Li_3(x) - S_{1,2}(x) + \Li_2(x)\right]
\nonumber\\
& & 
+ \frac{1}{2} (x-1) \ln^2(1-x) + \frac{1}{6} (1-x) \ln^3(1-x) 
\nonumber\\ 
& &
+(1-x)\ln(1-x) \Li_2(x)     
\\
%
\int_0^1 dy \int_0^1 dz \,\,\, \frac {y z^2  \ln(1-yz) \ln(1-xy)}{1-zy}
&=&  
x \zeta_3 - \frac{5}{4} + \frac{3}{4} \left(1- \frac{1}{x}\right) \ln(1-x) + \frac{3}{2} x \zeta_2
\nonumber\\ 
& & 
+ (1-x) \left[ \ln(1-x) \Li_2(x) + \frac{1}{6} \ln^3(1-x) \right] 
\nonumber\\ 
& & 
+ \left(-1 + \frac{3 x}{4} +\frac{1}{4 x} \right) \ln^2(1-x)
\nonumber\\ 
& &
+\left(\frac{1}{2x} -2 + \frac{3 x}{2} \right) \Li_2(x)
\nonumber\\ 
& &
+(1-x) \left[S_{1,2}(x) - \Li_3(x)\right]
\end{eqnarray}
\begin{eqnarray}
%
\int_0^1 dy \int_0^1 dz \,\,\, \frac {z \ln(y) \ln(1-xy) }{(1-zy)^2}
&=&   
x \left( \zeta_2 - \zeta_3 \right) + x \ln(1-x) \Li_2(x) 
\nonumber\\ 
& &
+ x \left[ S_{1,2}(x) - 2 \Li_3(x) \right] +x \Li_2(x) 
\nonumber\\ 
& &
- \frac{1}{2} (1-x) \ln^2(1-x) - x \zeta_2 \ln(1-x)
\\
%
\int_0^1 dy \int_0^1 dz \,\,\, \frac {z^2 \ln(y) \ln(1-xy) }{(1-zy)^2}
&=&  
\frac{3}{2} x + x^2 \ln(1-x) \left[\Li_2(x) - \zeta_2\right]
+ x\left(\frac{x}{2} +1\right) \Li_2(x) 
\nonumber\\ 
& &
+ x^2 \left[S_{1,2}(x) - 2 \Li_3(x) \right] + \frac{x-1}{2} \ln(1-x)
- x^2 \zeta_3 
\nonumber\\ 
& &
-x \zeta_2 \left(1-\frac{x}{2}\right) -\frac{1}{4} \left(1-x^2\right) \ln^2(1-x)
\\
%
\int_0^1 dy \int_0^1 dz \,\,\, \frac {z^3 \ln(y) \ln(1-xy) }{(1-zy)^2}
&=&  
\frac{5x}{8} + \frac{4x^2}{3} - x^3\zeta_3
+x^3 \left[S_{1,2}(x) - 2 \Li_3(x)\right]
\nonumber\\ 
& &
-\left( \frac{1}{2}-\frac{1}{3} x^2 - \frac{1}{6} x \right) \ln(1-x)
+ \left(\frac{x^3}{3}-\frac{x}{2} - x^2\right) \zeta_2 
\nonumber\\ 
& &
+\left(x^2 +\frac{1}{3}x^3+\frac{1}{2}x\right) \Li_2(x) + x^3 \ln(1-x) \Li_2(x)
\nonumber\\ 
& &
- \zeta_2 x^3 \ln(1-x) - \frac{1}{6} \left(1-x^3\right) \ln^2(1-x)
\\
%
\int_0^1 dy \int_0^1 dz \,\,\, \frac {y z^2 \ln(y) \ln(1-xy) }{(1-zy)^2}
&=&
-2 x \left(\zeta_3 - \zeta_2\right) -(1-2 x) \ln(1-x) \left[ \Li_2(x) -\zeta_2 \right]
\nonumber\\ 
& &
-2 (1-x) S_{1,2}(x)
+(1-4 x)\Li_3(x) 
+ 2 x \Li_2(x) 
\nonumber\\ 
& &
- (1-x) \ln^2(1-x)
\end{eqnarray}

The following integrals were also required. They are understood in the sense of an
analytic continuation and expanding in $\ep$ to the order needed in the present calculation.
\begin{eqnarray}
\int^1_0 dy \,\, y^{-2-\ep} \ln(1+uy) 
&=&
-\frac{u}{\ep}+u-(1+u)\ln(1+u) \nonumber \\ &&
+\ep \left[ u {\rm Li}_2(-u) +(1+u) \ln(1+u) -u \right] 
\\
%
\int^1_0 dy \,\, y^{-3-\ep} \ln(1+uy) 
&=&
\frac{u^2}{2\ep}-\frac{u^2}{4}-\frac{u}{2}-\frac{1}{2}(1-u^2)\ln(1+u) \nonumber \\ &&
+\ep \left[ \frac{u^2}{8}+\frac{3}{4}u-\frac{u^2}{2}{\rm Li}_2(-u)
+\frac{1}{4}(1-u^2) \ln(1+u) \right] \\
%
\int^1_0 dy \,\, y^{-4-\ep} \ln(1+uy) 
&=&
-\frac{u^3}{3\ep}+\frac{u^2}{3}-\frac{u}{6}+\frac{u^3}{9}-\frac{1}{3} (1+u^3) \ln(1+u)
\nonumber \\ &&
+\ep \left[ \frac{1}{3} u^3 {\rm Li}_2(-u)-\frac{u^3}{27}-\frac{4}{9} u^2+\frac{5}{36} u \right.
\nonumber \\ && \phantom{+\ep \left[ \right.}
\left. +\frac{1}{9} (1+u^3) \ln(1+u)
\right] 
\\
\int^1_0 dy \,\, y^{-2-\ep} \ln^2(1+uy) &=& -2 u {\rm Li}_2(-u) -(1+u) \ln^2(1+u) \\
\int^1_0 dy \,\, y^{-3-\ep} \ln^2(1+uy) &=&
-\frac{u^2}{\ep}+\frac{3}{2} u^2-u (1+u)\ln(1+u)
+u^2 {\rm Li}_2(-u) \nonumber \\ &&
-\frac{1}{2}(1-u^2) \ln^2(1+u) 
\end{eqnarray}
\begin{eqnarray}
%
\int^1_0 dy \,\, y^{-4-\ep} \ln^2(1+uy) &=&
\frac{u^3}{\ep}-\frac{1}{3} (1+u) (1-u+u^2) \ln^2(1+u)-\frac{1}{6} (2+7u) u^2
\nonumber \\ &&
-\frac{u}{3} (1+u) (1-3u) \ln(1+u)
-\frac{2}{3} u^3 {\rm Li}_2(-u) 
\\
\int^1_0 dy \,\, y^{-3-\ep} \ln(1+u_1y) \ln(1+u_2y) &=&
-\frac{1}{2\ep} u_1 u_2+\frac{3}{4} u_1 u_2
-\frac{1}{4} \ln(1+u_1) \ln(1+u_2)
\nonumber \\ &&
+\frac{u_1^2}{2} \left[ {\rm Li}_2(-u_2)
-{\rm Li}_2 \left( \frac{u_1}{u_1-u_2} \right)
+{\rm Li}_2 \left( \frac{u_1 (1+u_2)}{u_1-u_2} \right) \right.
\nonumber \\ && \phantom{+ \,\,\,\, \frac{u_1^2}{2}} \left.
+\ln(1+u_2) \ln \left( \frac{u_2 (1+u_1)}{u_2-u_1} \right) \right]
\nonumber \\ &&
-\frac{u_2}{2} (1+u_1) \ln(1+u_1)
+\{u_1 \leftrightarrow u_2\} \\
%
\int^1_0 dy \,\, y^{-4-\ep} \ln(1+u_1y) \ln(1+u_2y) &=&
\frac{1}{2\ep} u_1^2 u_2 -\frac{1}{6} u_2 u_1^2
-\frac{1}{6} \ln(1+u_1) \ln(1+u_2)
\nonumber \\ &&
-\frac{u_1^2}{3} \left[ u_2 -(1+u_2) \ln(1+u_2) \right]
-\frac{u_1^3}{3} {\rm Li}_2(-u_2) 
\nonumber \\ &&
-\frac{u_1}{6} \left[ u_2+(1-u_2^2) \ln(1+u_2)+\frac{u_2^2}{2} \right]
\nonumber \\ &&
+\frac{u_1^3}{3} {\rm Li}_2 \left( \frac{u_1}{u_1-u_2} \right)
-\frac{u_1^3}{3} {\rm Li}_2 \left( \frac{u_1 (1+u_2)}{u_1-u_2} \right)
\nonumber \\ &&
-\frac{u_1^3}{3} \ln(1+u_2) \ln \left( \frac{u_2 (1+u_1)}{u_2-u_1} \right)
+\{u_1 \leftrightarrow u_2 \}
\nonumber\\
\end{eqnarray}

\vspace*{-5mm}
\begin{eqnarray}
\int_0^1 dy~y(1-y)~\Li_2\left(-\frac{x}{(1-x)^2 y (1-y)}\right) &=& 
- \frac{4}{9} \frac{x}{(1-x)^2} 
+ \frac{5}{9} \ln(1-x) 
- \frac{1}{3} \ln^2(1-x)
\nonumber\\ &&
+\frac{1}{3} \ln(x) \ln(1-x) 
+ \frac{1}{9} \frac{x(5 x^2 -12 x +3)}{(1-x)^3} \ln(x)~.
\nonumber\\
\end{eqnarray}
\subsection{Results for the Feynman integrals}
\label{Feyn_Int_Res_Section}

We will now present the results for all of the integrals appearing in
equations (\ref{DiagramA}) to (\ref{DiagramH}). We give the results up to
$O(\ep^0)$ in $x$-space. Logarithms, polylogarithms and Nielsen functions appear 
repeatedly in the expressions, so in order to save space, we use the following 
shorthand notation:

\restylefloat{table}
\begin{center}
\renewcommand{\arraystretch}{1.2}
\begin{tabular}{llll}
$L_1=\ln(x)$,          & $L_2=\ln(1-x)$,           & $L_3=\ln(1+x)$,            & $P_1=\ln(x)$,       \\
$P_2={\rm Li}_2(1-x)$, & $P_3 = \Li_2(-x)$,        & $R_1 = \Li_3(x)$,          & $R_2 = \Li_3(1-x)$, \\ 
$R_3={\rm Li}_3(-x)$,  & $R_4 = {\rm S}_{1,2}(x)$, & $R_5 = {\rm S}_{1,2}(1-x)$,& $R_6 = {\rm 
S}_{1,2}(-x)$~.\\
\end{tabular}
\end{center}
\renewcommand{\arraystretch}{1} 
 
\noindent
Let us define for $X=A,B,C,E$ 
\begin{eqnarray}
X^{a,b}_{\nu_1 \nu_2 \nu_3 \nu_4 \nu_5} &=&
\int^1_0 dx \,\, x^n 
(m^2)^{4-\nu_{12345}+\ep} \left( \Delta \cdot p \right)^{a+b}
{\tilde X}^{a,b}_{\nu_1 \nu_2 \nu_3 \nu_4 \nu_5}  \\
\text{and}~~~~~~F^{a,b}_{\nu_1 \nu_2 \nu_3 \nu_4 \nu_5} &=&
\int^1_0 dx \,\, \left( x^n-1 \right) 
(m^2)^{4-\nu_{12345}+\ep} \left( \Delta \cdot p \right)^{a+b}
{\tilde F}^{a,b}_{\nu_1 \nu_2 \nu_3 \nu_4 \nu_5}~.
\end{eqnarray}


The $A$-type integrals are

\begin{eqnarray} 
{\tilde A}^{0,n}_{01111} &=&  
{2 \over \ep^2} L_1 
- {1 \over \ep} \left[ 4 P_1-4 \zeta_2-{1 \over 2} L_1^2 \right] 
+{1 \over 2} \zeta_2 L_1
-6 \zeta_3
-{1 \over 12} L_1^3 -2 R_1 + 8 R_4 
\\
{\tilde A}^{1,n}_{01111} &=& 
{2 \over \ep^2} \left[ 1-x +(1+x) L_1 \right]
- {1 \over \ep} \BLB  (1+x) \BLP 4 P_1 -L_1 
 -{1 \over 2} L_1^2 -4 \zeta_2 \BRP
\nonumber \\
&&
      +(1-x) \left( 2 -4 L_2 \right) 
\BRB 
+{1 \over 4} (3 x -1) L_1^2
+(1+x) \BLB {1 \over 2} \zeta_2 L_1 -6 \zeta_3 -2 P_1
\nonumber \\
&&
 +2 \zeta_2 -L_1 -2 R_1 +8 R_4 -{1 \over 12} L_1^3
   \BRB
+(1-x) \BLB 4 L_2^2 -4 L_2 -{3 \over 2} \zeta_2 +2 \BRB
\\
{\tilde A}^{0,n}_{02111} &=&
P_2 +{1 \over 2} L_1^2+\zeta_2 
\\
{\tilde A}^{0,n}_{10111} &=& 
-{4 \over \ep^2} (1-x)
-{8 \over \ep} (1-x) \left[ L_2 -1 \right]
+(1-x) \left( 3 \zeta_2 + 16 L_2 -8  L_2^2 -16 \right)
\nonumber \\
&&
+
(-1)^n \BLB
-{4 \over \ep^2} (1-x) +{8 \over \ep} \left( 1 -x +x L_1 \right)
+8 (1+x) \left( P_3 +L_1 L_3 \right)
\nonumber \\
&&
+(3+5 x) \zeta_2 +4 x L_1^2 -16 x L_1 
-16 (1-x)
\BRB
\\
{\tilde A}^{0,n}_{11011} &=& (-1)^n \BLB
{2 \over \ep^2} L_1 +{5 \over 2 \ep} L_1^2
+4 R_1+8 R_3 +2 \zeta_3  
-4 L_1 P_3 -2 L_1 P_1
\nonumber \\ && \phantom{ (-1)^n \BLB} 
+{7 \over 12} L_1^3 +{1 \over 2} \zeta_2 L_1
\BRB
\\
{\tilde A}^{1,n}_{11011} &=& (-1)^n \BLCB
-{2 \over \ep^2} \left[ x L_1 +1 -x \right]  
-{1 \over \ep} \BLB 2 x -2 +{5 \over 2} x L_1^2 
+(1 -3 x) L_1 \BRB
\nonumber \\
&&
+{1 \over 4} (1 +9 x) L_1^2 
+4 x L_1 P_3
-{1 \over 2} (1-x) \zeta_2
+2 x L_1 P_1 -{1 \over 2} x \zeta_2 L_1
\nonumber \\
&&
-{7 \over 12} x L_1^3 
-2 \zeta_3 x
-4 x R_1-8 x R_3+(1 -3 x) L_1 +2 x-2
\BRCB
\\
{\tilde A}^{0,n}_{12011} &=& (-1)^n \left[
{3 \over 2} L_1^2 -P_1-2 P_3
+2 {L_1 \over 1+x}
-L_1 L_2
-2 L_1 L_3 \right]
\\
{\tilde A}^{0,n}_{11101} &=& {\tilde A}^{0,n}_{11011}
\\
{\tilde A}^{1,n}_{11101} &=& (-1)^n \BLCB
{2 \over  \ep^2} \left[ 1 -x +L_1 \right] 
-{1 \over \ep} \left[ 2 -2 x -{5 \over 2} L_1^2 
+ (3 x-1) L_1 \right]
+{1 \over 2} \zeta_2 L_1 
\nonumber \\
&&
-{1 \over 4} (1 + 9 x) L_1^2 
+{1 \over 2} (1-x) \zeta_2
+2 \zeta_3 
+{7 \over 12} L_1^3 
-4 L_1 P_3 -2 L_1 P_1
\nonumber \\
&&
+4 R_1
+8 R_3 +(3 x-1) L_1 +2 -2 x
\BRCB
\\
{\tilde A}^{0,n}_{12101} &=& {\tilde A}^{0,n}_{12011}
\\
{\tilde A}^{0,n}_{11110} &=& 
{2 \over \ep^2} L_1
-{1 \over \ep} \left[ 4 P_1-4 \zeta_2 -{1 \over 2} L_1^2 \right]
-6 \zeta_3
+{1 \over 2} \zeta_2 L_1 
-{1 \over 12} L_1^3 -2 R_1 +8 R_4 
\\
{\tilde A}^{1,n}_{11110} &=&
-{2 \over \ep^2} (1-x)
-{1 \over \ep} [ (1+x) L_1 + 4 (1-x) L_2
 -2 +2 x ]
+{1 \over 4} (1 -3 x) L_1^2
\nonumber \\
&&
-(1-x) \left( 4 L_2^2 - 4 L_2 +2 \right)
+ (1+x) \left( L_1 +2 P_1 \right)
-{1 \over 2} (1 +7 x) \zeta_2
\end{eqnarray}


\begin{eqnarray}
{\tilde A}^{0,n}_{12110} &=& 
P_2+{1 \over 2} L_1^2 +\zeta_2 
\\
{\tilde A}^{0,n}_{22110} &=& 
-\left( {1 \over 2 \ep^2} -{1 \over 2 \ep} -{1 \over 2} +{5 \over 8} \zeta_2
\right) \delta(1-x) 
-\left( {1 \over \ep} -1 \right) {\cal D}_0(x) 
\nonumber \\
&&
-2 {\cal D}_1(x)
+{L_1 \over 1-x} 
\\
{\tilde A}^{0,n}_{21110} &=& 
-{2 \over \ep^2} 
-{2 \over \ep} \left[ 2 L_2 - L_1 \right]
+2 P_2 -4 L_2^2+4 L_1 L_2-{1 \over 2} \zeta_2 
\\
{\tilde A}^{1,n}_{21110} &=& 
-{2 \over \ep^2}
-{4 \over \ep} L_2
+\left( {1 \over 2} -x \right) \zeta_2
-4 L_2^2 
-(1+x) P_2 
-{1 \over 2} x L_1^2
\\
{\tilde A}^{n,0}_{01111} &=& 
{2 \over \ep^2} L_2
-{1 \over \ep} \left[ 4 P_2 -4 \zeta_2 - {1 \over 2} L_2^2 \right]
+{5 \over 2} \zeta_2 L_2 
-2 \zeta_3
-L_2 P_2
+{1 \over 12} L_2^3
\nonumber \\
&&
-2 R_2 +4 R_5
\\
{\tilde A}^{n,1}_{01111} &=& 
-{2 \over \ep^2} \left[ x +(1-x) L_2 \right]
-{1 \over \ep} \BLB (2-x) L_2
+(1-x) \left( {1 \over 2} L_2^2 +4 \zeta_2 -4 P_2 \right)
\nonumber \\
&&
+4 x L_1
-2 x \BRB 
 -4 L_1 L_2
-{1 \over 4} (2-x) L_2^2
-2 x L_1^2
+{x \over 2} \zeta_2
+(3 x -2) L_2
\nonumber \\
&&
-2 x
-(1-x) \BLB L_2 P_1 
+{3 \over 2} \zeta_2 L_2
-2 \zeta_3
+ L_1 L_2^2 
+{1 \over 12} L_2^3 +4 R_5
-2 R_2 
\BRB 
\nonumber \\
&&
+(x-4) P_1
\\
{\tilde A}^{n,0}_{10111} &=& 
-{4 \over \ep^2} -{4 \over \ep} \left[ L_2 -1 \right]
-4 -2 L_2^2 
+4 L_2 -\zeta_2 
\\
{\tilde A}^{n,0}_{20111} &=& 
-\left(
{2 \over \ep^2} -{2 \over \ep}  +2 +{1 \over 2} \zeta_2 
\right) \delta(1-x)
-{2 \over \ep} {\cal D}_0(x) 
+2 {\cal D}_0(x) -2 {\cal D}_1(x) 
\\
{\tilde A}^{n,0}_{11011} &=& 
{2 \over \ep^2} L_1 +{5 \over 2 \ep} L_1^2 
+4 R_1+8 R_3+2 \zeta_3
-4 L_1 P_3
-2 L_1 P_1 +{7 \over 12} L_1^3 +{1 \over 2} \zeta_2 L_1
\\
{\tilde A}^{n,1}_{11011} &=& 
{2 \over \ep^2} \left[ x L_1 +1-x \right]
-{1 \over \ep} \BLB (3 x -1) L_1 -{5 \over 2} x L_1^2 
+2-2 x \BRB 
+{1 \over 2} (1-x) \zeta_2 
\nonumber \\
&&
+{x \over 2} \zeta_2 L_1  -2 x L_1 P_1
+2 \zeta_3 x +8 x R_3 +(3 x-1) L_1 +4 x R_1 +{7 \over 12} x L_1^3 
\nonumber \\
&&
-{1 \over 4} (1+9 x) L_1^2
-4 x L_1 P_3 +2-2 x
\\
{\tilde A}^{n,0}_{21011} &=& 
P_2-2 P_3-\zeta_2
+{3 \over 2} L_1^2+2 {L_1 \over 1+x}
-2 L_1 L_3
\\
{\tilde A}^{n,0}_{11101} &=&
-{2 \over \ep^2} 
-{1 \over \ep} \left[ 4 L_2 -L_1 -2 \right]
-{1 \over 2} \zeta_2 -{1 \over 4} L_1^2 
-{1 +3 x \over 1-x} L_1 -2 L_2^2 -2 
\\
{\tilde A}^{n,1}_{11101} &=& 
{1-x \over \ep^2}
-{1+x \over \ep} \left[ {1 \over 2} L_1 -2 L_2 +1 \right]
+(1-x) \BLB L_2^2 +{1 \over 8} L_1^2 +{1 \over 4} \zeta_2 +1 \BRB
\nonumber \\
&&
+{1 \over 2} (1 +3 x) L_1 
\\
{\tilde A}^{n,0}_{21101} &=& 
-\left( {2 \over \ep^2}+{\zeta_2 \over 2} \right) \delta(1-x)
-{2 \over \ep} {\cal D}_0(x)-2 {\cal D}_1(x)+1
+{1+x \over 1-x} \left( 1+{L_1 \over 1-x} \right)
\\
{\tilde A}^{n,0}_{11110} &=&  
-{2 \over \ep^2} -{1 \over \ep} \left( 3 L_2 -2 \right)
-{7 \over 4} L_2^2 +{1 \over 2} \zeta_2 
+P_1+3 L_2 -2
\end{eqnarray}


\begin{eqnarray}
{\tilde A}^{n,1}_{11110} &=&
-{x \over \ep^2} 
-{x \over \ep} \left( {3 \over 2} L_2 -1 \right)
+{x \over 4} \zeta_2 -{7 \over 8} x L_2^2 
+{1 \over 2} (3 x +1) L_2
+{x \over 2} P_1 -x 
\\
{\tilde A}^{n,0}_{21110} &=& 
-\left( {2 \over \ep^2} -{2 \over \ep} +2 +{1 \over 2} \zeta_2 
\right) \delta(1-x)
-{2 \over \ep} {\cal D}_0(x)
+2 {\cal D}_0(x)
-2 {\cal D}_1(x)
\nonumber \\ &&
-P_1 
-{1 \over 2} L_2^2 -\zeta_2  
\\
{\tilde A}^{n,0}_{31110} &=& 
 \left( {1 \over 2 \ep^2}-{1 \over \ep}-{1 \over 2}
+1-{3 \over 8} \zeta_2 \right) \delta(1-x)
-{\cal D}_0(x)-{\cal D}_1(x)
\nonumber \\
&&
-(1-x)^{-2+\ep} \left[ {3 \over 2}-{1 \over \ep}
-\left( {3 \over 2}+{1 \over 4} \zeta_2 \right) \ep \right]
+\zeta_2+{1 \over 2} L_2^2+P_1
\\
{\tilde A}^{n,1}_{21110} &=& 
-\left( {1 \over \ep^2}-{3 \over 2 \ep}
+{7 \over 4}+{1 \over 4} \zeta_2 \right) \delta(1-x)
-{1 \over \ep} \left( {\cal D}_0(x)-1 \right)
+{3 \over 2} {\cal D}_0(x)
\nonumber \\
&&
-{\cal D}_1(x)
-{3 \over 2}-x \zeta_2-{x \over 2} L_2^2+{1 \over 2} L_2-x P_1
\\
{\tilde A}^{0,n}_{12111} &=& 
-{1 \over 3 \ep} \delta(1-x)
-{2 \over 3} {\cal D}_0(x)
-{2 \over 3} P_2 
-{1 \over 3} L_1^2
-{2 \over 3} \zeta_2
-{2 \over 3 (1-x)} L_1 
\nonumber \\
&&
+{1 \over 3} (1-x)^{-3+2\varepsilon} L_1^2
+{2 \over 3} (1-x)^{-2+2\varepsilon} L_1 
- \left( {1 \over 2} -\zeta_2 \right) \delta(1-x)
\nonumber \\
&&
-(-1)^n \BLB
{4 \over 3} P_3 
- {2 \over 3} \left( 1 - {2 \over (1+x)^3} \right) P_2 
+ {2 \over 3} \zeta_2
- \left( 1 - {1 \over (1+x)^3} \right) L_1^2
\nonumber \\
&&
+ {4 \over 3} L_1 L_3 + {4 \over 3 (1+x)^2}
+ {2 \over 3 (1+x)} \left( {2 \over (1+x)^2} - {1 \over 1+x} - 1 \right) 
L_1 
\nonumber \\
&&
- {2 \over 3 (1+x)}  \BRB 
\\
{\tilde A}^{n,0}_{21111} &=& 
\left( -{1 \over 2 \ep^2} +{3 \over 8} \zeta_2 +{1 \over 2} \right)
\delta(1-x) +{\cal D}_1(x)
+\left( {1 \over (1-x)^2}-{3 \over 2} \right) \zeta_2
-{1 \over 4} L_2^2
\nonumber \\
&&
+\left( {2 \over (1-x)^2}-1 \right) \left( L_1 L_3+P_3 \right)
+\left( 1-{1 \over 2 (1-x)^2} \right) P_2
-{L_2 \over 2 (1-x)}
\nonumber \\
&&
+{1 \over (1-x)^2} \BLB
{3 \over 4} x (x-2) L_1^2 
+{x (x-3) \over 2 (1+x)} L_1
+{x \over 2} (x-2) L_1 L_2 
\BRB
\end{eqnarray}

\noindent
The integrals $A^{0,n}_{11111}$, $A^{n,0}_{11111}$ and $A^{n,1}_{11111}$ are finite and appear 
multiplied by a factor of $D-4=\ep$ in the expressions of diagrams \ref{DiagramsI}a 
and \ref{DiagramsI}c. Therefore they only contribute to those diagrams starting at 
$O(\ep)$. For this reason we have chosen not to present them here, although they are needed
in Eq. (\ref{E_IBP}).

The $B$-type integrals are given by


\begin{eqnarray}
{\tilde B}^{0,n+1}_{12011} &=& (-1)^n \BLCB
-{1-x \over \ep^2}
-{1 \over \ep} \BLB {1 \over 2} (5-x) L_1 +2 -2 x \BRB
+2 x-2
-{1 \over 4} (5-x) \zeta_2
\nonumber \\
&&
+P_2
-{1 \over 8} (7+x) L_1^2
+2 {x^2-1-x \over 1+x} L_1
-2 L_1 L_3
-2 P_3
\BRCB
\\
{\tilde B}^{0,n+1}_{12101} &=& {\tilde B}^{0,n+1}_{12011}
\end{eqnarray}
\begin{eqnarray}
{\tilde B}^{n+1,0}_{21011} &=& 
{x-1 \over \ep^2} 
-{1 \over 2 \ep} \left[ (5-x) L_1 +5 (1-x) \right]
+{x-5 \over 4} \zeta_2 
-{7+x \over 8} L_1^2 
\nonumber \\
&&
-{1 \over 4} (7 +11 x) {1-x \over 1+x} L_1
+ P_2 
-2 P_3 -2 L_1 L_3 
-{3 \over 2} (1-x) 
\\
{\tilde B}^{n+1,0}_{21101} &=& 
-{x \over 2 \ep^2}
-{1 \over 4 \ep} \left( 4 x L_2 -x L_1 +1 -3 x \right)
-{x \over 16} L_1^2 
-{x \over 8} \zeta_2
\nonumber \\
&&
+{1 - 9 x^2 \over 8 (1-x)} L_1 -{x \over 2} L_2^2
+{1 \over 4} (1-3 x)
\\
{\tilde B}^{0,n+1}_{12110} &=& 
-{1-x \over \ep^2}
-{1 \over \ep} \left[ {1 \over 2} \left( 3 x +1 \right) L_1 
+2 (1-x) L_2 \right]
-{1 \over 4} (1-x) \zeta_2
\nonumber \\
&&
+{1 \over 8} (1-x) L_1^2 
-3 x L_1 L_2
-2 x P_2
-2 (1-x) L_2^2
+P_1 
\\
{\tilde B}^{n+1,0}_{21110} &=& 
-\left( {1 \over \ep^2}-{3 \over 2 \ep}
+{7 \over 4}+{1 \over 4} \zeta_2 \right) \delta(1-x)
-{x \over 2 \ep^2}
-{1 \over \ep} \left( {\cal D}_0(x)+{3 \over 4} x L_2-{x \over 2}-1 \right)
\nonumber \\
&&
+{3 \over 2} {\cal D}_0(x)
-{\cal D}_1(x)
-{3 \over 2}-{7 \over 8} x \zeta_2-{15 \over 16} x L_2^2
+{3 \over 4} (1+x) L_2-{3 \over 4} x P_1-{x \over 2} 
\end{eqnarray}

The $E$-type integrals are

\begin{eqnarray}
{\tilde E}^{0,0}_{01111} &=& 
{4 \over \ep^2} \left[ L_2 -L_1 \right]
-{1 \over \ep} \BLB 3 L_1^2 -3 L_2^2 -2 P_1 
+2 P_2 \BRB
+\zeta_2 L_1+\zeta_2 L_2 -{1 \over 2} L_1^3 
\nonumber \\
&&
+5 R_2 -2 L_2 L_1^2 -2 L_1 P_2 
-{5 \over 2} L_1 L_2^2 
+{11 \over 6} L_2^3
-3 R_5 -7 L_2 P_2 +3 \zeta_3
\\
{\tilde E}^{1,0}_{01111} &=& 
{2 \over \ep^2} \left[ 2 x L_2 -(1 +2 x) L_1 \right]
-{1 \over \ep} \BLB -4 (1+x) P_1 
+\left( {1 \over 2} +3 x \right) L_1^2 -3 x L_2^2
\nonumber \\
&&
-2 x L_1 L_2 
+(4 +2 x) \zeta_2 \BRB 
+(3 x+4) L_1 L_2^2
-{3 \over 2} x L_1^2 L_2  
+(3 x -2) \zeta_3
\nonumber \\
&&
-3 x R_4 -x L_1 P_1 
-\left( x +{1 \over 2} \right) \zeta_2 L_1 
+2 (x+4) R_2+(2+3 x) R_1 
\nonumber \\
&&
+4 (x+2) L_2 P_1
+{1 \over 12} (1 -6 x) L_1^3
+{11 \over 6} x L_2^3 -(8 +3 x) \zeta_2 L_2 
\\
{\tilde E}^{0,1}_{01111} &=& 
-{2 \over \ep^2} \left[ (1-2 x) L_2 +2 x L_1 \right]
-{1 \over \ep} \BLB 4 (1-x) P_1 
+(4-2 x) L_1 L_2 +3 x L_1^2
\nonumber \\
&&
+\left( {1 \over 2} -3 x \right) L_2^2 
+2 x \zeta_2  \BRB
-\left( {3 \over 2} x +2 \right) L_1^2 L_2
-(x+4) L_1 P_1 -x \zeta_2 L_1
\nonumber \\
&&
+\left( {9 \over 2} x -1 \right) L_1 L_2^2
+(7 x-1) L_2 P_1 
-\left( {3 \over 2} +6 x \right) \zeta_2 L_2 -{x \over 2} L_1^3
-2 \zeta_3
\nonumber \\
&&
-{1 -22 x \over 12} L_2^3+(3 x+4) R_1
+(5 x+2) R_2 
\\
{\tilde E}^{1,1}_{01111} &=& 
{2 x \over \ep^2} \left[ (2 x-1) L_2 -(2 x+1) L_1 \right]
-{1 \over \ep} \BLB 2 x (x+2) \zeta_2 
+ {x \over 2} (6 x+1) L_1^2 
\nonumber \\
&&
+2 x (2-x) L_1 L_2 -4 x^2 P_1 
-{x \over 2} (6 x-1) L_2^2 \BRB
+{x \over 2} (3+2 x) \zeta_2 L_1 
\nonumber \\
&&
+x (2+3 x) \zeta_3
-{x \over 2} (5-2 x) \zeta_2 L_2 
-{x \over 12} (1-22 x) L_2^3 
-x (1+2 x) L_1^2 L_2
\nonumber 
\end{eqnarray}
\begin{eqnarray}
&& -{x \over 2} (8+5 x) L_1 L_2^2 
-x (1+x) \left[ 2 L_1 +7 L_2 \right] P_2
 +{x \over 12} (1-6 x) L_1^3
\nonumber \\
&&
+5 x (2+x) R_2-3 x (2+x) R_5
\\
{\tilde E}^{0,2}_{01111} &=& 
{2 \over \ep^2} \left[ x -2 x^2 L_1 +(2 x^2 -2 x +1) L_2 \right]
-{1 \over \ep} \BLB
2 x+4 x L_1 L_2+3 x^2 L_1^2 
\nonumber \\
&&
-\left( 3 x^2 -x +{1 \over 2} \right) L_2^2 +2 x (2-x) P_1 
+(x-2) L_2 +(2 x^2 -4 x +4) P_2 
\nonumber \\
&&
-4 x L_1 -4 \zeta_2 (1-x)
\BRB
-{x \over 2} \zeta_2 -\left( 2 x^2 +3 x -{3 \over 2} \right) \zeta_2 L_2
-3 x^2 \zeta_2 L_1 +2 x 
\nonumber \\
&&
+{x^2 \over 2} L_1^2 L_2 
+\left( 1 -2 x +{1 \over 2} x^2 \right) L_1 L_2^2
+(3 x^2 -2 x +1) P_1 L_2
+x^2 P_1 L_1 
\nonumber \\
&&
+4 L_1 L_2 -4 x^2 L_2 P_2 +2 x^2 P_2 L_1
+2 x L_1^2 +{2-x \over 4} L_2^2 -{x^2 \over 2} L_1^3
\nonumber \\
&&
+{1 \over 12} (22 x^2-2 x+1) L_2^3
+(4-8 x) R_5 +3 x^2 R_1+(5 x^2+4 x-2) R_2
\nonumber \\
&&
+(4-x) P_1+(2-3 x) L_2
+(4 x-2) \zeta_3
\\
{\tilde E}^{2,0}_{01111} &=&
{2 \over \ep^2} \left[ 2 x^2 L_2 -(1+2 x +2 x^2) L_1 +x-1 \right]
-{1 \over \ep} \BLB 2 x -2 -2 x^2 L_1 L_2 
\nonumber \\
&&
-3 x^2 L_2^2 
-4 (1+x)^2 P_1 +\left( {1 \over 2} +x +3 x^2 \right) L_1^2
+4 (1-x) L_2 +(1+x) L_1 
\nonumber \\
&&
+(4+8 x+2 x^2) \zeta_2
\BRB
-{1+7 x \over 2} \zeta_2 -x^2 \zeta_2 L_2 
-\left( {1 \over 2} +x +x^2 \right) \zeta_2 L_1
\nonumber \\
&&
+2 x -2 -{3 \over 2} x^2 L_1^2 L_2
+2 x^2 L_1 L_2^2 +2 x^2 L_2 P_1 -x^2 L_1 P_1 
+{1-3 x \over 4} L_1^2 
\nonumber \\
&&
-4 (1-x) L_2^2
+\left( {1 \over 12} +{x \over 6} -{x^2 \over 2} \right) L_1^3
+{11 \over 6} x^2 L_2^3 -(8+16 x+5 x^2) R_4
\nonumber \\
&&
+(3 x^2+4 x+2) R_1 +(1+x) ( L_1 +2 P_1) +4 (1-x) L_2 
\nonumber \\
&&
+(5 x^2+12 x+6) \zeta_3
\\
{\tilde E}^{0,0}_{10111} &=&
{4 \over \ep^2} L_2 -{1 \over \ep} \left( 4 \zeta_2 -6 L_2^2 \right)
+8 \zeta_3 +{14 \over 3} L_2^3
-7 \zeta_2 L_2
\nonumber \\
&&
+(-1)^n \BLB -{4 \over \ep^2} L_1 
-{2 \over \ep} \left( 4 P_3 +4 L_1 L_3 +L_1^2 +2 \zeta_2 \right)
-{2 \over 3} L_1^3 -16 R_6 
\nonumber \\
&&
-16 L_3 P_3 -8 L_1 L_3^2 -8 \zeta_2 L_3
+8 \zeta_3 +8 R_3 -4 L_1^2 L_3 -8 L_1 P_3 -\zeta_2 L_1
\BRB
\\
{\tilde E}^{1,0}_{10111} &=&
{4 \over \ep^2} \left( x L_2 +1-x \right)
-{1 \over \ep} \left[ x \left( 4 \zeta_2 -6 L_2^2 \right)
+8 (1-x) \left( 1 -L_2 \right) \right]
+8 \zeta_3 x
\nonumber \\
&& 
+(1-x) \left( 8 L_2^2 -3 \zeta_2 -16 L_2 +16 \right)
+{14 \over 3} x L_2^3 -7 x \zeta_2 L_2
\nonumber \\
&&  
+(-1)^n \BLB {4 \over \ep^2} \left( x L_1 +1-x \right)
-{2 \over \ep} \left( 4 x L_1 -4 x L_1 L_3 -4 x P_3 -2 x \zeta_2 
\phantom{L_1^2} \right.
\nonumber \\
&&  \left.
-x L_1^2 -4 x +4 \right)
-(3+5 x) \zeta_2 + \zeta_2 x L_1 +8 x \zeta_2 L_3 +16 x L_1
\nonumber \\
&&  
-8 (1+x) L_1 L_3 +8 x L_1 L_3^2 +4 x L_3 L_1^2 +16 x L_3 P_3
+8 x L_1 P_3 -4 x L_1^2 
\nonumber \\
&&  
+{2 \over 3} x L_1^3 -8 (1+x) P_3 -8 x R_3
+16 x R_6 -8 \zeta_3 x +16 -16 x
\BRB  
\end{eqnarray}
\begin{eqnarray}
{\tilde E}^{0,1}_{10111} &=&
{4 \over \ep^2} \left( x L_2 +1 \right)
-{2 \over \ep} \left( 2 x \zeta_2 -2 L_2 -3 x L_2^2 +2 \right)
+\zeta_2 -7 x \zeta_2 L_2 +8 \zeta_3 x +4
\nonumber \\
&&
+{14 \over 3} x L_2^3 +2 L_2^2 -4 L_2
+(-1)^n x \BLB {4 \over \ep^2} L_1
+{2 \over \ep} \left( 2 \zeta_2 +L_1^2 +4 L_1 L_3 +4 P_3 \right)
\nonumber \\
&&
+ \zeta_2 \left( L_1 +8 L_3 \right) 
-8 \zeta_3 +{2 \over 3} L_1^3
+4 L_1^2 L_3 
+8 L_1 \left( P_3 +L_3^2 \right) +16 L_3 P_3
\nonumber \\
&&
 -8 R_3 +16 R_6
\BRB
\\
{\tilde E}^{1,0}_{11011} &=&
-{2 \over \ep^2} (1-x) 
-{2 \over \ep} \left[ (2-x) L_1 +1-x \right]
-2+2 x +(2 x-4) L_1 -{x+4 \over 2} L_1^2
\nonumber \\
&&
+(1-x) \BLP P_2 -2 P_3 -{3 \over 2} \zeta_2 -2 L_1 L_3 \BRP
+(-1)^n \BLB -{2 \over \ep^2} \left( L_1 +1-x \right)
\nonumber \\
&&
-{1 \over \ep} \left( {5 \over 2} L_1^2 +(4-2 x) L_1 -2 x+2 \right)
-{1 \over 2} \zeta_2 L_1
-2 \zeta_3 -{7 \over 12} L_1^3 -{x+4 \over 2} L_1^2
\nonumber \\
&&
+(2 x-4) L_1 +2 L_1 P_1
-(1-x) \left( 2 L_1 L_3 + L_2 L_1+2 P_3+P_1 +{1 \over 2} \zeta_2 \right)
\nonumber \\
&&
+4 L_1 P_3 -8 R_3 -4 R_1
+2 x -2
\BRB
\\
{\tilde E}^{0,1}_{11011} &=& (-1)^n {\tilde E}^{1,0}_{11011}
\\
{\tilde E}^{1,1}_{11011} &=& \left( 1-(-1)^n \right) x \BLB 
-{2 \over \ep^2} \left( L_1 +1-x \right)
-{1 \over \ep} \left( {5 \over 2} L_1^2 +(4-2 x) L_1 -2 x+2 \right)
\nonumber \\
&&
-{1 \over 2} \zeta_2 L_1 -2 \zeta_3 -{7 \over 12} L_1^3 +(2 x-4) L_1+2 L_1 P_1
+4 L_1 P_3 -8 R_3 -4 R_1 
\nonumber \\
&&
-{x+4 \over 2} L_1^2
-(1-x) \left( {1 \over 2} \zeta_2 +2 L_1 L_3 + L_1 L_2 +2 P_3 +P_1 +2 \right)
\BRB
\\
{\tilde E}^{2,0}_{11011} &=& x \BLB
-{2 \over \ep^2} (1-x) 
-{2 \over \ep} \left[ (2-x) L_1 +1-x \right]
+2 (x-2) L_1 -{x+4 \over 2} L_1^2
\nonumber \\
&&
+(1-x) \left( P_2 -2 P_3 -{3 \over 2} \zeta_2 -2 L_1 L_3 -2 \right)
\BRB
\nonumber \\
&&
+(-1)^n \BLB {2 \over \ep^2} \left( 2 x L_1 +1-x^2 \right)
+{1 \over \ep} \BLP 5 x L_1^2 +(1+2 x) (1-x) L_1 
\nonumber \\
&&
-2 (1-x)^2 \BRP
 -5 x \zeta_2 L_1 +2 (1-x^2)
+2 x R_5 +10 x R_1 
+16 x R_3
\nonumber \\
&&
+(1-x) \left( {1 \over 2} (1+3 x) \zeta_2 -x P_2 
+2 x P_3 +2 x L_1 L_3 -{1 \over 4} (1+2 x) L_1^2 \right)
\nonumber \\
&&
+2 \zeta_3 x -(1+2 x^2-7 x) L_1
+x L_1 \left( 6 P_2 - 8 P_3 \right)
+5 x L_1^2 L_2 +{7 \over 6} x L_1^3 
\BRB
\\
{\tilde E}^{0,2}_{11011} &=& -(-1)^n {\tilde E}^{2,0}_{11011}
\\
{\tilde E}^{0,0}_{11101} &=& 
{1 \over 1-x} \BLB {2 \over \ep^2} L_1
-{1 \over \ep} \left( {1 \over 2} L_1^2 -4 L_1 L_2 -2 P_2 \right)
+{1 \over 2} \zeta_2 L_1 +{1 \over 12} L_1^3 +2 L_1 L_2^2 
\nonumber \\
&&
+2 R_2 -R_5 \BRB
+{(-1)^n \over 1+x} \BLB
-{4 \over \ep^2} L_1
-{1 \over \ep} \left( 2 \zeta_2 +4 P_3 +3 L_1^2 +4 L_1 L_3 \right)
\nonumber \\
&&
-2 \zeta_2 L_3 - \zeta_2 L_1 +3 \zeta_3 -4 L_3 P_3 -2 L_1 P_3
-{1 \over 2} L_1^3 -3 L_1^2 L_3 +2 L_1 P_1 
 \nonumber 
\end{eqnarray}
\begin{eqnarray}
&&
-2 L_1 L_3^2 -2 R_3 -4 R_1 -4 R_6
\BRB
\\
{\tilde E}^{1,0}_{11101} &=&
{x \over 1-x} \BLB {2 \over \ep^2} L_1
-{1 \over \ep} \left( {1 \over 2} L_1^2 -4 L_1 L_2 -2 P_2 \right)
+{1 \over 2} \zeta_2 L_1 +{1 \over 12} L_1^3 +2 L_1 L_2^2 
\nonumber \\
&&
+2 R_2 -R_5 \BRB
+{(-1)^n \over 1+x} \BLB  -{2 -2 x \over \ep^2} L_1
+{1 \over \ep} \BLP 2 \zeta_2 x +{1 \over 2} (x-5) L_1^2
\nonumber \\
&&
+4 x L_1 L_3 +4 x P_3 \BRP
-{1-x \over 2} \zeta_2 L_1 +2 x \zeta_2 L_3 -(5 x+2) \zeta_3
+2 L_1 P_1 
\nonumber \\
&&
-{7+x \over 12} L_1^3 +3 x L_1^2 L_3
+2 (3 x+2) L_1 P_3 +2 x L_1 L_3^2 +4 x L_3 P_3
-4 R_1
\nonumber \\
&&
+4 x R_6-2 (3 x+4) R_3
\BRB
\\
{\tilde E}^{0,1}_{11101} &=& 
{2 \over \ep^2} \left( 1 + {x \over 1-x} L_1 \right)
-{1 \over \ep} \BLB {x \over 1-x} \BLP {1 \over 2} L_1^2 -2 P_2 
-4 L_1 L_2 \BRP +2 -4 L_2 +L_1 \BRB  
\nonumber \\
&&
+{\zeta_2 \over 2} +{L_1^2 \over 4} +2 L_2^2
+2 +{x \over 1-x} \left( {1 \over 2} \zeta_2 L_1 
+2 L_1 L_2^2 +{1 \over 12} L_1^3 -R_5 +2 R_2 \right)
\nonumber \\
&&
+{1 +3 x \over 1-x} L_1
+(-1)^n {x \over 1+x} \BLB
{4 \over \ep^2} L_1
+{1 \over \ep} \left( 2 \zeta_2 +3 L_1^2 +4 P_3  
+4 L_1 L_3  \right)
\nonumber \\
&&
+ \zeta_2 L_1 +2 \zeta_2 L_3 +{1 \over 2} L_1^3
+3 L_1^2 L_3 +2 L_1 P_3 +2 L_1 L_3^2
-2 L_1 P_1 +4 L_3 P_3
\nonumber \\
&&
+2 R_3 +4 R_1 +4 R_6 -3 \zeta_3 
\BRB
\\
{\tilde E}^{1,1}_{11101} &=& 
{2 \over \ep^2} \left( x + {x^2 \over 1-x} L_1 \right)
-{x \over \ep} \BLB  L_1 -4  L_2 +2  
+{x \over 2 (1-x)} \BLP L_1^2 -8 L_1 L_2 
-4 P_2 \BRP \BRB
\nonumber \\
&&
+{x^2 \over 1-x} \BLB {1 \over 2} \zeta_2 L_1 
+{1 \over 12} L_1^3 +2 L_1 L_2^2 +2 R_2 -R_5 \BRB
+{x \over 2} \zeta_2 +{x \over 4} L_1^2 +2 x L_2^2 
\nonumber \\
&&
+x {1 +3 x \over 1-x} L_1
+2 x 
+{(-1)^n \over 1+x} \BLB
{2 \over \ep^2} x (1-x) L_1 -{1 \over \ep} \BLP
 2 x^2 \zeta_2 +4 x^2 P_3 
\nonumber \\
&&
+4 x^2 L_1 L_3 -{5-x \over 2} x L_1^2 \BRP
+{x \over 2} (1-x) \zeta_2 L_1 +(2 +5 x) x \zeta_3
+{x \over 12} (7+x)  L_1^3
\nonumber \\
&&
-x^2 \left( 2 \zeta_2 L_3 +3 L_1^2 L_3 
+2 L_1 L_3^2 +4 L_3 P_3 +4 R_6 \right)
-2 x L_1 P_1 +4 x R_1
\nonumber \\
&&
+2 x (4 +3 x) R_3 -2 x (2 +3 x) L_1 P_3
\BRB
\\
{\tilde E}^{2,0}_{11101} &=& 
{x^2 \over 1-x} \BLB
{2 \over \ep^2} L_1
-{1 \over \ep} \left( {1 \over 2} L_1^2 -4 L_1 L_2 -2
  P_2 \right) 
+{1 \over 2} \zeta_2 L_1 +{1 \over 12} L_1^3 +2 L_1 L_2^2
\nonumber \\
&&
+2 R_2 -R_5 \BRB
+(-1)^n \BLB
-{2 \over \ep^2} \left(
1-x +{1+x^2 \over 1+x} L_1 \right)
-{1 \over \ep} \BLP
{5 +x^2 \over 2 (1+x)} L_1^2 
\nonumber \\
&&
+(1 -3 x) L_1 
+{x^2 \over 1+x} \left( 2 \zeta_2 +4 P_3 +4 L_1 L_3 \right)
-2 (1-x) \BRP
-{1 \over 2} (1-x) \zeta_2
\nonumber \\
&&
+(1 -3 x) L_1
+{1+9 x \over 4} L_1^2
-2 {4 -3 x^2 \over 1+x} R_3
-{7 -x^2 \over 12 (1+x)} L_1^3
-{2 -5 x^2 \over 1+x} \zeta_3
\nonumber 
\end{eqnarray}
\begin{eqnarray}
&&
-{x^2 \over 1+x} \left( 4 R_6 +2 \zeta_2 L_3
+2 L_1 L_3^2 +4 L_3 P_3 +3 L_1^2 L_3 \right)
+2 {2 -3 x^2 \over 1+x} L_1 P_3
\nonumber \\
&&
-{1+x^2 \over 2 (1+x)} \zeta_2 L_1
+{2 \over 1+x} \left( L_1 P_1 -2 R_1 \right)
-2 +2 x
\BRB 
\\
{\tilde E}^{1,0}_{11110} &=&
{2 \over \ep^2} \left( L_2 - L_1 \right)
-{1 \over \ep} \left( 4 \zeta_2 -{7 \over 2} L_2^2 -2 P_1 
+2 L_1 L_2 +{1 \over 2} L_1^2 \right)
-{9 \over 2} \zeta_2 L_2 +7 \zeta_3
\nonumber \\
&&
-{1 \over 2} \zeta_2 L_1 -2 L_1 L_2^2 -P_1 L_2 
 +{35 \over 12} L_2^3 -R_5 +2 R_1 +{1 \over 12} L_1^3-6 R_4
\\
{\tilde E}^{0,1}_{11110} &=&
{2 \over \ep^2} \left( 1 + L_2 \right)
-{1 \over \ep} \left( 2 +2 \zeta_2 -3 L_2 -{7 \over 2} L_2^2 -2 P_2 \right)
-{1 \over 2} \zeta_2 -{13 \over 2} \zeta_2 L_2 +3 \zeta_3
\nonumber \\
&&
+{35 \over 12} L_2^3 
+4 L_2 P_2 -2 R_2 +L_1 L_2^2 +P_1 L_2 -R_5 -P_1 -3 L_2 +{7 \over 4} L_2^2 +2
\\
{\tilde E}^{1,1}_{11110} &=&
{2 x \over \ep^2} \left( L_2 -L_1 +1 \right)
+{x \over \ep} \left( 2 P_2 +{7 \over 2} L_2^2 +3 L_2 
+4 P_1 -6 \zeta_2 -{1 \over 2} L_1^2 -2 \right)
\nonumber \\
&&
- x \BLB
{1 \over 2} \zeta_2 L_1 +{13 \over 2} \zeta_2 L_2 +{1 \over 2} \zeta_2
-9 \zeta_3 -{1 \over 12} L_1^3 - L_1 L_2^2 -4 L_2 P_2 -{35 \over 12} L_2^3 
\nonumber \\
&&
-{7 \over 4} L_2^2 -L_2 P_1 +3 L_2 +P_1 +2 R_2 -2 R_1 +8 R_4 +R_5 -2
\BRB
\\
{\tilde E}^{2,0}_{11110} &=&
{2 \over \ep^2} \left( x L_2 -x L_1 +1-x \right)
-{1 \over \ep} \BLB  2-2 x -4 x P_1 +6 x \zeta_2 -4 (1-x) L_2
\nonumber \\
&&
-2 x P_2 -(1+x) L_1 +{x \over 2} L_1^2 -{7 \over 2} x L_2^2 \BRB
+(1-x) \left( 4 L_2^2 -4 L_2 -{3 \over 2} \zeta_2 +2 \right)
\nonumber \\
&&
+(1+x) \left( 2 P_2 -L_1 +2 L_1 L_2 \right)
-8 x R_4 +2 x R_1 -x R_5 -2 x R_2
+9 \zeta_3 x 
\nonumber \\
&&
-{1-3 x \over 4} L_1^2 +{35 \over 12} x L_2^3 
+{x \over 12} L_1^3 -{x \over 2} \zeta_2 L_1 +3 x L_2 P_2
-{11 \over 2} x \zeta_2 L_2
\\
{\tilde E}^{1,1}_{21110} &=&
{2 \over \ep^2} \left( {\cal D}_0(x)+\delta(1-x)-1 \right)
-{2 \over \ep} \BLB 3 L_2 +{x \over 1-x} L_1
-3 {\cal D}_1(x)-{\cal D}_0(x)
\nonumber \\ &&
+1+\left( 1+\zeta_2 \right) \delta(1-x) \BRB
+{x^2 \over 1-x} P_2
+\left( {7 \over 2}+3 x \right) \zeta_2
+(x-7) L_2^2 
\nonumber \\ &&
+ x P_1-2 L_2
-x {4-x \over 1-x} L_1 L_2 +2
+\left( 2-{3 \over 2} \zeta_2+2 \zeta_3 \right) \delta(1-x)
\nonumber \\ &&
-\left( 2+{7 \over 2} \zeta_2 \right) {\cal D}_0(x)
+2 {\cal D}_1(x)+7 {\cal D}_2(x)
\\
{\tilde E}^{1,0}_{11111} &=& {1 \over 1-x} \BLB 
2 x P_3 -2 P_2 +2 x L_1 L_3 +{x^2 \over 1-x} L_1^2 -(2-3 x) \zeta_2
\BRB
\nonumber \\
&&
+2 (\zeta_2+\zeta_3) \delta(1-x)
+{(-1)^n \over 1+x} \BLB
-2 (x+2) P_3 -2 {1-x \over 1+x} P_2
\nonumber \\
&&
-2 (x+2) L_1 L_3 -(x+2) \zeta_2 +x {4+x \over 1+x} L_1^2
\BRB
\\
{\tilde E}^{0,1}_{11111} &=&
{1 \over 1-x} \BLB 2 (1+2 x) \left( L_1 L_3 +P_3 \right)
-(1-3 x) L_1 L_2 -(1-4 x) P_1 \BRB
-{1 \over 2} L_2^2 
\nonumber \\
&&
-{x (5-7 x) \over 2 (1-x)^2} L_1^2
+2 (\zeta_2+\zeta_3) \delta(1-x)
+{(-1)^n \over 1+x} \BLB {x (1-2 x) \over 1+x} L_1^2 +\zeta_2 x 
\nonumber \\
&&
+2 x L_1 L_3 +{2 x (1-x) \over 1+x} P_2 +2 x P_3
\BRB
\end{eqnarray}
\begin{eqnarray}
{\tilde E}^{1,1}_{11111} &=&
{1 \over 1-x} \BLB x (1+x) L_1 L_2 
+x (1+2 x) \left( 2 P_3 +2 L_1 L_3 +P_1 \right)
\BRB
-{x \over 2} L_2^2 
\nonumber \\
&&
+{x^2 (5 x-3) \over 2 (1-x)^2} L_1^2 -4 \zeta_2 x
+2 (\zeta_2+\zeta_3) \delta(1-x)
+
{(-1)^n \over 1+x} \BLB
{2 x (1-x) \over 1+x} P_2 
\nonumber \\
&&
-{x^2 (4+x) \over 1+x} L_1^2
+x (2+x) \left( 2 P_3 +2 L_1 L_3 +\zeta_2 \right)
\BRB
\\
{\tilde E}^{2,0}_{11111} &=&
{1 \over 1-x} \BLB (x^2 -2 x -1) P_2
+2 x^2 L_1 L_3 +(4 x^2-2 x-1) \zeta_2
+{x (1+x^2) \over 2 (1-x)} L_1^2
\nonumber \\
&&
+2 x^2 P_3
\BRB
+2 (\zeta_2+\zeta_3) \delta(1-x)
+{(-1)^n \over 1+x} \BLB 
{x (3-8 x-5 x^2) \over 2 (1+x)} L_1^2
\nonumber \\
&&
+(2 x^2+2 x-1) \left( 2 P_3 +\zeta_2 +2 L_1 L_3 \right)
+(x^2+2 x-1) {1-x \over 1+x} P_2
\BRB
\end{eqnarray}


The $F$-type integrals are given by

\begin{eqnarray}
{\tilde F}^{0,0}_{01111} &=&
{1 \over (1-x)} \BLB
-{2 \over \ep^2} L_2 
-{1 \over \ep} \left( {1 \over 2} L_2^2-4 P_2+4 \zeta_2 \right)
-{5 \over 2} \zeta_2 L_2 
+2 \zeta_3
\nonumber \\
&&
-{1 \over 12} L_2^3+L_2 P_2+2 R_2-4 R_5
\BRB 
\\
{\tilde F}^{1,0}_{01111} &=& x {\tilde F}^{0,0}_{01111}
\\
{\tilde F}^{0,1}_{01111} &=&
{2 \over \ep^2} \left( L_2 +{x \over 1-x} \right)
-{1 \over \ep} \BLP 4 P_2 
-{1 \over 1-x} \left( 4 x L_1+(2-x) L_2-2 x \right)
-4 \zeta_2
\nonumber \\
&&
-{1 \over 2} L_2^2 \BRP
+{5 \over 2} \zeta_2 L_2-2 \zeta_3+{1 \over 12} L_2^3+4 R_5-2 R_2-L_2 P_2
-{4-x \over 1-x} P_2
\nonumber \\
&&
+{1 \over 1-x} \BLP {8-3 x \over 2} \zeta_2 +{2-x \over 4} L_2^2
+(2-3 x) L_2+x \left( L_1 L_2+2 L_1^2+2 \right) \BRP
\\
{\tilde F}^{1,1}_{01111} &=& x {\tilde F}^{0,1}_{01111}
\\
{\tilde F}^{2,0}_{01111} &=& x^2 {\tilde F}^{0,0}_{01111}
\\
{\tilde F}^{0,2}_{01111} &=&
-{1 \over \ep^2} \left[ 2 (1-x) L_2+{x (2-3 x) \over 1-x} \right]
-{1 \over \ep} \BLB (1-x) \left( {1 \over 2} L_2^2+4 \zeta_2-4 P_2 \right)
\nonumber \\
&&
+{3 x^2-10 x+6 \over 2 (1-x)} L_2+{2 x (2-3 x) \over 1-x} L_1
+{x (3 x-1) \over 1-x} \BRB
-{2 x \over 1-x} L_1
\nonumber \\
&&
-{3 x^2-10 x+6 \over 8 (1-x)} L_2^2
-{9 x^2-12 x+5 \over 2 (1-x)} L_2 
-{9 x^2-38 x+24 \over 4 (1-x)} \zeta_2
\nonumber \\
&&
-{x (2-3 x) \over 1-x} \left( L_1^2+{1 \over 2} L_1 L_2 \right)
+{3 (x^2-6 x+4) \over 2 (1-x)} P_2
-{x (1-6 x) \over 2 (1-x)}
\nonumber \\
&&
+(1-x) \BLB 2 \zeta_3-{5 \over 2} \zeta_2 L_2
-{1 \over 12} L_2^3
+L_2 P_2+2 R_2-4 R_5 \BRB 
\\
{\tilde F}^{0,0}_{10111} &=&
{1 \over 1-x} \left( {4 \over \ep^2}-{4 \over \ep} \left( 1-L_2 \right)
+2 L_2^2-4 L_2+\zeta_2+4 \right) 
\\
{\tilde F}^{1,0}_{10111} &=& x {\tilde F}^{0,0}_{10111}\\
{\tilde F}^{0,1}_{10111} &=&
{1 \over 1-x} \BLB -{2-4 x \over \ep^2} 
-{1 \over \ep} \left( (2-4 x) L_2+4 x-1 \right)
-{1-2 x \over 2} \zeta_2
\nonumber 
\end{eqnarray}
\begin{eqnarray}
&&
-(1-2 x)L_2^2
+(1-4 x) L_2
+4 x-{1 \over 2}
\BRB 
\\
{\tilde F}^{1,0}_{11011} &=&
{x \over 1-x} \BLB -{2 \over \ep^2} L_1-{5 \over 2 \ep} L_1^2
-{1 \over 2} \zeta_2 L_1-2 \zeta_3+2 L_1 P_1
-{7 \over 12} L_1^3
\nonumber \\
&&
+4 L_1 P_3-8 R_3-4 R_1 \BRB 
\\
{\tilde F}^{0,1}_{11011} &=&
-{2 \over \ep^2} \left[ 1+{x \over 1-x} L_1 \right]
-{1 \over \ep} \left[ {5 x \over 2 (1-x)} L_1^2
+{1-3 x \over 1-x} L_1-2 \right]
\nonumber \\
&&
+{x \over 1-x} \BLB 2 L_1 P_1+4 L_1 P_3-4 R_1-8 R_3
-{1 \over 2} \zeta_2 L_1-2 \zeta_3-{7 \over 12} L_1^3 \BRB
\nonumber \\
&&
-{1 \over 2} \zeta_2
+{1+9 x \over 4 (1-x)} L_1^2
+{1-3 x \over 1-x} L_1-2 
\\
{\tilde F}^{1,1}_{11011} &=& x {\tilde F}^{0,1}_{11011}
\\
{\tilde F}^{2,0}_{11011} &=& x {\tilde F}^{1,0}_{11011}
\\
{\tilde F}^{0,2}_{11011} &=&
{1 \over \ep^2} \left( 1-3 x-{2 x^2 \over 1-x} L_1 \right)
-{1 \over \ep} \left( {5 x^2 \over 2 (1-x)} L_1^2 
-{9 x^2-4 x+1 \over 2 (1-x)} L_1-3 x \right)
\nonumber \\
&&
+{27 x^2+4 x-1 \over 8 (1-x)} L_1^2
-{9 x^2-3 x+1 \over 2 (1-x)} L_1
+{x^2 \over 1-x} \BLP 4 L_1 P_1+4 L_1 P_3
\nonumber \\
&&
+L_1^2 L_2
-6 R_1-8 R_3-2 R_5-{1 \over 2} \zeta_2 L_1
-{7 \over 12} L_1^3 \BRP
+{1-3 x \over 4} \zeta_2
-{1 \over 2}-3 x
\\
{\tilde F}^{0,0}_{11101} &=& 
{1 \over 1-x} \left[ {2 \over \ep^2}
-{1 \over \ep} \left( L_1-4 L_2+2 \right)
+{1 \over 2} \zeta_2+2 L_2^2+{1 \over 4} L_1^2
+{1+3 x \over 1-x} L_1+2 \right]
\\
{\tilde F}^{1,0}_{11101} &=& x {\tilde F}^{0,0}_{11101}
\\
{\tilde F}^{0,1}_{11101} &=& -{1 \over 2} (1-x) {\tilde F}^{0,0}_{11101}
\\
{\tilde F}^{1,1}_{11101} &=& x {\tilde F}^{0,1}_{11101}
\\
{\tilde F}^{2,0}_{11101} &=& x^2 {\tilde F}^{0,0}_{11101}
\\
{\tilde F}^{1,0}_{11110} &=&
{x \over 1-x} \left( {2 \over \ep^2}
-{1 \over \ep} \left( 2-3 L_2 \right)
-{1 \over 2} \zeta_2+{7 \over 4} L_2^2-3 L_2-P_1 +2\right)
\\
{\tilde F}^{1,1}_{11110} &=& {x \over 2} {\tilde F}^{1,0}_{11110} -{x \over 2 (1-x)} 
L_2
\\
{\tilde F}^{2,0}_{11110} &=& x {\tilde F}^{1,0}_{11110}
\\
{\tilde F}^{0,1}_{11110} &=& {1 \over 2} {\tilde F}^{1,0}_{11110}-{1 \over 2 (1-x)} L_2
\\
{\tilde F}^{1,0}_{21110} &=&
-x (1-x)^{-2+\ep} \left[ -{2 \over \ep}+2-\left( {\zeta_2 \over 2}+2 \right) \ep \right]
+{x \over 1-x} \left( \zeta_2+P_1+{1 \over 2} L_2^2 \right)
\\
{\tilde F}^{1,0}_{11111} &=&
{x \over 1-x} \BLB -{1 \over 2} L_2^2
+2 {1+x \over 1-x} \left( L_1 L_3+P_3 \right)
+{1 \over 1-x} \BLP (3 x-1) \zeta_2-x L_1 L_2
\nonumber \\
&&
-{3 \over 2} x L_1^2+(1-2 x) P_2 \BRP \BRB
\\
{\tilde F}^{0,1}_{11111} &=& {\tilde F}^{1,0}_{11111}-{1 \over 1-x} L_2-{x \over (1-x)^2} L_1
\\
{\tilde F}^{1,1}_{11111} &=& x {\tilde F}^{0,1}_{11111}
\\
{\tilde F}^{2,0}_{11111} &=& x {\tilde F}^{1,0}_{11111}
\end{eqnarray}


\end{document}